\documentclass[11pt,a4paper]{article}
\usepackage[left=2.35cm,top=3cm,bottom=3cm,right=2.3cm]{geometry}

\usepackage{authblk}

\usepackage[normalem]{ulem}

\usepackage{amsmath}
\usepackage{amssymb}
\usepackage{xcolor}
\usepackage{cite}
\usepackage{graphicx}
\usepackage{subcaption}
\usepackage{slashed}
\usepackage{mathtools}
\usepackage{multirow}
\usepackage{threeparttable}

\usepackage{makecell}
\usepackage{booktabs}
\newcommand{\pdforweb}[2]{#1}

\usepackage{tikz}
\usetikzlibrary{shapes.geometric}
\usetikzlibrary{decorations.pathmorphing,positioning,decorations.pathreplacing,shapes,calc}
\usetikzlibrary{decorations.pathreplacing}
\usetikzlibrary{decorations.markings}
\usetikzlibrary{arrows}
\tikzset{
    photon/.style={decorate, decoration={snake,amplitude=1.5pt,segment length=4pt}},
}

\allowdisplaybreaks

\usepackage[colorlinks=true
,urlcolor=blue
,anchorcolor=blue
,citecolor=blue
,filecolor=blue
,linkcolor=blue
,menucolor=blue
,linktocpage=true
,pdfproducer=medialab
]{hyperref}
\usepackage{orcidlink}

\usepackage{listings}

\DeclareUnicodeCharacter{03B3}{\ensuremath{\gamma}}

\newcommand{\afkqed}{{\sc AfkQed}}
\newcommand{\babayaga}{{\sc BabaYaga@NLO}}
\newcommand{\mcmule}{{\sc McMule}}
\newcommand{\phokhara}{{\sc Phokhara}}
\newcommand{\Sherpa}{{\sc Sherpa}}
\newcommand{\Comix}{{\sc Comix}}
\newcommand{\Amegic}{{\sc Amegic}}
\newcommand{\Photos}{{\sc Photos}}

\newcommand{\vp}{\mathsf{p}}  
\newcommand{\cO}{\mathcal{O}}
\newcommand{\cM}{\mathcal{M}}
\newcommand{\cA}{\mathcal{A}}

\newcommand{\GeV}{\mathrm{GeV}}
\newcommand{\MeV}{\mathrm{MeV}}
\newcommand{\rad}{\mathrm{rad}}

\newcommand{\dd}{\mathrm{d}}
\renewcommand{\Re}{\mathrm{Re}}
\renewcommand{\Im}{\mathrm{Im}}

\newcommand{\figref}[1]{Figure~\ref{#1}}
\newcommand{\tabref}[1]{Table~\ref{#1}}
\newcommand{\secref}[1]{Section~\ref{#1}}

\newcommand{\minidiagSize}[2]{\begin{minipage}{#2} \includegraphics[width=#2]{figs/#1} \end{minipage}}

\newcommand{\mpi}{m_\pi}

\numberwithin{equation}{section}

\title{ Radiative corrections and Monte Carlo tools for \\[5pt]
low-energy hadronic cross sections in $e^+ e^-$ collisions}

\date{}

\usepackage{authblk}

    \author[1]{RadioMonteCarLow 2 Working Group: Riccardo~Aliberti\,\orcidlink{0000-0003-3500-4012}\,}
    \author[2]{Paolo~Beltrame\,\orcidlink{0000-0001-9523-6128}\,}
    \author[3,4]{Ettore~Budassi\,\orcidlink{0009-0001-0999-0878}\,}
    \author[4]{Carlo~M.~Carloni~Calame\,\orcidlink{0000-0002-7315-0638}\,}
    \author[5]{Gilberto~Colangelo\,\orcidlink{0000-0003-3954-9503}\,}
    \author[2]{Lorenzo~Cotrozzi\,\orcidlink{0000-0002-0375-0611}\,}
    \author[1]{Achim~Denig\,\orcidlink{0000-0001-7974-5854}\,}
    \author[6,7]{Anna~Driutti\,\orcidlink{0000-0003-0771-5642}\,}
    \author[8]{Tim~Engel\,\orcidlink{0000-0003-2794-9032}\,}
    \author[2,9]{Lois~Flower\,\orcidlink{0000-0002-2786-228X}\,}
    \author[3,6,7]{Andrea~Gurgone\,\orcidlink{0000-0003-3700-4948}\,}
    \author[5]{Martin~Hoferichter\,\orcidlink{0000-0003-1113-9377}\,}
    \author[2]{Fedor~Ignatov\,\orcidlink{0000-0001-7061-6060}\,}
    \author[10,11]{Sophie~Kollatzsch\,\orcidlink{0000-0002-8560-1619}\,}
    \author[12]{Bastian~Kubis\,\orcidlink{0000-0002-1541-6581}\,}
    \author[13,14,*]{Andrzej~Kup\'s\'c\,\orcidlink{0000-0003-4937-2270}\,}
    \author[11,10]{Fabian~Lange\,\orcidlink{0000-0001-8531-5148}\,}
    \author[15,7]{Alberto~Lusiani\,\orcidlink{0000-0002-6876-3288}\,}
    \author[16]{Stefan~E.~M\"uller\,\orcidlink{0000-0001-6273-7102}\,}
    \author[2]{J\'er\'emy~Paltrinieri\,\orcidlink{0000-0003-4226-7056}\,}
    \author[2]{Pau~Petit~Ros\`as\,\orcidlink{0009-0009-8824-5208}\,}
    \author[4]{Fulvio~Piccinini\,\orcidlink{0000-0003-4378-7870}\,}
    \author[17]{Alan~Price\,\orcidlink{0000-0002-0372-1060}\,}
    \author[7,15]{Lorenzo~Punzi\,\orcidlink{0009-0000-5058-839X}\,}
    \author[10,18]{Marco~Rocco\,\orcidlink{0000-0002-2561-1209}\,}
    \author[19,20]{Olga~Shekhovtsova\,\orcidlink{0000-0002-4237-8170}\,}
    \author[17]{Andrzej~Si\'odmok\,\orcidlink{0000-0001-9614-7856}\,}
    \author[10,11,*]{Adrian~Signer\,\orcidlink{0000-0001-8488-7400}\,}
    \author[21]{Giovanni~Stagnitto\,\orcidlink{0000-0002-9513-5914}\,}
    \author[10,11]{Peter~Stoffer\,\orcidlink{0000-0001-7966-2696}\,}
    \author[2]{Thomas~Teubner\,\orcidlink{0000-0002-0680-0776}\,}
    \author[2]{William~J.~Torres~Bobadilla\,\orcidlink{0000-0001-6797-7607}\,}
    \author[3,4]{Francesco~P.~Ucci\,\orcidlink{0009-0005-4969-452X}\,}
    \author[2,5,*]{Yannick~Ulrich\,\orcidlink{0000-0002-9947-3064}\,}
    \author[2,7,*]{Graziano~Venanzoni\,\orcidlink{0000-0002-3525-476X}\,}
    \affil[1]{Institute for Nuclear Physics, Johannes Gutenberg University Mainz, Germany}
    \affil[2]{University of Liverpool, Liverpool L69 3BX, U.K.}
    \affil[3]{Dipartimento di Fisica, Universit\`a di Pavia, Via A. Bassi 6, 27100, Pavia, Italy}
    \affil[4]{INFN, Sezione di Pavia, Via A. Bassi 6, 27100, Pavia, Italy}
    \affil[5]{Albert Einstein Center for Fundamental Physics, Institute for Theoretical Physics, University of Bern, Sidlerstrasse 5, 3012 Bern, Switzerland}
    \affil[6]{Dipartimento di Fisica, Universit\`a di Pisa, Largo B. Pontecorvo 3, 56127, Pisa, Italy}
    \affil[7]{INFN, Sezione di Pisa, Largo B. Pontecorvo 3, 56127, Pisa, Italy}
    \affil[8]{Albert-Ludwigs-Universit\"at Freiburg, Physikalisches Institut, D-79104 Freiburg, Germany}
    \affil[9]{Institute for Particle Physics Phenomenology, Durham University, South Road, Durham, DH1 3LE, U.K.}
    \affil[10]{PSI Center for Neutron and Muon Sciences, 5232 Villigen PSI, Switzerland}
    \affil[11]{Physik-Institut, Universit\"at Z\"urich, 8057 Z\"urich, Switzerland}
    \affil[12]{Helmholtz-Institut f\"ur Strahlen- und Kernphysik (Theorie) and Bethe Center for Theoretical Physics, Universit\"at Bonn, 53115 Bonn, Germany}
    \affil[13]{Department of Physics and Astronomy, Uppsala University, Box 516, SE-75120 Uppsala, Sweden}
    \affil[14]{National Centre for Nuclear Research, Pasteura 7, 02-093 Warsaw, Poland}
    \affil[15]{Scuola Normale Superiore, Pisa, Italy}
    \affil[16]{HZDR Dresden, Germany}
    \affil[17]{Jagiellonian University, ul. prof. Stanislawa \L{}ojasiewicza 11, 30-348 Krak\'ow, Poland}
    \affil[18]{Universit\`a degli Studi di Torino \& INFN, Via Pietro Giuria 1, Torino 10125, Italy}
    \affil[19]{NSC KIPT Institute for Theoretical Physics, Kharkov, Ukraine}
    \affil[20]{INFN Sezione di Perugia, 06123, Perugia, Italy}
    \affil[21]{Universit\`a degli Studi di Milano-Bicocca \& INFN, Piazza della Scienza 3, Milano 20126, Italy}
    \affil[*]{Coordinator and corresponding author}

\begin{document}

\begin{titlepage}
\clearpage\maketitle

\thispagestyle{empty}
\begin{abstract}
We present the results of Phase~I of an ongoing review of Monte Carlo tools relevant for low-energy hadronic cross sections.  This includes a detailed comparison of Monte Carlo codes for electron--positron scattering into a muon pair, pion pair, and electron pair, for scan and radiative-return experiments. After discussing the various approaches that are used and effects that are included, we show differential cross sections obtained with \afkqed, \babayaga, KKMC, MCGPJ, \mcmule, \phokhara, and \Sherpa, for scenarios that are inspired by experiments providing input for the dispersive evaluation of the hadronic vacuum polarisation.
\end{abstract}
\end{titlepage}

\newpage

\tableofcontents


\setcounter{footnote}{0}

\section{Introduction}\label{sec:intro}

Monte Carlo codes are essential tools for the analysis of low-energy scattering experiments at electron--positron colliders. Accordingly, there is an extensive and long-standing effort by the community to provide and improve codes that are able to produce fully differential predictions for processes related to $e^+\,e^-\to\, \text{hadrons}$ at centre-of-mass energies up to a few GeV. Through the community effort described in this article we collect such tools and facilitate their access and usage. In addition, we present a comprehensive comparison of the physical effects included, methods used, and their likely effect on the accuracy of the theoretical predictions. We update the report of the \textit{Working Group on Radiative Corrections and Monte Carlo Generators for Low Energy}~\cite{WGRadCor:2010bjp} and highlight the developments of the past ten years. However, the report presented here is more focused. We restrict ourselves to the processes listed in \eqref{intro:scan} and \eqref{intro:return}, augmented by some remarks about $3\pi$ production and contrary to~\cite{WGRadCor:2010bjp} do not consider issues related to luminosity measurements. 

As in~\cite{WGRadCor:2010bjp} we adopt the concept of {\it tuned comparisons} when presenting the results of different Monte Carlo programmes. Such tuned comparisons use the same set of input parameters and sometimes codes are compared without vacuum-polarisation corrections. These comparisons must also use the same experimental cuts. For the latter, we include realistic acceptance selection, while we refrain from taking into account additional kinematic selection (like kinematic fit) and detector effects. Correspondingly, we do not compare to experimental data at all. At this stage, this is a purely theoretical effort and the presented results require careful interpretation. We consider the current article to be Phase~I of an ongoing community effort~\cite{Abbiendi:2022liz, Durham22, Zurich23, Mainz24} and plan for a long-term continuation of the programme.

One of the main reasons to consider these low-energy processes is of course their impact on the determination of the hadronic vacuum polarisation (HVP) corrections $a_\mu^\text{HVP}$ to the anomalous magnetic moment of the muon (also called muon $g-2$). We hope that a coordinated effort from the Monte Carlo community will help to shed light on possible shortcomings or future improvements of these codes and, hence, to their impact on data-driven calculations of the HVP contribution to the muon $g-2$. Thus, there is a close link to the Muon $(g-2)$ Theory Initiative~\cite{Aoyama:2020ynm}. However, there is also a justified interest in these processes as such. After all, the situation regarding the process $e^+\,e^-\to\pi^+\,\pi^-$, which provides the dominant contribution to $a_\mu^\text{HVP}$, is very unclear~\cite{Stoffer:2023gba}. There are large tensions among different experiments but also between most experiments and lattice results. A critical assessment of the Monte Carlo tools will be an important component for a better understanding of these fundamental issues.

There has been considerable progress in the evaluation of radiative corrections to scattering processes since~\cite{WGRadCor:2010bjp}. New groups have entered the field, either providing new tools for low-energy scattering processes or maintaining and further developing existing codes. In this article we consider in detail \afkqed, {\sc{Babayaga@NLO}}, {\sc{KKMCee}}, MCGPJ, \mcmule, \phokhara, and \Sherpa. There are other codes that can play an important role for the processes under consideration and we encourage their future inclusion. The codes considered here use partly overlapping and partly complementary approaches. This offers a rich environment to obtain a better understanding of the applicability and reliability of different approximations.

The core purpose of this work is to assess the importance of various contributions in the theoretical description of fully differential cross sections. More concretely, in Phase~I we are concerned with the $2\to{2}$ processes
\begin{subequations}\label{intro:scan}
\begin{align}
 e^+\,e^- &\to \pi^+\,\pi^-   \label{intro:scanPi}\,,  \\
 e^+\,e^- &\to \mu^+\,\mu^-  \label{intro:scanMu}\,, \\
 e^+\,e^- &\to e^+\,e^- \label{intro:scanE} \, ,
\end{align}
\end{subequations}
relevant for energy scan experiments and the $2\to{3}$ processes
\begin{subequations}\label{intro:return}
\begin{align}
 e^+\,e^- &\to \pi^+\,\pi^-\,\gamma  \label{intro:returnPi}\,,  \\
 e^+\,e^- &\to \mu^+\,\mu^-\,\gamma  \label{intro:returnMu}\,, \\
 e^+\,e^- &\to e^+\,e^-\, \gamma \label{intro:returnE} \, ,
\end{align}
\end{subequations}
relevant for radiative-return experiments. We refer to \eqref{intro:return} as radiative processes (sometimes they are also called initial-state radiation (ISR) processes) and it is understood that the photon is hard, i.e., it has sufficient energy to be detectable. Of course, the inclusion of higher-order corrections implies the consideration of additional -- possibly soft -- photons in the final state. For all six processes, we consider a collection of observables within several scenarios. The latter consist of particular centre-of-mass energies and acceptance cuts. 
Since we do not include detector effects nor additional kinematic selection, the results presented here are meant to illustrate a comparison between different theory approaches. They should not be used to compare directly with experimental data.

Our results can also serve as benchmark for future theory developments and as toolbox for experimental collaborations. In the spirit of open science, through\pdforweb{~\cite{repo}}{}
\begin{quote}     
\url{https://radiomontecarlow2.gitlab.io}
\end{quote} 
we make publicly available all source codes that have been used for this report. To ensure reproducibility, the exact configurations that have been used to obtain the results are also given. This includes the precise definition of the observables,  the input parameters (usually through a run card), the Monte Carlo version, and analysis and plotting pipelines. It is foreseen that this repository will be updated continuously and possibly extended with additional Monte Carlo codes and additional processes. We hope this facilitates the use of the tools by experimental collaborations as well as the theory community.

This article is not meant to be a complete review of all activities related to low-energy hadronic cross sections. It is only meant to provide the foundation for a detailed analysis of the current theory status for the processes listed in \eqref{intro:scan} and \eqref{intro:return} at energies up to $10\,\GeV$. \pdforweb{Accordingly, in \secref{sec:exp} we start with a brief overview over the colliders and experiments that are most relevant for these processes, as well as for $3\pi$ production.}{} The different computational approaches that are used for these processes are described in \secref{sec:comp}. This includes fixed-order QED calculations up to next-to-next-to-leading order (NNLO), approaches to include logarithmically enhanced higher-order terms through parton showers or soft-photon resummation, and the inclusion of non-perturbative hadronic effects due to insertions of vacuum polarisation (VP) or the hadronic light-by-light (HLbL) four-point function. The treatment of pions in the final state is rather delicate and, hence, receives particular attention. \pdforweb{With these methods in hand, \secref{sec:gen} provides a brief description of which contributions are included in the various Monte Carlo tools. \secref{sec:mcc} contains our main results. For five scenarios with different centre-of-mass energies and acceptance cuts we present detailed Monte Carlo comparisons. This is done for selected observables with electron, muon, and pion final states. A more comprehensive list of observables can be found in the repository~\cite{repo}. If two codes include the same higher-order corrections, the comparisons are used for the technical validation of the codes. Where different approaches have been taken, the comparison is used to study the importance of particular components of higher-order corrections. An executive summary of our findings and an outlook towards further improvement of the codes is given in \secref{sec:outlook}.}{}

\section{Experiments at electron--positron colliders}\label{sec:exp}

There are two classes of experiments at electron--positron colliders related to the processes \eqref{intro:scan} and \eqref{intro:return}.
Energy scan  experiments access cross sections of the processes \eqref{intro:scan} by setting the beam energies of a collider to a given centre-of-mass (c.m.) energy, $\sqrt{s}$, which can be achieved with good precision. The observed number of events for a given integrated luminosity is corrected for the acceptance and radiative effects due to, for example, soft photon emission and virtual corrections. To collect data at other $\sqrt{s}$ requires changes and adjustments of accelerator settings.
Radiative-return (also called ISR) experiments use \eqref{intro:return} to extract the energy dependence of the processes \eqref{intro:scan} with just a single working point $\sqrt{s}$ of the collider. The emission of a hard photon from initial state allows these experiments to collect events with continuously distributed invariant masses $w$ of the final-state system from threshold to the c.m. energy of the collider. The mass $w$ is determined by a measurement of the momenta of the final particles.

\begin{figure}[th]
\centering
\includegraphics[width=1.\linewidth]{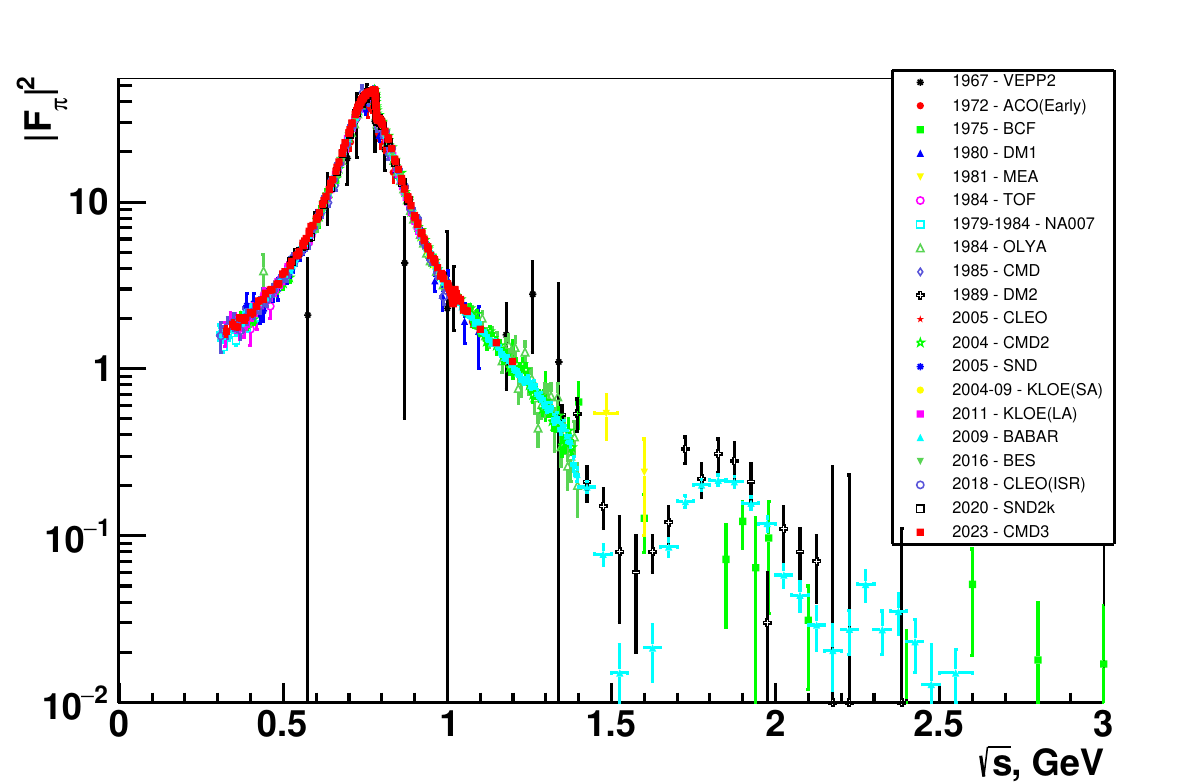}
\caption{Status of the $e^+e^-\to\pi^+\pi^-$ cross-section measurement
  with different experiments contributing over the years. Plot updated from~\cite{Ignatov:2018fss}.}
\label{fig-fpiyears}
\end{figure}

Many experiments have contributed to the measurements of the $e^+e^-\to\pi^+\pi^-$ channel over the years, as shown in \figref{fig-fpiyears}. Their use for a data-driven determination of the HVP is described in detail in~\cite{Aoyama:2020ynm}. 
Information about two-pion cross-section measurements is collected in the database
\begin{quote}
\url{https://precision-sm.github.io/2pi-db}.
\end{quote}
\tabref{table2pi} gives a summary of the experiments included in the database including the references and links to  the HEPData\footnote{\url{https://www.hepdata.net/}}  records.

\begin{table}[hp]
  \centering 
  
 \begin{tabular}{llclr}
\bf Accelerator  &\bf Exp.&\bf Year&\bf References                            & \bf HEPData                                                  \\\hline\hline
BEPC             & BESIII & 2016   & \cite{BESIII:2015equ}                    & \href{https://www.hepdata.net/record/ins1385603}{ins1385603} \\
(Beijing)                 &        &                                          &                                                              \\\hline
SLAC             & BaBar  & 2012   & \cite{BaBar:2012bdw}                     & \href{https://www.hepdata.net/record/ins1114155}{ins1114155} \\
(Standford U.)   &        &        &                                          &                                                              \\\hline
CESR             & CLEO   & 2018   & \cite{Xiao:2017dqv}                      & \href{https://www.hepdata.net/record/ins1643020}{ins1643020} \\
(Cornell U.)     &        & 2013   & \cite{Seth:2012nn}                       & \href{https://www.hepdata.net/record/ins1189656}{ins1189656} \\
                 &        & 2005   & \cite{CLEO:2005tiu}                      & \href{https://www.hepdata.net/record/ins693873}{ins693873}   \\\hline
DAPHNE           & KLOE   & 2017   & \cite{KLOE-2:2017fda}                    &                                                              \\
(LNF)            &        & 2012   & \cite{KLOE:2012anl}                      &                                                              \\
                 &        & 2010   & \cite{KLOE:2010qei}                      & \href{https://www.hepdata.net/record/ins859660}{ins859660}   \\
                 &        & 2008   & \cite{KLOE:2008fmq}                      & \href{https://www.hepdata.net/record/ins797438}{ins797438}   \\
                 &        & 2004   & \cite{KLOE:2004lnj}                      & \href{https://www.hepdata.net/record/ins655225}{ins655225}   \\\hline
Adone            & MEA    & 1980   & \cite{Esposito:1980bz}                   & \href{https://www.hepdata.net/record/ins158283}{ins158283}   \\
(LNF)            &        & 1977   & \cite{Esposito:1977xg}                   & \href{https://www.hepdata.net/record/ins124109}{ins124109}   \\
                 & BCF    & 1975   & \cite{Bollini:1975pn}                    & \href{https://www.hepdata.net/record/ins100180}{ins100180}   \\\hline
CERN             & NA007  & 1984   & \cite{Amendolia:1983di}                  & \href{https://www.hepdata.net/record/ins195944}{ins195944}   \\\hline
ACO              &        & 1976   & \cite{Cosme:1976ft}                      & \href{https://www.hepdata.net/record/ins109771}{ins109771}   \\
(Orsay)          &        & 1972   & \cite{Benaksas:1972ps}                   & \href{https://www.hepdata.net/record/ins73648}{ins73648}     \\\hline
DCI              & DM2    & 1989   & \cite{DM2:1988xqd}                       & \href{https://www.hepdata.net/record/ins267118}{ins267118}     \\
(Orsay)          & DM1    & 1978   & \cite{Quenzer:1978qt}                    & \href{https://www.hepdata.net/record/ins134061}{ins134061}   \\\hline
VEPP-2000   & CMD-3 & 2023 & \cite{CMD-3:2023alj} &     \\
(Novosibirsk) &  SND   & 2021   & \cite{SND:2020nwa}                       & \href{https://www.hepdata.net/record/ins1789269}{ins1789269} \\\hline
   VEPP-2M          &  SND      & 2005   & \cite{Achasov:2005rg}; erratum:~\cite{Achasov:2006vp} &\href{https://www.hepdata.net/record/ins686349}{ins686349}  \\
           
        (Novosibirsk)             & CMD-2   & 2007   & \cite{CMD-2:2006gxt}                     & \href{https://www.hepdata.net/record/ins728302}{ins728302}   \\
                 &        & 2006   & \cite{Aulchenko:2006dxz}                 & \href{https://www.hepdata.net/record/ins728191}{ins728191}   \\
                 &        & 2005   & \cite{CMD-2:2005mvb}                     & \href{https://www.hepdata.net/record/ins712216}{ins712216}   \\
                 &        & 2002   & \cite{CMD-2:2001ski}; erratum:~\cite{CMD-2:2003gqi}                     & \href{https://www.hepdata.net/record/ins568807}{ins568807}   \\
                 &  OLYA  & 1984   & \cite{Barkov:1985ac}                     & \href{https://www.hepdata.net/record/ins221309}{ins221309-Table1}   \\
                 &   CMD     & 1983   & \cite{Barkov:1985ac}                   & \href{https://www.hepdata.net/record/ins221309}{ins221309-Table2}   \\
                 & TOF    & 1981   & \cite{Vasserman:1981xq}                   & \href{https://www.hepdata.net/record/ins167191}{ins167191}   \\
                 
 \hline                
 VEPP-2   & VEPP-2    & 1972   &  \cite{Balakin:1972vg}                  &  \href{https://www.hepdata.net/record/ins75634}{ins75634}    \\
(Novosibirsk) &  & 1971   & \cite{Balakin:1971zh}                    & \href{https://www.hepdata.net/record/ins69313}{ins69313}     \\
      &        & 1969   & \cite{Auslander:1969ak}                 & \href{https://www.hepdata.net/record/ins57008}{ins57008}     \\
                 &        & 1967   &   \cite{Auslander:1967xma} &\href{https://www.hepdata.net/record/ins1392895}{ins1392895} \\
      \hline
       \hline
\end{tabular}

  \caption{Summary of the published $e^+e^- \rightarrow \pi^+\pi^-$ measurements with links to the datasets.}\label{table2pi}
\end{table}

\begin{table}[b]
\renewcommand{\arraystretch}{1.5}
\centering
\begin{tabular}{c||c|c}
 & Detected ``photon"    & Undetected ``photon"\\
\hline\hline
\multirow{3}{*}{Normalisation to $e^+e^-$} & KLOE10 ($1.0\,$GeV, $\pi^+\pi^-$) & KLOE08 ($1.020\,$GeV, $\pi^+\pi^-$) \\
 & BES-III ($3.773\,$GeV, $\pi^+\pi^-$) & BaBar ($10.580\,$GeV, $p\overline{p}$)\\
 & BaBar ($10.580\,$GeV, most channels) & \\
\hline
\multirow{3}{*}{Normalisation to $\mu^+\mu^- \gamma$} & BaBar ($10.580\,$GeV, $\pi^+\pi^-$) & KLOE12 ($1.020\,$GeV, $\pi^+\pi^-$)\\
 & CLEO-c ($3.671\,$GeV, $\pi^+\pi^-$)  &  \\
\end{tabular}
\caption{
Summary of the choices made by radiative-return experiments, comparing the choice of normalisation process and whether a calorimeter-detected photon is required. All energies are given in the c.m. system.}
\label{tab:ISRexp}
\end{table}
There are variations of the radiative-return experiments which have different sensitivity to the radiative  corrections. The configurations adopted by the different experiments are shown in  \tabref{tab:ISRexp}. All experiments fall into one of two categories regarding the process used for normalising the $e^+ e^- \to \pi^+ \pi^-$ cross section: either $e^+ e^- \to e^+ e^-$ is used, or $e^+ e^- \to \mu^+ \mu^- \gamma$. The table also details whether each experiment requires explicit detection of a photon in the calorimeter, which will impact the sensitivity to radiative corrections.

In the following we give an overview of the main experiments ordered by collider c.m. energy while for more details we refer to \cite{WGRadCor:2010bjp} and the literature.

\subsection{BINP, Novosibirsk: VEPP-2M,VEPP-2000 and SND, CMD-2, CMD-3}

The VEPP-2M electron--positron collider in Novosibirsk~\cite{Tumaikin:1977su,Shatunov:2000zc} has been in
use for more than 25 years, starting from 1974, with several generations of detectors.
The latest cycle of experiments from 1992 to 2000 was performed by the CMD-2~\cite{Anashkin:2006zz} and SND~\cite{Achasov:1999ju} detectors, installed in two interaction
regions of the VEPP-2M collider. The peak luminosity reached by the collider was $3\times 10^{30}~\text{cm}^{-2}\text{s}^{-1}$,
and the two detectors together collected about 60 pb$^{-1}$ of the total
integrated luminosity, covering the energy range from 0.36\,GeV up to 1.38\,GeV.
Both detectors are general purpose detectors which include central tracking systems, calorimetry and
additional auxiliary subsystems such as a muon veto. CMD-2 was operated
in a 1\,T magnetic field, while SND was non-magnetic and was not
able to distinguish charges of particles. Instead, the SND
detector has the advantage in detecting neutral modes, with a more comprehensive 3-layer spherical electromagnetic calorimeter.
These scan experiments produced data on $e^+e^-$
annihilation to hadrons, covering the production from
$\pi^0\gamma$, $\pi^+\pi^-$ to $4\pi$, and $K\bar K$ final states~\cite{Ignatov:2008zz}.
The two-pion measurements were based
on about $1.1\times10^6$ and $4.5\times10^6$ selected $\pi^+\pi^-$
events by CMD-2 and SND, respectively. The achieved precision on
the combined total hadronic cross section $R(s)$ was about 1\% at c.m. energies around the $\rho$-resonance and
3.5\% at 1.38\,GeV. The systematic uncertainties in the main $e^+e^-\to\pi^+\pi^-$ channel
were 0.6\% for the CMD-2 measurement and 1.3\% for SND.
The statistics of the two experiments were limited and contributed almost as much as the systematic uncertainty to the overall accuracy
of the dispersive integral for $a_\mu^{\mathrm{had}}$.
As a result of both experiments, the accuracy of the value $a_\mu^{\text{HVP}}$ at that time was improved by a factor of 3.

The VEPP-2M collider was decommissioned in 2000 to make way for the
new storage ring VEPP-2000~\cite{Shatunov:2016bdv,Shwartz:2016fhe}, which started to provide data for
new experiments in 2010. The machine covers the wider c.m. energy range from $\sqrt{s}$ = 0.32\,GeV to 2.0\,GeV. It employs the novel technique of round beams to reach a
luminosity of up to $10^{32}~\text{cm}^{-2}\text{s}^{-1}$ at 2\,GeV, a world record of single
bunch luminosity at these low energies.
Today, VEPP-2000 is the only collider able to scan energies below 2\,GeV for the
measurement of exclusive $e^+e^-\to \text{hadrons}$ channels.  A special
system based on the Compton backscattering of laser photons is used to measure the beam
energy with relative accuracy better than $10^{-4}$~\cite{Abakumova:2012pn,Abakumova:2013fsa}.
Two new generation detectors, \mbox{CMD-3}~\cite{Aulchenko:2001je,Khazin:2008zz}
and SND~\cite{Achasov:2009zza}, are installed at opposite interaction regions of
the collider. A major upgrade of all subsystems was performed, including
completely new modernised electronics and more elaborate triggers in
comparison to previous experiments.
For example, in the case of the CMD-3 detector, a new drift
chamber, which provides higher efficiency and more than twice better the momentum resolution, and
a new LXe calorimeter, with multi-layer tracking capabilities and shower profile measurement, were constructed.

The main goals of experiments at VEPP-2000 include the high-precision measurement of cross sections of various
modes of $e^+e^- \to \text{hadrons}$ in the whole available c.m. energy range up to 2\,GeV.
All major channels are under analysis with final states of up to 7 pions, or 2 kaons and
3 pions~\cite{Ignatov:2019omb}. Many results have
already been published by the CMD-3 and SND experiments, but many more are
still under analysis.

The most demanding final state, due to the required precision, is the  $e^+e^-\to \pi^+\pi^-$ process.
The first energy scan below 1\,GeV for the $\pi^+\pi^-$ measurement was performed at the VEPP-2000 collider in 2013, which collected an integrated luminosity of 17.8\,pb$^{-1}$.
In 2014--2016, there was a long shutdown for the collider and detector upgrades.
In particular, a new electron and positron injector facility was commissioned,
which allowed a significant increase in luminosity.
The next energy scan in the $\rho$-meson c.m. energy region was carried out during
the 2017--2018 data-taking season, when about 45.4\,pb$^{-1}$ were collected.
At the end of 2019, an additional 1\,pb$^{-1}$ data sample was
collected near the threshold region at c.m. energies $\sqrt{s}<0.6$\,GeV.
The latest scan below 1\,GeV was performed in the first
half of 2024 in the energy range between the $\omega$ and $\phi$ resonances.

The first measurement of the $e^+e^-\to \pi^+\pi^-$ cross section
at VEPP-2000 was presented by the SND experiment, covering the energy
range $0.525<\sqrt{s}<0.883\,\GeV$ with a systematic
uncertainty of about 0.8\%~\cite{SND:2020nwa}. This result was based
on partial data of the first $\rho$ scan, corresponding to about 10\% of the total collected statistics below 1\,GeV.
The CMD-3 experiment has recently performed the full statistics analysis from 0.32 to 1.2\,GeV
with the conservative estimation of systematic uncertainty 0.7\% in the
dominant $\rho$-resonance region~\cite{CMD-3:2023alj,CMD-3:2023rfe}.
The CMD-3 analysis was based on the largest ever dataset at the $\rho$-resonance region, with $34\times10^{6}$ selected $\pi^+\pi^-$ events at $\sqrt{s}<1\,\GeV$.
The large statistics were crucial to study various systematic effects in detail.
The main features of the analysis include three independent
procedures for measuring the number of detected $\pi^+\pi^-$ events:
using momentum distributions of two particles measured in the tracking
system, or using detected energy depositions in the LXe calorimeter, or using the
polar angular distribution. Two of these methods, using momenta and angles of tracks, relied heavily on
differential cross section predictions from Monte Carlo generators.
The collected statistics allowed to perform comparisons of
measured momentum distributions with the predicted spectra from the Monte Carlo,
where a discrepancy in tails was observed. This was traced to the limitations of
a collinear jet approximation used in the MCGPJ generator, and
required upgrading the generator to take into account the angular distributions of photons in jets.
Another feature of the $\pi^+\pi^-$ analysis by CMD-3 was a comprehensive
study of the detector acceptance systematic uncertainty due to the
determination of the polar angle.
The forward--backward charge asymmetry of $\pi^+\pi^-$  was measured with an integrated
statistical precision of about 0.025\%. A 1\% level deviation of
the data from the theoretical predictions was observed. This highlighted a limitation of the commonly used
sQED approach for the calculation of radiative
corrections from the pion final state~\cite{Ignatov:2022iou,Colangelo:2022lzg}.

The overall collected integrated luminosity per detector surpassed the
projected value of 1\,fb$^{-1}$ in spring 2024.
The experiments will continue to collect data in the current
configurations for the next few years. After this, a moderate
upgrade of the detector subsystems is expected.
In the longer term, further collider upgrades
are being discussed to achieve even higher luminosity in the threshold region, and new dedicated detectors are foreseen with the potential to improve
the systematic accuracies of the cross-section measurement.

\subsection{LNF-INFN, Frascati: DA\texorpdfstring{$\Phi$}{Ph}NE and KLOE} \label{sec:exp-frascati}
DA$\Phi$NE at Frascati LNF-INFN is an electron--positron collider optimised to run at the c.m. energy corresponding to the $\phi(1020)$ meson mass.
KLOE is a multipurpose detector for the DA$\Phi$NE collider. Its
main component is a cylindrical drift chamber of 3.3\,m length
and 2\,m diameter, with an internal radius of 25\,cm. Together
with a lead-scintillating fibre electromagnetic calorimeter the
chamber is embedded in the 0.52\,T field of a superconducting
solenoid. For a more detailed description of the detector we
refer to~\cite{Adinolfi:2002uk, Adinolfi:2002zx,
KLOE:2002mvh,Adinolfi:2002me,Aloisio:2004ig}. KLOE took data from
2000 to 2006 and acquired 2.5\,fb$^{-1}$ of data at the $\phi$
peak, plus a further 250\,pb$^{-1}$ at other energies.
Reviews of the KLOE experimental results using data collected until 2005, including analyses finished until 2008, are given in~\cite{Franzini:2006aa,Bossi:2008aa}, while the physics programme for KLOE-2 was laid out in~\cite{Amelino-Camelia:2010cem}.

KLOE made use of the radiative-return method for the measurement of the hadronic cross section, focusing in particular on the $\pi\pi$ hadronic channel. 
 KLOE published four hadronic cross section measurements \cite{KLOE:2004lnj,KLOE:2008fmq,KLOE:2010qei,KLOE:2012anl}, which for convenience are called KLOE05\footnote{This measurement was superseded by KLOE08.}, KLOE08, KLOE10, and KLOE12, respectively. The KLOE08 analysis measures the differential cross section for $e^+e^-\rightarrow \pi^+\pi^-\gamma$ as a function of the $\pi^+\pi^-$ invariant mass, $s_{\pi\pi}$, for radiative events. The di-pion cross section $\sigma(e^+e^-\rightarrow \pi^+\pi^-)$, which for shorthand we write as $\sigma_{\pi\pi}$, is derived using
\begin{equation}
    \label{eq:radiator}
    s\frac{d\sigma(ee\to\pi\pi\gamma)}{ds_{\pi\pi}}\Bigr|_{\text{ISR}} =
    \sigma_{\pi\pi}(s_{\pi\pi})H(s_{\pi\pi},s)\, ,
\end{equation}
where $H(s_{\pi\pi},s)$ is the radiator function. In the KLOE08 analysis, $H(s_{\pi\pi},s)$ was obtained with the \phokhara{} (Version 5) Monte Carlo generator.

KLOE08 was performed using 240.0\,pb$^{-1}$ of on-peak data ($\sqrt{s} = 1019.48\,\MeV$) corresponding to about 3 million events. Small-angle selection cuts were applied in the analysis: photons were restricted to a cone of $\theta_\gamma<15^\circ$ ($>165^\circ$) around the beam line. The two charged tracks, on the other hand, were detected in the range $50^\circ < \theta^\pm < 130^\circ$ and any photons in this region were not detected.

The measurement of the $\pi\pi\gamma$ cross section was normalised to the DA$\Phi$NE luminosity using large-angle Bhabha scattering (using the \babayaga{} Monte Carlo generator \cite{Balossini:2006wc}), with 0.3\% total systematic uncertainty. The pion vector form factor (VFF) $F_\pi(s)$ and $a_\mu^{\text{HVP}}$ were derived using 60 points in the $s_{\pi\pi}$ region between 0.35 and 0.95\,GeV$^2$.

KLOE10 used so-called large-angle cuts: both the photon and the charged tracks are detected at large angles $50^\circ < \theta^\pm , \theta_{\gamma} < 130^\circ$. With this selection it is possible to reach the dipion threshold, but at the price of reduced signal yield and enhancing FSR and $\phi\to\pi^+\pi^-\pi^0$ backgrounds. Therefore the KLOE10 analysis was performed using 232.6 pb$^{-1}$ of data taken at the c.m. energy $\sqrt{s} = 1\,\GeV$, corresponding to 0.6 million events. The analysis of  $a_\mu^{\pi\pi}$ has 75 points in the region (0.10--0.85)\,GeV$^2$.

KLOE08 and KLOE10 were both normalised to the DA$\Phi$NE luminosity, and used the radiator function to obtain the pion VFF and $a_\mu^\text{HVP}$. KLOE12, on the other hand, was normalised with respect to the muon radiative differential cross section. In this approach many systematic effects cancel out, including some related to theoretical uncertainties. KLOE12 was published using the same 240\,pb$^{-1}$ data sample from 2002 used in KLOE08, and implemented identical small-angle cuts. Like KLOE08, the differential cross section was measured across 60 points in the energy region between 0.35 and 0.95\,$\GeV^2$.

While KLOE08 and KLOE12 used KLOE on-peak data from 2002, there exists around 1.7\,fb$^{-1}$ of on-peak data from 2004 and 2005 which has never been analysed for the measurement of the $2\pi$ cross section. A new experimental effort has begun with the aim of measuring $a_\mu^{\pi\pi}$ with greater precision by using the full 2004--2005 dataset.  In order to minimise biases from the published analysis, the new analysis will be conducted blindly.
In addition, KLOE is analysing the three-pion cross section using the radiative return method with 1.7\,fb$^{-1}$ of data collected at the $\phi$ meson mass~\cite{Cao:2020jus}.

\subsection{IHEP, Beijing: BEPCII and BESIII }

The BESIII experiment~\cite{BESIII:2009fln} is located at Institute for High Energy Physics (IHEP) in Beijing. It records symmetric e$^{+}$e$^{-}$ collisions provided by the BEPC-II collider with c.m. energies in the range between 1.8 and 5\,GeV and luminosity exceeding $10^{33}$\,cm$^{-2}$s$^{-1}$.
The BESIII detector adopts an onion-shape structure, which allows to cover 93\% of the full solid angle. It consists of a spectrometer, based on a multilayer drift chamber and (starting from fall 2024) three cylindrical gas electron multiplier layers~\cite{BESIII:CGEM2022ktq} in the inner region; a time-of-flight system made of plastic scintillators; a CsI(Tl) electromagnetic calorimeter; and a muon identification system provided by resistive plate chambers, which instrument the return-flux joke of the superconducting magnet, enclosing the detector and providing a 1\,T magnetic field.
Starting from 2008, BESIII has collected the world's largest datasets of electron--positron collisions in the $\tau$-charm energy region, accounting for about 50\,fb$^{-1}$.
Highlights of the BESIII datasets are:
\begin{itemize}
    \item $10^{10}$ $J/\psi$ decays,
    \item $2.7 \times 10^9$ $\psi(3686)$ decays,
    \item $20$\,fb$^{-1}$ at the $\psi(3770)$ resonance,
    \item 130 scan points between 2 and 4.6\,GeV.
\end{itemize}
In summer 2024, the BEPCII accelerator underwent an upgrade, in order to extend the energy range over which peak luminosity can be reached. Profiting from the increased luminosity, the BESIII Collaboration plans to collect large data samples, mainly in the energy region above 4\,GeV~\cite{BESIII:2020nme}.

The BESIII Collaboration is very active in hadronic cross section measurements for the muon $(g-2)$ effort.
In 2015, a first measurement of the $e^+e^- \to \pi^+\pi^-$ cross section with 0.9\% systematic uncertainty has been achieved~\cite{BESIII:2015equ}, based on a data sample of 2.9\,fb$^{-1}$ collected at 3.773\,GeV.
The measurement used the radiative return technique to access the energy region 600--900\,MeV, thus including the dominant contribution from the $\rho$ resonance.
In this analysis, the selection cuts required both pion tracks to be reconstructed in the drift chamber, and for the hard photon to be detected at large angle in the electromagnetic calorimeter.
A kinematic fit was then performed to constrain the four-momentum of the reconstructed $\pi^+\pi^-\gamma$ final state to the centre-of-mass energy.
For this result, the event yield was normalised to the integrated luminosity, obtained by measuring Bhabha events, and the radiator function in \eqref{eq:radiator} was obtained using the \phokhara{} event generator.

Preliminary radiative-return results have also been presented for the three-pion channel~\cite{BESIII:2019gjz}. Furthermore, a series of cross section measurements of multi-hadronic states have been carried out by energy scan above 2\,GeV.
Most notably, the $e^+e^- \to \pi^+\pi^-\pi^0$ cross section has been measured in the energy range between 2.0 and 3.08\,GeV~\cite{BESIII:2024okl}. In addition, a precision measurement of the total hadronic cross section has recently been measured with the world's highest accuracy via an energy scan between 2.23 and 3.08\,GeV~\cite{BESIII:2021wib}.

In the future, a new precision analysis of the channel $e^+e^- \to \pi^+\pi^-$ using the $20$\,fb$^{-1}$ dataset at the $\psi(3770)$ resonance is foreseen. The statistics of this data sample are sufficient to allow normalisation to the di-muon sample, which will improve the systematic uncertainty as the error on the radiator function and the luminosity decreases. Additional radiative-return analyses concern the processes $e^+e^- \to K^+K^-$ and $e^+e^- \to K^0 {\bar K^0}$. Exploiting the high-statistics energy scan sample between 2.0 and 4.6\,GeV, the uncertainty of the total hadronic cross section measurement will be further improved. A detailed description of the BESIII programme is given in~\cite{BESIII:2020nme}.

\subsection{SLAC, Stanford: PEP-II and BABAR}

The BABAR experiment~\cite{BaBar:2001yhh} operated a general-purpose
quasi-full-solid-angle detector at the SLAC PEP-II asymmetric
electron--positron collider from 1999 to 2008, collecting data for
$e^+e^-$ collisions at and around the $\Upsilon(4S)$ resonance for a
total integrated luminosity of 424\,fb$^{-1}$.

BABAR measured the cross section of the process $e^+ e^- \to
\text{hadrons} (\gamma)$ for a large number of exclusive channels, using
the radiative return method~\cite{Arbuzov:1998te}, which provides good
experimental access to hadronic final states with invariant masses from
the production threshold to typically $3{-}5\,\GeV$. The reported
measurements entirely cover energies up to $1.8{-}2.0\,\GeV$, the energy
beyond which the number and multiplicity of the exclusive hadronic modes
and the small size of the total hadronic cross section make inclusive
measurements and theory predictions more convenient for the calculation
of the $a_\mu^\text{HVP}$ dispersive integral. BABAR measured all
relevant hadronic channels except for the part of the $\pi^+\pi^-4\pi^0$
channel that does not proceed through $\eta 3\pi$. Using isospin
symmetries, it has been estimated that this unmeasured contribution
amounts to only $(0.016\pm0.016)\%$~\cite{Davier:2019can}.

Almost all BABAR measurements select events requiring a detected
large-angle hard photon as ISR photon candidate (with a ${\sim}10\%$
efficiency)~\cite{BaBar:2009wpw, BaBar:2012bdw, BaBar:2013jqz,
  BaBar:2004ytv,BaBar:2005dch, BaBar:2012sxt, BaBar:2007qju,
  BaBar:2006vzy, BaBar:2017zmc, BaBar:2007ceh, BaBar:2007ptr,
  BaBar:2011btv, BaBar:2005pon, BaBar:2013ves, BaBar:2007fsu}, relying
on the relative large energy of the ISR photon over the whole range of
the measured invariant mass of the hadronic system at the BABAR beam
energies. By constraining the whole event reconstructed candidates to
the well known position, invariant mass and momentum of the colliding
electron-positron system, this analysis strategy suppresses background
contamination, and also permits the measurement of extra higher-order
radiation consisting of additional detected photons and either one
undetected large-angle photon or two opposite-angle beam-collinear
photons~\cite{BaBar:2009wpw, BaBar:2012bdw, BaBar:2013jqz}. Furthermore,
requiring a hard photon that is well contained in the detector increases
the probability that also the hadronic recoil system is fully detected
on the other side, improving the quality of particle identification, and
reducing systematic uncertainties related to the simulation of the
selection and reconstruction efficiency, particularly for high
multiplicity channels. Due to the relatively high involved hadronic
invariant mass of the measured final state, no detected ISR photon is
required for one $p \bar{p}$ channel measurement~\cite{BaBar:2013ukx}.

Most BABAR measurements~\cite{BaBar:2004ytv,BaBar:2005dch,
  BaBar:2012sxt, BaBar:2007qju, BaBar:2006vzy, BaBar:2017zmc,
  BaBar:2007ceh, BaBar:2007ptr, BaBar:2011btv, BaBar:2005pon,
  BaBar:2013ves, BaBar:2007fsu} normalise the measured yields using a
radiator function Monte Carlo simulation at NLO QED and the estimated
integrated luminosity (typically order 1\% precise) of the analysed
sample. Signal and background ISR processes are simulated with an event
generator based on EVA, additional ISR photons that are collinear to the
beams are simulated with the structure function
method~\cite{Caffo:1997yy}, and  additional FSR photons are simulated
with \Photos{}. For checks and additional studies, BABAR uses the generators \phokhara{} and \afkqed.

The most precise BABAR measurements, on
$\pi^+\pi^-(\gamma)$~\cite{BaBar:2009wpw, BaBar:2012bdw} and
$K^+K^-(\gamma)$~\cite{BaBar:2013jqz}, measure the ratio between the
hadronic and the $\mu^+\mu^-(\gamma)$ cross sections, which is not
affected by the uncertainties on the integrated luminosity. Furthermore,
this approach permits using loose event selections that include
higher-order radiative events in both the measurement and the
normalisation channel, since these contributions mostly cancel in the
ratio, greatly reducing systematic uncertainties related to the
simulation of higher-order radiative events.

Using a data sample corresponding to an integrated luminosity of
232\,fb$^{-1}$, BABAR measured the $e^+e^-\to\pi^+\pi^-(\gamma)$
cross section from threshold to $3.0\,\GeV$, performing on the same
sample a simultaneous measurement of the $e^+e^- \to \mu^+\mu^-
(\gamma)$ cross section as normalisation~\cite{BaBar:2009wpw,
BaBar:2012bdw}. The corresponding $a_\mu^{\text{HVP}}$ contribution has
a precision of 0.74\%, the best of all BABAR measurements. The $e^+e^-
\to \mu^+\mu^- (\gamma)$ cross section agrees over the whole range with
the NLO QED prediction normalised with the integrated luminosity
measured by BABAR, within a 1.1\% uncertainty, which is dominated by the
luminosity uncertainty. In the $\rho$ peak region, the estimated
systematic uncertainty is 0.5\%. The reported measurement of the
$K^+K^-(\gamma)$ channel, dominated by the $\phi$ resonance, has been
performed with the same analysis strategy~\cite{BaBar:2013jqz}.
Regarding the $e^+e^-\to\pi^+\pi^-(\gamma)$ channel, there is an
on-going effort to complete an improved measurement that will use the
whole collected statistics and will statistically separate pions from
muons using their kinematic distributions in order to reduce the
systematic uncertainties related to particle identification.

BABAR has precisely measured the 3-pion and 4-pion final states, which
correspond to the largest multi-hadronic cross sections below $2\,\GeV$:
$\pi^+\pi^-\pi^0$~\cite{BaBar:2004ytv},
$2\pi^+2\pi^-$~\cite{BaBar:2005dch,BaBar:2012sxt} and $\pi^+ \pi^-
2\pi^0$~\cite{BaBar:2017zmc}. The measurements reported by BABAR also
include
the proton--antiproton final state~\cite{BaBar:2005pon, BaBar:2013ves, BaBar:2013ukx},
final states with kaons $K_S^0 K_L^0$, $K \bar{K} + n~\rm{pions}$
with $n=1,2$, $K \bar{K} K^+K^-$~\cite{BaBar:2017nrz,
BaBar:2007ceh, BaBar:2014uwz, BaBar:2017pkz, BaBar:2007ptr,
BaBar:2011btv}, final states including $\eta$ mesons,  $\eta \pi^+
\pi^-$~\cite{BaBar:2007qju, BaBar:2018erh}, $\eta 2\pi^+
2\pi^-$~\cite{BaBar:2007qju}, $\eta \pi^+ \pi^-
2\pi^0$~\cite{BaBar:2018rkc}, and final states with up to 6 quasi-stable
hadrons~\cite{BaBar:2006vzy}.

\subsection{KEK, Tsukuba: SuperKEKB and Belle-II}

The Belle-II experiment \cite{Belle-II:2010dht} is located at the SuperKEKB collider~\cite{Akai:2018mbz}, an upgrade of the KEKB accelerator, running at a c.m. energy of 10.58\,GeV (near the $\Upsilon(4S)$ resonance) with  7\,GeV electron and 4\,GeV positron beams. The Belle-II experimental programme \cite{Belle-II:2018jsg} is mainly related to $B$-meson physics.
 The design peak luminosity is  $6\times 10^{35}\,\text{cm}^{-2}\text{s}^{-1}$.
The experiment started in 2019 and plans to collect 50\,ab$^{-1}$ of data by 2032~\cite{Aihara:2024zds}. Radiative-return measurements were not possible at the previous Belle experiment~\cite{Belle:2012iwr} running between 1999 and 2010 due to trigger efficiency, where one of the reasons was that the Bhabha veto condition used only polar angle information~\cite{Crnkovic:2013bva}. Belle-II implements Bhabha veto based on both polar and azimuthal angles.
Recent Belle-II results on the $e^+e^-\to\pi^+\pi^-\pi^0\gamma$ channel \cite{Belle-II:2024msd} use 191\,fb$^{-1}$ of data. In the reconstruction the polar angle of the detected (ISR) photon in the c.m. system is in the range from 48$^\circ$ to 135$^\circ$. The invariant mass of the $\pi^+\pi^-\pi^0\gamma$ system is required to be greater than 8.0\,GeV to suppress the effect of extra emitted photons.
\phokhara{}9.1 is used to simulate the signal~\cite{Czyz:2005as}, with a claimed systematic uncertainty for radiative corrections of 0.5\%~\cite{Rodrigo:2001kf}. In the future, a measurement of the $e^+e^-\to\pi^+\pi^-\gamma$ cross section is planned, with a target precision of 0.5\% for $a_\mu^{\pi\pi}$. The measurement will follow BaBar methods as a baseline, but compared to BaBar it will use a much larger integrated luminosity to control systematic effects, which dominate the uncertainty.


\pdforweb{
    \newcommand{\ampamp}[2]{
        \setlength{\tabcolsep}{0cm}
        \begin{tabular}{m{0.185\textwidth}m{2.3ex} m{0.185\textwidth}}
            \includegraphics[width=\linewidth]{#1} &
            \Large $\times$ &
            \reflectbox{\includegraphics[width=\linewidth]{#2}}
        \end{tabular}
    }
}{
    \newcommand{\ampamp}[2]{
        \setlength{\tabcolsep}{0cm}
        \begin{tabular}{ccc}
            \includegraphics[width=\linewidth]{#1} &
            \Large $\times$ &
            \reflectbox{\includegraphics[width=\linewidth]{#2}}
        \end{tabular}
    }
}

\section{Computational setup}\label{sec:comp}

In this section we provide a conceptual overview of the various contributions that are potentially required to obtain a precise theoretical prediction, with a special focus on $e^+\,e^-\to\,X^+\,X^-$ and the radiative processes $e^+\,e^-\to\,X^+\,X^-\,\gamma$. If $X\in\{e,\mu\}$, the only effects that cannot be computed within standard QED perturbation theory are due to internal light quark loops leading to hadronic loops. If this hadronic loop is attached to two photons, we get HVP contributions. In the case of a loop attached to four photons, we have  HLbL contributions. However, the latter only contribute beyond NNLO. For $X=\pi$ the situation is more delicate and requires adapted techniques for different parts of a cross section. As a result, complete higher-order corrections to processes~\eqref{intro:scan} and \eqref{intro:return}, taking into account all terms of a particular order in $\alpha$, are the exception rather than the rule. Instead, partial corrections are assembled to obtain the best possible theoretical description.

\begin{figure}[b]
    \centering  
    \includegraphics[width=0.25\textwidth]{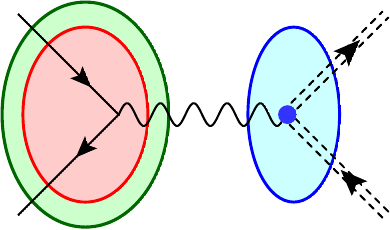}
\qquad\quad
    \includegraphics[width=0.25\textwidth]{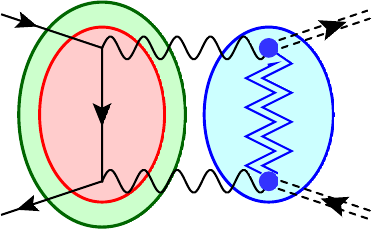}
\qquad\quad
    \includegraphics[width=0.25\textwidth]{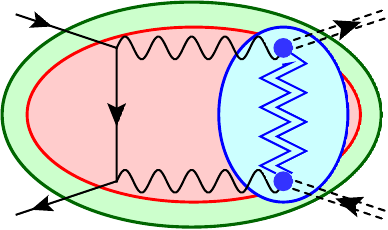}
\\[10pt]
    \caption{Illustration of various categories of corrections to $e^+\,e^-\to\,\pi^+\,\pi^-\,(+\gamma)$. The red blob stands for fixed-order QED corrections, while the green blob indicates potentially resummed approximate QED radiation effects. Since the pion is not pointlike, effects beyond standard QED perturbation theory are present, as indicated by the blue blob. The emission of additional real photons is not depicted, but understood.}
    \label{fig:comp-WP}
\end{figure}

The split of a cross section into different contributions can be dictated by the applied techniques or by physical considerations. We broadly identify three different techniques, illustrated in \figref{fig:comp-WP}. Fixed-order QED contributions are depicted as the red blob.  Tremendous progress has been made in the past years in the ability to perform such computations. An overview of the required parts and the current status will be given in \secref{sec:comp-fo}. The green blob represents approximate higher-order radiative corrections and includes potential resummation of multiple collinear and soft photon emission. This is the topic of  \secref{sec:comp-bfo}. Finally, the blue blob involves techniques beyond pure QED, as it has to take into account the nonpointlike structure of pions. They will be described \secref{sec:comp-fshad}. The HVP and HLbL effects will be discussed in \secref{sec:comp-inthad}.

From a physical point of view, it is also possible to split contributions of a cross section into initial-state corrections (ISC), final-state corrections (FSC), and mixed corrections (interference initial-final) parts. Such a split is partly motivated by technical aspects (different techniques used, different computational complexity) but also by different numerical impact. Looking at the left panel of \figref{fig:comp-WP}, where there is a clear separation into ISC (red and green) and FSC (blue), the ISC are often dominant. They include higher-order corrections enhanced by large collinear logarithms $\alpha^n L_c^n\equiv \alpha^n \log^n(Q^2/m_e^2)$ with the typical scale of the process $Q\sim \sqrt{s}$ much larger than the electron mass $m_e$. Since in our case the initial state always consists of an electron--positron pair, ISC can be systematically improved through QED perturbation theory. For ISC, the initial state is linked to the final state through the exchange of a single photon and we will adopt the term one-photon exchange (1PE) contribution familiar from lepton--proton scattering for this topology. 

In a pure QED process such as \eqref{intro:scanMu}, FSC can be obtained by recycling the ISC computation. If the final state contains hadrons, different techniques are required, as mentioned above. The mixed corrections are technically much more demanding even for purely leptonic processes. They involve two-photon exchange (2PE) and at NNLO even three-photon exchange (3PE) contributions, i.e., contributions where the inital and final states are connected by two or at NNLO even three photons. Only recently the computation of such corrections at NNLO for $2\to 2$ processes including mass effects became feasible. 

In the following subsections we will go through the various techniques mentioned above. We will not give a detailed description of how the computations are done, but refer to the literature for such aspects. However, we will describe in more detail what contributions exist and set up a more precise notation for them. This will facilitate the description of what precisely is included in the various codes and the relative numerical importance of these partial corrections.


\subsection{Fixed-order QED} \label{sec:comp-fo}

In this section we focus on fixed-order QED contributions. The impact of heavy electroweak gauge bosons is typically negligible for the low-energy processes we are considering here. In order to establish our notation and present the basics of the computational framework, we focus on the process $e^+\,e^- \to \mu^+\,\mu^-$. At the end of the subsection we briefly comment on the complications with pions in the final state, as a preparation for \secref{sec:comp-fshad}.

We start with the tree-level amplitude, denoted by $\cA_{mm}^{(0)}(q_e\,q_m)\sim q_e\,q_m$. The subscript indicates the final state ($m$ for muon), suppressing the initial state that is always $e^+\,e^-$ in this paper. As the argument we have given the power of the electron charge $q_e$ and the muon charge $q_m$, respectively. We formally distinguish between these charges to identify various gauge-invariant contributions in more complicated situations. As generic symbol we use $q_\ell$. The (differential) LO cross section $\dd\sigma_{mm}^{(0)}(q_e^2\,q_m^2)$ is then obtained by integrating the tree-level squared matrix element 
\begin{align}\label{comp:born}
 \cM_{mm}^{(0)}(q_e^2\,q_m^2) &= \big|\cA_{mm}^{(0)}(q_e\,q_m)\big|^2
\end{align}
over the two-body phase space. 

For cross sections that correspond to realistic experimental situations the phase-space integration needs to be done through numerical Monte Carlo methods. Some Monte Carlo codes are so-called integrators, where directly histograms for arbitrary IR safe observables are produced. Experimental cuts are taken into account during the numerical integration. More common are so-called generators, where in a first step generic events are generated. These events are then used in a second analysis step to obtain differential distributions. Generators offer a greater flexibility to include detailed detector simulations. For this to be efficient, however, it is important to keep the number of events with negative weight under control.

\begin{figure}
    \centering  
    \includegraphics[width=0.2\textwidth]{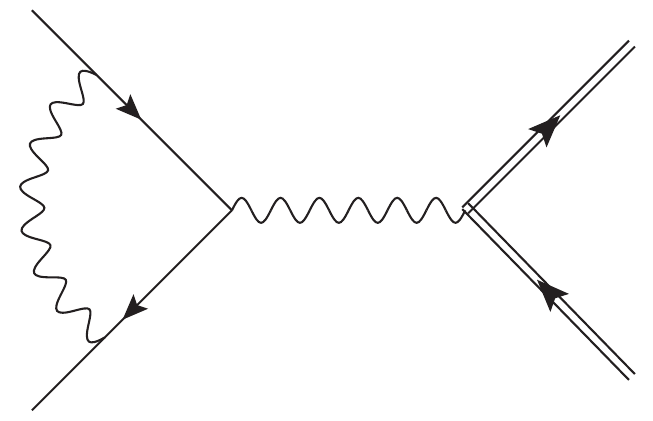}
\qquad
    \includegraphics[width=0.2\textwidth]{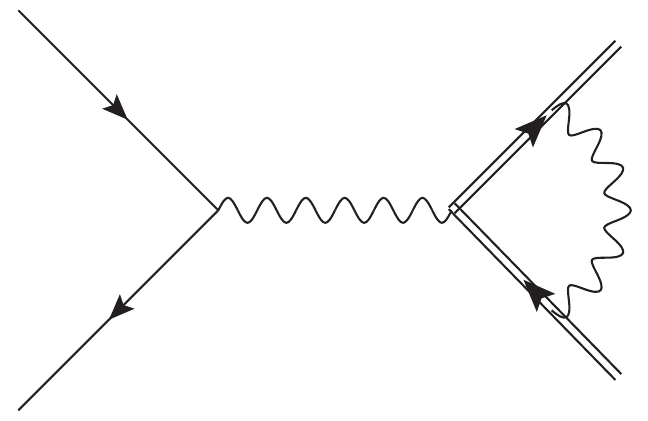}
\qquad
    \includegraphics[width=0.2\textwidth]{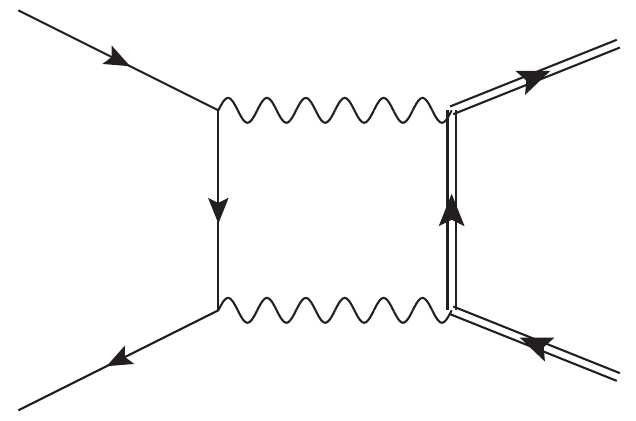}
\qquad
    \includegraphics[width=0.2\textwidth]{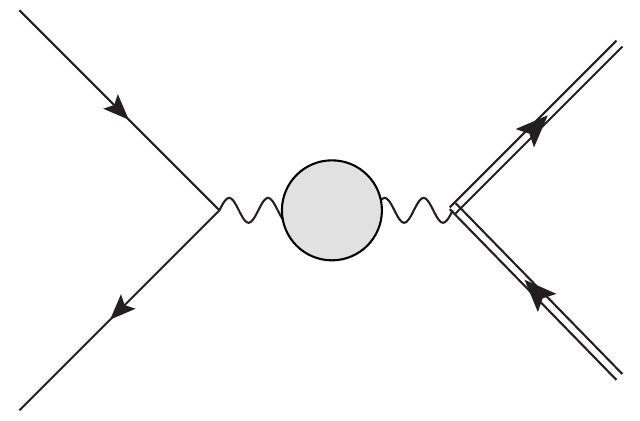}
\\[10pt]
    \caption{Some diagrams for the NLO amplitude of  $e^+\,e^- \to \mu^+\,\mu^-$, corresponding to the four gauge-invariant parts $\cA_{mm}^{(1)}(q^3_e\,q_m)$, $\cA_{mm}^{(1)}(q_e\,q^3_m)$,  $\cA_{mm}^{(1)}(q^2_e\,q^2_m)$, and $\cA_{mm}^{(1)}(q_e\,q_m\,\Pi^{(1)})$ of \eqref{comp:olA}.}
    \label{fig:comp-Aeemm}
\end{figure}

Going beyond LO, we need the one-loop amplitude that we split into four parts
\begin{align} \label{comp:olA}
 \cA_{mm}^{(1)}(q_e\,q_m\,q_\ell^2)&=    
 \cA_{mm}^{(1)}(q^3_e\,q_m) 
 + \cA_{mm}^{(1)}(q_e\,q^3_m) 
 + \cA_{mm}^{(1)}(q^2_e\,q^2_m) + \cA_{mm}^{(1)}(q_e\,q_m\,\Pi^{(1)})
\end{align}
as illustrated in \figref{fig:comp-Aeemm}. The first (second) term on the right corresponds to corrections to the electron (muon) line with three couplings of the photon to the electron (muon) and only one coupling to the muon (electron). The third term is due to box diagrams with two couplings each to the electron and muon. Together they make up the photonic corrections. Finally, the last term is due to an insertion of the (one-loop leptonic and hadronic) VP, denoted by $\Pi^{(1)}=\Pi^{(1)}_\ell+\Pi_h$ with $\Pi^{(1)}_\ell\sim q_\ell^2$.  We will call fermionic corrections all corrections that contain a closed lepton or hadron loop.

As alluded to above, the insertion of $\Pi_h$ is a contribution beyond strict QED and the counting in terms of LO, NLO, etc, is not fully applicable any longer. Related to this, the VPC are often treated by resumming VP insertions. In this approach, the (complex) VP can be evaluated to the highest possible precision. 
Up to $n=2$ the leptonic VP can be split into $\Pi^{(n)}_\ell(q_\ell^{2n})=\Pi^{(n)}_e+\Pi^{(n)}_m+\Pi^{(n)}_\tau$ and it can be obtained with full mass dependence from~\cite{Djouadi:1993ss, Laporta:2024aok}. Three-loop results also exist in different forms of mass-term expansions~\cite{Chetyrkin:1996cf,Steinhauser:1998rq}.
However, beyond $n=2$, hadronic contributions mix into the leptonic part and a strict separation is not possible any longer. The determination of $\Pi_h$ from experimental data will be addressed in \secref{sec:comp-inthad}.  In \secref{sec:mccVP} we will discuss the determination of the full VP from experimental data and compare various packages available in the literature.

The complete virtual NLO corrections to $e^+\,e^- \to \mu^+\,\mu^-$  are obtained by integrating
\begin{align} \label{comp:olv}
  \cM_{mm}^{(1)}(q_e^2 q^2_m q_\ell^2)&= 2\,\Re
   \big(\cA_{mm}^{(1)}(q_e\,q_m\,q_\ell^2)\,  \cA_{mm}^{(0)\,*}(q_e\,q_m)\big)
\end{align}
over the two-body phase space. Here we introduced the notation 
$\cM_{xx}^{(n)}$ to include all terms of the squared matrix elements with a power $\cO(\alpha^n)$ relative to the LO term. After removing the UV singularities in \eqref{comp:olv} through renormalisation (usually in the on-shell scheme) there are still IR singularities. As discussed below, they cancel for physical (IR-safe) observables if the virtual corrections are combined with real corrections. The latter are obtained by integrating
\begin{align}\label{comp:olr}
   \cM^{(0)}_{mm\gamma}(q_e^2\,q_m^2\,q_\ell^2)&= \big|\cA^{(0)}_{mm\gamma}(q_e\,q_m\,q_\ell)\big|^2
   = \big|\cA^{(0)}_{mm\gamma}(q_e^2\,q_m)+\cA^{(0)}_{mm\gamma}(q_e\,q^2_m)\big|^2
\end{align}
over the three-body phase space. Here, $\cA^{(0)}_{mm\gamma}(q_e^2\,q_m)$ and 
$\cA^{(0)}_{mm\gamma}(q_e\,q_m^2)$ are the gauge-invariant part of the tree-level amplitude for 
$e^+\,e^- \to \mu^+\,\mu^-\,\gamma$ where the photon is radiated off the initial and final state, respectively. 

Keeping finite fermion masses, there are no collinear singularities in QED. Instead, the already mentioned collinear logarithms of the form $L_c=\log(Q^2/m_\ell^2)$
arise in the angular integration of radiation degrees of freedom around the emitting fermion lines. For large gaps between the two scales $Q^2$ and $m_\ell^2$, as is typically the case for $\ell=e$, such logarithms enhance the higher-order terms of the perturbative series at fixed order.
In Sections~\ref{sec:comp-bfoPDF} and \ref{sec:comp-bfoPS} the approximations able to resum the collinear
logarithms to all orders are discussed.

Even with finite fermion masses there are still soft singularities. They can be regularised either by introducing a fictitious photon mass $m_\gamma$ or by using dimensional regularisation also for the IR case. The regularised singularities then appear as $\log (Q^2/m_\gamma^2)$ or $1/\epsilon$ poles in intermediate expressions. The extraction of the IR singularities of the real corrections can be done in a process- and observable-independent way, using either the slicing method or a subtraction method. A short description of the two can be found, e.g., in~\cite{Afanasev:2023gev}. Once real and virtual corrections are combined, it is safe to remove the regulator $m_\gamma\to 0$ or $\epsilon\to 0$. While not strictly necessary, typically the slicing method is combined with photon-mass regularisation, whereas the subtraction method is used with dimensional regularisation. 

The core property that lies at the heart of the IR cancellation is the fact that it is impossible in principle to distinguish an electron (or any charged lepton) from an electron accompanied by a cloud of arbitrarily soft photons. Even the best detector has a finite resolution below which photons remain undetected. From a more formal point of view, the IR divergences of the scattering matrix stem from the violation of a basic assumption of scattering theory: due to the long-range nature of the Coulomb interaction, the electrons are never free, not even in the asymptotic past or future.

From a practical point of view, this is acceptable as long as the observables considered are defined accordingly. They must not depend on whether or not additional arbitrarily soft photons are emitted. For such so-called IR-safe observables the cancellation between real and virtual IR singularities is guaranteed to all orders in perturbation theory. However, it is often the case that as a remnant of this cancellation there are (large) soft logarithms. Indeed, if the observable is defined such that the real emission of additional photons is restricted to have energies $\Delta{E}$, there are logarithms of the form $\alpha^n\log^n(Q^2/\Delta{E}^2)$. The precise form of the soft logarithms depends on the definition of the observable. For small $\Delta{E}$, the soft logarithms can be numerically enhanced, potentially leading to a bad convergence of the perturbative expansion. The inclusion of these logarithms beyond fixed order is discussed in \secref{sec:comp-bfoYFS}.

As for \eqref{comp:olA}, the NLO corrections to the cross section, $\dd\sigma_{mm}^{(1)}(q_e^2\,q_m^2\,q_\ell^2)$, can also be split into four parts. If we were to restrict ourselves to initial-state photonic corrections only, we would use 
\begin{align}\label{comp:olvi}
 \cM_{mm}^{(1)}(q_e^4\,q_m^2) &= 2\,\Re
   \big(\cA_{mm}^{(1)}(q^3_e\,q_m)\,  \cA_{mm}^{(0)\,*}(q_e\,q_m)\big)\,, \\
   \label{comp:olri}
  \cM^{(0)}_{mm\gamma}(q_e^4\,q_m^2)&= \big|\cA^{(0)}_{mm\gamma}(q^2_e\,q_m)\big|^2\,, 
\end{align}
and integrate \eqref{comp:olvi} and \eqref{comp:olri} over the two-body and three-body phase space, respectively, to obtain $\dd\sigma^{(1)}_{mm}(q_e^4\,q_m^2)$. Analogously, we can obtain  $\dd\sigma^{(1)}_{mm}(q_e^2\,q_m^4)$, the final-state NLO corrections to $e^+\,e^- \to \mu^+\,\mu^-$. Finally, there are the mixed corrections consisting of $\dd\sigma^{(1)}_{mm}(q_e^3\,q_m^3)$ (interference between emission of initial state and final state) and $\dd\sigma_{mm}^{(1)}(q_e^2\,q_m^2\,\Pi^{(1)})$ (VP corrections). The sum of these four parts is denoted by \begin{align}\label{comp:olS}
    \dd\sigma^{(1)}_{mm}(q_e^2\,q_m^2\,q_\ell^2)&=
    \underbrace{\dd\sigma^{(1)}_{mm}(q_e^4\,q_m^2)}_{\text{ISC}} 
    + \underbrace{\dd\sigma^{(1)}_{mm}(q_e^2\,q_m^4)}_{\text{FSC}} 
    + \underbrace{\dd\sigma^{(1)}_{mm}(q_e^3\,q_m^3) }_{\text{mixed}}
    + \underbrace{\dd\sigma^{(1)}_{mm}(q_e^2\,q_m^2\,\Pi^{(1)})}_{\text{VPC}} 
\end{align}    
and corresponds to the integration of the full matrix elements \eqref{comp:olv} and \eqref{comp:olr} over the two- and three-body phase space. In \eqref{comp:olS}, the ISC and VPC make up the red blob on the left panel of \figref{fig:comp-WP} for the $\mu^+\,\mu^-$ final state. 

Of course, we can also integrate \eqref{comp:olr} and explicitly require a resolved photon, i.e., a photon with energy above a certain threshold such that in principle it could be detected. In this case we obtain 
\begin{align} \label{comp:loR}
\dd\sigma^{(0)}_{mm\gamma}(q_e^2\,q_m^2\,q_\ell^2) &=
\underbrace{\dd\sigma^{(0)}_{mm\gamma}(q_e^4\,q_m^2)}_{\text{ISC}}
+\underbrace{\dd\sigma^{(0)}_{mm\gamma}(q_e^2\,q_m^4)}_{\text{FSC}}
+\underbrace{\dd\sigma^{(0)}_{mm\gamma}(q_e^3\,q_m^3)}_{\text{mixed}}\,  , 
\end{align}
the LO cross section to the process $e^+\,e^- \to \mu^+\,\mu^-\,\gamma$. This process is often called an ISR process, a terminology that is somewhat misleading. The photon can be from ISC, but also from FSC or interference (mixed contribution). Hence, we will refer to such a process simply as a radiative process. How important the non-ISC terms are compared to the ISC part for a cross section like \eqref{comp:loR} depends crucially on the precise definition of the observable and is a question to be addressed \pdforweb{in \secref{sec:mcc}}{later}.

The split of fixed-order corrections to  $e^+\,e^-\to X^+\,X^-$ into ISC, FSC, mixed, and VPC can be extended to any order. At NNLO, double virtual corrections have to be combined with real-virtual and double real in order to obtain a meaningful IR finite result. The corrections $\dd\sigma_{mm}^{(2)}(q_e^2\,q_m^2\,q_\ell^4)$ contain at least two powers of $q_e$ and $q_m$. We call VPC those corrections that contain precisely a factor $q_e^2$ and a factor $q_m^2$. They are obtained by VP insertions to the photon propagator present in the Born diagram, namely one insertion of the two-loop VP, $\dd\sigma_{mm}^{(2)}(q_e^2\,q_m^2\,\Pi^{(2)})$, and two insertions of the one-loop VP, $\dd\sigma_{mm}^{(2)}(q_e^2\,q_m^2\,(\Pi^{(1)})^2)$, including those from the one-loop amplitude squared, 
$|\cA_{mm}^{(1)}(q_e\,q_m\,\Pi^{(1)})|^2$. 
The VPC have only double virtual corrections. These corrections are a subset of the resummed VPC discussed above. 

\begin{figure}[t]
    \centering
    \ampamp{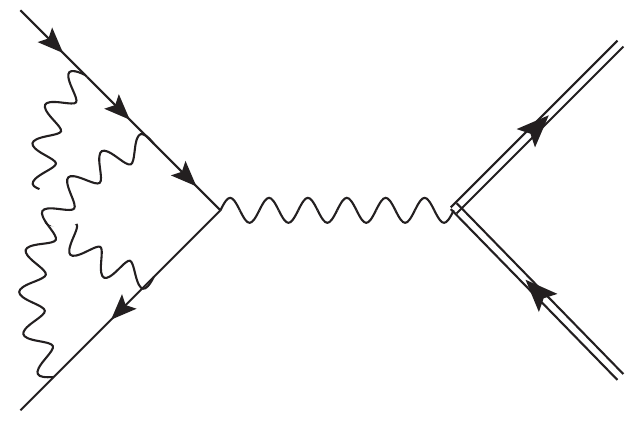}{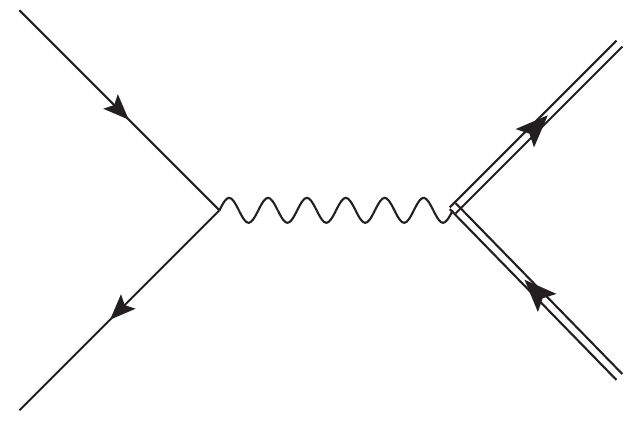}
    \qquad
    \ampamp{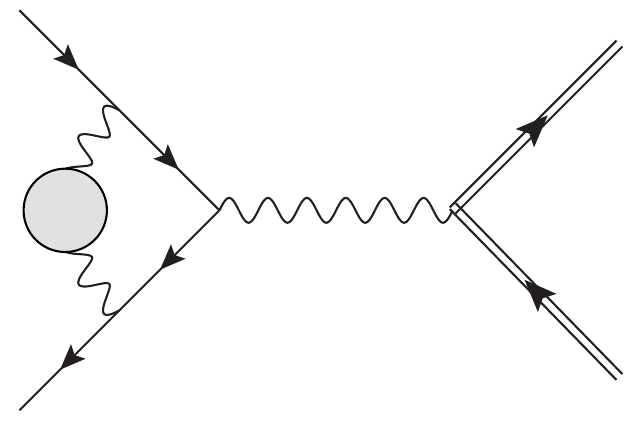}{figs/comp/mm-LO.pdf}
    \\[10pt]
    \ampamp{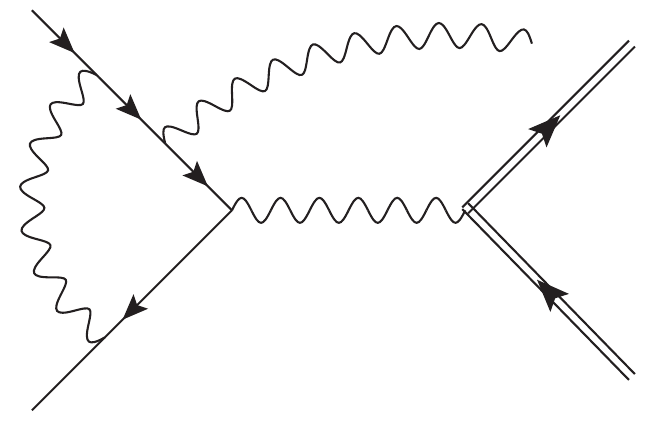}{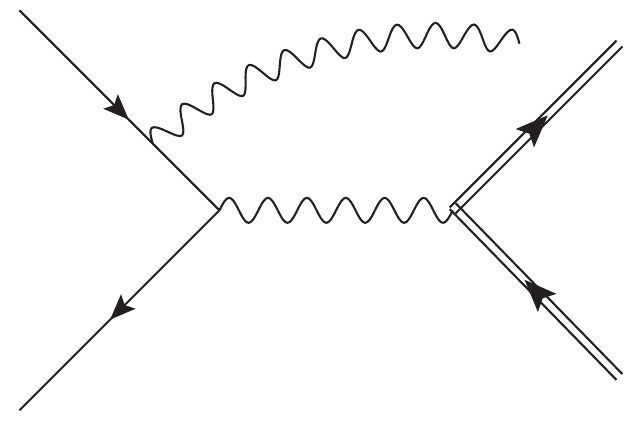}
    \qquad
    \ampamp{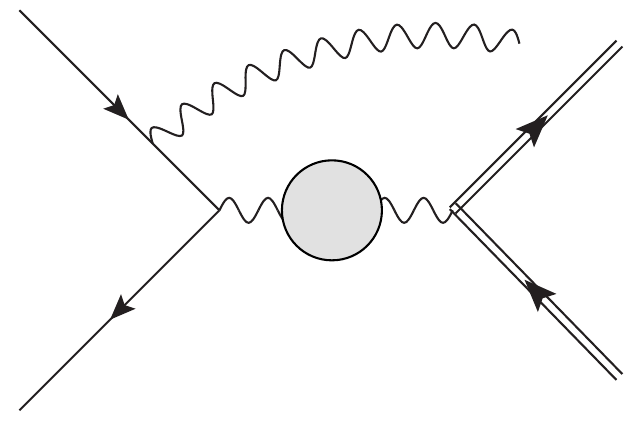}{figs/comp/mm-ISR.pdf}
    \\[10pt]
    \ampamp{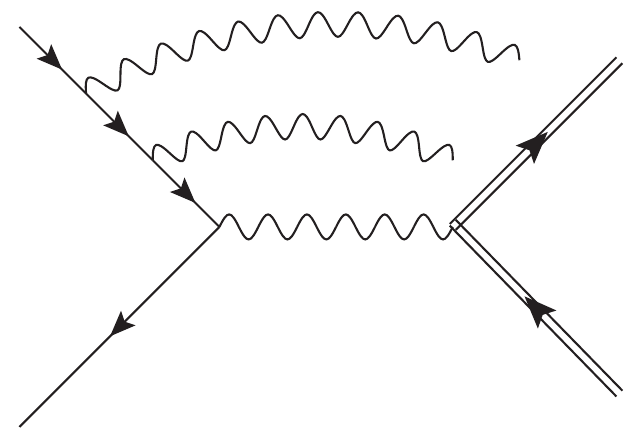}{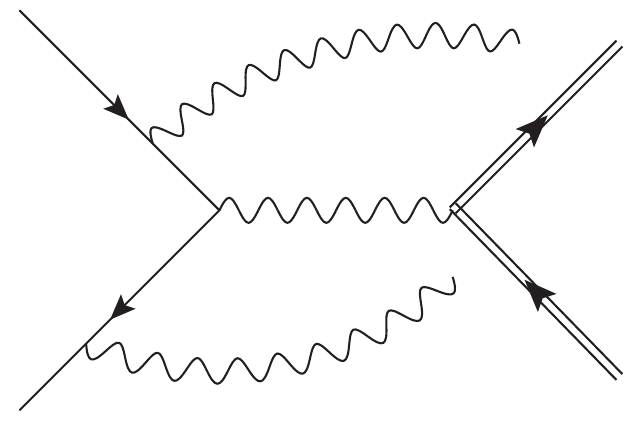}
    \qquad
    \ampamp{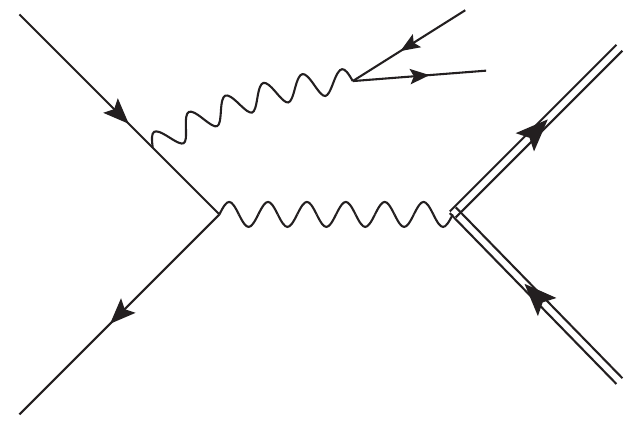}{figs/comp/mm-ISRee.pdf}
    \\[10pt]
    \caption{Representative diagrams contributing to $\dd\sigma_{mm}^{(2)}(q_e^6\,q_m^2)$ (left panels) and $\dd\sigma_{mm}^{(2)}(q_e^4\,q_m^2\,\Pi^{(1)})$ (right panels), the pure ISC of the NNLO corrections for $e^+\,e^- \to \mu^+\,\mu^-$. The double real VP contribution at the bottom of the right panel corresponds to a measurably different process  $e^+\,e^- \to \mu^+\,\mu^-\,e^+\,e^-$.}
    \label{fig:comp-eemmISR}
\end{figure}

ISC also contain precisely a power $q_m^2$ at NNLO, i.e., the minimal possible coupling to the final state (the muon in this case), but they contain $q_e^n$ with $n>2$. Some representative examples for the squared matrix elements are shown in \figref{fig:comp-eemmISR}. In the top row, there are double virtual corrections. Apart from purely photonic corrections $\dd\sigma_{mm}^{(2)}(q_e^6\,q_m^2)$ (example in left panel) there are also VP contributions to ISC, $\dd\sigma_{mm}^{(2)}(q_e^4\,q_m^2\,\Pi^{(1)})$ (right panel). Not shown are terms due to the one-loop amplitude squared, $|\cA_{mm}^{(1)}(q_e^3\,q_m)|^2$. Also for the real-virtual corrections illustrated in the middle row there are photonic contributions (left panel) and VP contributions (right panel). Finally, there are double real contributions with examples in the bottom row. The associated VP contributions are related to the process with an additional lepton pair in the final state. Keeping the mass of this lepton different from zero, this results in a separately finite contribution and a measurably different process. Still, the cross section for such a process is logarithmically enhanced (as a remnant of the would-be IR singularities in the limit of massless leptons) and, hence, its impact has to be carefully considered.

All ISC at NNLO are by construction 1PE contributions between the incoming leptonic state and the final state. This makes them simpler from a technical point of view. The required two-loop amplitudes are covered by the two-loop form factor for massive particles~\cite{Bonciani:2003ai} requiring two-loop integrals with at most three external legs. For the real-virtual part one-loop box diagrams are sufficient. The same holds for 
the FSC. They can be obtained by recycling the ISC computation, exchanging the roles of $q_e$ and $q_m$.  

\begin{figure}[bt]
    \centering
    \includegraphics[width=0.25\textwidth]{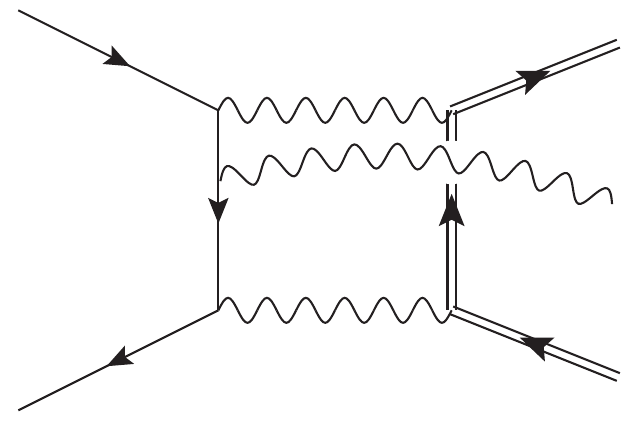}
    \hspace*{0.15\textwidth}
    \includegraphics[width=0.25\textwidth]{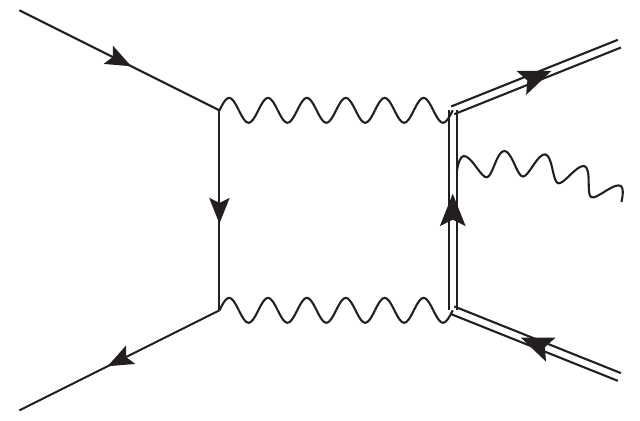}
    \caption{Representative 2PE diagrams contributing to 2PE-ISR $\cA^{(1)}_{mm\gamma}(q_e^3\,q_m^2)$ (left) and 2PE-FSR $\cA^{(1)}_{mm\gamma}(q_e^2\,q_m^3)$ (right) as part of the NLO amplitude for the process
    $e^+\,e^- \to \mu^+\,\mu^-\,\gamma$.}
    \label{fig:comp-eemmy2PE}
\end{figure}

\begin{figure}[t]
    \centering
    \ampamp{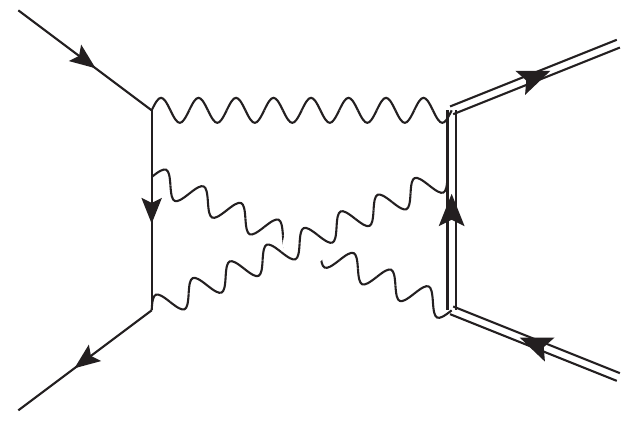}{figs/comp/mm-LO.pdf}
    \qquad
    \ampamp{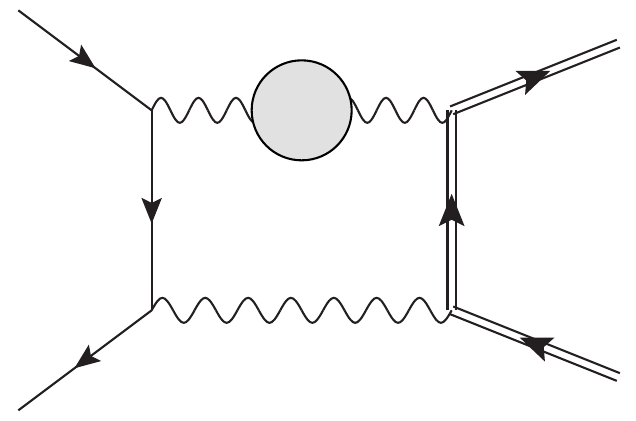}{figs/comp/mm-LO.pdf}
    \\[10pt]
    \ampamp{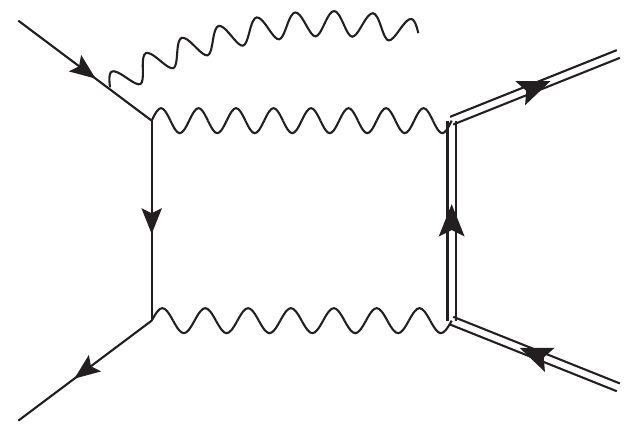}{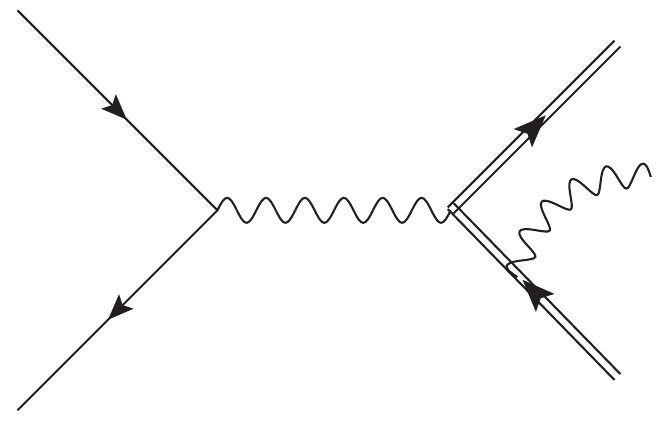}
    \qquad
    \ampamp{figs/comp/mm-ISR-VP.pdf}{figs/comp/mm-FSR.pdf}
    \\[10pt]
    \ampamp{figs/comp/mm-ISR-ISR-4.pdf}{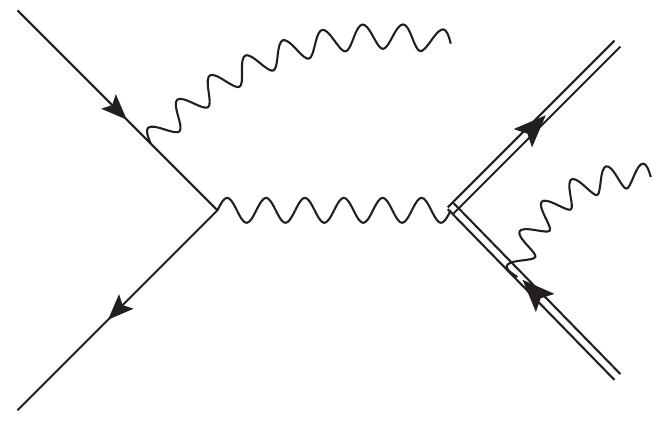}
    \qquad
    \ampamp{figs/comp/mm-ISRee.pdf}{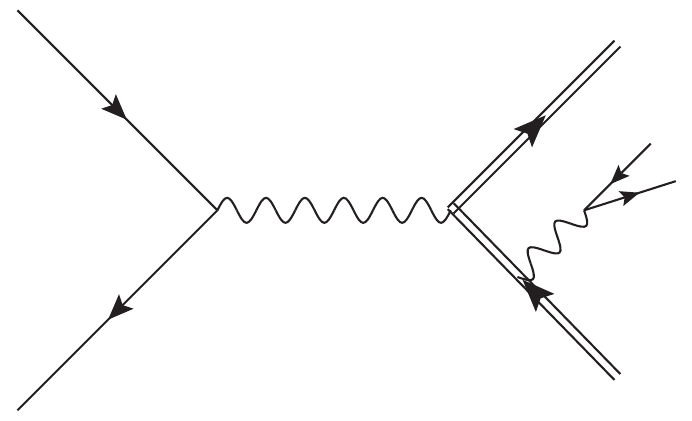}
    \caption{Representative diagrams contributing to mixed NNLO corrections 
    $\dd\sigma_{mm}^{(2)}(q_e^i\,q_m^f\,q_\ell^x)$ (left panels) and 
    $\dd\sigma_{mm}^{(2)}(q_e^i\,q_m^f\,\Pi^{(1)})$ (right panels), with $i>2$ and $j>2$ for 
    $e^+\,e^- \to \mu^+\,\mu^-$. The double real VP contribution at the bottom of the right panel corresponds to a measurably different process  $e^+\,e^- \to \mu^+\,\mu^-\,e^+\,e^-$.}
    \label{fig:comp-eemmX}
\end{figure}

Finally, the mixed corrections are all those that are neither VPC, ISC, or FSC. In principle they can be disentangled further according to powers of $q_e$ and $q_m$~\cite{Banerjee:2020tdt}. For example, the 2PE part of the one-loop amplitudes for the radiative process $e^+\,e^-\to\mu^+\,\mu^-\,\gamma$ can be split into additional emission from the electron line, $\cA^{(1)}_{mm\gamma}(q_e^3\,q_m^2)$, and muon line $\cA^{(1)}_{mm\gamma}(q_e^2\,q_m^3)$, as illustrated in \figref{fig:comp-eemmy2PE}. The former (latter) we call 2PE-ISR (2PE-FSR) contributions. For processes with pointlike particles this split is mainly done for computational reasons. Indeed, for the mixed corrections, the computations are much more involved than for ISC, even for pointlike final states. As depicted in the first row of \figref{fig:comp-eemmX} two-loop integrals with four external legs are required. This is a serious complication, in particular for massive fermions. Only recently all two-loop four-point integrals with a massive muon and massless electron have been computed~\cite{Mastrolia:2017pfy, DiVita:2018nnh}. The integrals with two equal nonvanishing masses involve elliptic functions~\cite{Delto:2023kqv}. The real-virtual corrections, illustrated in the second row of \figref{fig:comp-eemmX}, now involve pentagon one-loop diagrams that often lead to numerical complications. Fortunately, there is by now a lot of experience how to deal with them~\cite{Denner:2016kdg, Buccioni:2017yxi, Buccioni:2019sur}.

With the NNLO contributions discussed above, we also have all ingredients for $\dd\sigma_{mm\gamma}^{(1)}$, the NLO cross section of the radiative process $e^+\,e^-\to\mu^+\,\mu^-\,\gamma$. Of course, the double virtual part, i.e., the two-loop and 
one-loop squared diagrams, are not required in this case. However, the split into ISC, FSC, and mixed contributions carries through. As illustrated in \figref{fig:comp-eemmy2PE}, a further separation of mixed contributions into 2PE-ISR and 2PE-FSR is sometimes useful. 

For pointlike final states, the situation described so far is roughly speaking the current state of the art. Next steps that are expected to be completed in the foreseeable future are N$^3$LO ISC for
$e^+\,e^-\to X^+\,X^-$ and NNLO corrections to the radiative process $e^+\,e^-\to\mu^+\,\mu^-\,\gamma$. For the former, the three-loop triple virtual corrections are available~\cite{Fael:2022rgm, Fael:2022miw, Fael:2023zqr}. However, a consistent combination with the various real corrections is still a formidable task. In particular the two-loop amplitudes with an additional photon are currently only known for massless electrons~\cite{Badger:2023xtl, Fadin:2023phc}. For the latter, the two-loop amplitude is the key ingredient that requires five-point two-loop integrals with an internal mass~\cite{Badger:2022hno, FebresCordero:2023pww, Badger:2024fgb}.

We end this subsection with a few comments regarding processes with a $\pi^+\,\pi^-$ pair in the final state. For the ISC there are no conceptual difficulties in their evaluation, since in this case the pions couple through the VFF with a single photon. Care has to be taken to avoid double counting of the VPC, as they are included in the pion VFF. However, FSC and mixed corrections are much more delicate. From a technical point of view it is possible to compute all contributions in scalar QED, order by order in perturbation theory, but the phenomenological reliability of this procedure is questionable. Including form factors at pion--photon vertices, i.e., using FsQED, can improve the situation is specific cases. For example, the 2PE-ISR contribution depicted in the left panel of \figref{fig:comp-eemmy2PE} can be reasonably approximated through FsQED. Additional photon emission from the pions is, however, more delicate. The 2PE-FSR part illustrated in the right panel of \figref{fig:comp-eemmy2PE} for instance cannot be described well through FsQED. This leads to serious limitations in our ability to describe processes with pions in the final state. In particular, the effect of higher-order perturbative corrections on the leptonic side might be drowned in nonperturbative hadronic uncertainties appearing already at lower orders in $\alpha$. A more detailed discussion of processes with pions in the final state is given in  \secref{sec:comp-fshad}.


\subsection{QED beyond fixed order} \label{sec:comp-bfo}

As detailed in the previous section, to a first approximation, the most logical way to perform calculations of observables is \textit{order-by-order}, that is all the contributions beyond a fixed order are exactly equal to zero. In this section, a description of the methods to provide theoretical predictions beyond fixed order in QED will be given. 

\begingroup
\renewcommand{\arraystretch}{1.2}
\begin{table}[b]
\centering
    \begin{tabular}{r|ccccc}
     & LL & NLL & NNLL & N$^3$LL & $\cdots$ \\[0.1 em]
    \hline
    LO &   1 & \multicolumn{4}{c}{}\\[0.1 em]
    NLO &    $\alpha L_c$ & $\alpha$ & \multicolumn{3}{c}{}\\[0.1 em]
    NNLO&    $\alpha^2 L_c^2$ & $\alpha^2 L_c$ & $\alpha^2$ & \multicolumn{2}{c}{}\\[0.1 em]
    N$^3$LO& $\alpha^3 L_c^3$ & $\alpha^3 L_c^2$ & $\alpha^3 L_c$ & $\alpha^3$& \\[-0.1 em]
    \multicolumn{1}{c|}{\vdots}& \vdots & \vdots & \vdots & \vdots & $\ddots$ \\
\end{tabular}
\caption{Contributions of the QED perturbative series, where $\alpha$ is the expansion constant and $L_c=\log\left(Q^2/m_e^2\right)$ is the collinear logarithm at the typical scale of the process $Q\sim\sqrt{s}$, which is much larger than the electron mass. The power counting of $\alpha$ is normalised to the LO. Rows represent corrections in increasing perturbative orders. Columns represent corrections in increasing logarithmic approximations.}
\label{tab:logtower}
\end{table}
\endgroup

It is possible to write the complete set of QED perturbative leading and subleading corrections as in \tabref{tab:logtower}. The \textit{order-by-order} criterion corresponds to the sequential calculation of rows in \tabref{tab:logtower} and accounts for increasing powers of the perturbative expansion constant $\alpha$. For example, the LO contribution $\dd\sigma_{xx}^{(0)}$ includes the only term that is present in the first row, whereas the NLO contribution $\dd\sigma_{xx}^{(1)}$ accounts for the terms proportional to $\alpha L_c$ and to $\alpha$.
However, since for $s\gg m_e^2$ the collinear logarithm can be large, $L_c=\log\left(Q^2/m_e^2\right)\sim{\cal O}(10)$, the terms proportional to powers of $L_c$ break the perturbative expansion in $\alpha$. Moreover, corrections in the LL column give relevant results, phenomenology-wise. Thus, a resummation procedure of the leading logarithms on top of a fixed-order calculation is crucial to reach high-precision Monte Carlo predictions. In formulae, the merging of the LL approximation on top of the LO calculation -- corresponding to the first column of \tabref{tab:logtower} -- is
\begin{equation}
    \dd \sigma_{xx}^{(0)} + \dd\sigma_{xx}^{(LL\ge1)}\,.
\end{equation}
Presently, Monte Carlo event generators which are used at flavour factories rely on exact fixed-order QED corrections and the LL approximation of higher-order effects, together with their consistent matching.

The most used techniques to account for multiple photon radiation are the QED structure function approach and the YFS exponentiation. \pdforweb{As more thoroughly explored in \secref{sec:gen}, the}{The} first approach is used in the {\sc BabaYaga@NLO} (Monte Carlo parton shower algorithm), MCGPJ (analytical QED structure functions in the collinear approximation) and \afkqed{}~\cite{Caffo:1994dm,Caffo:1997yy} (similarly to MCGPJ) generators, whereas the second approach is used in KKMC and \Sherpa.

\subsubsection{QED parton distribution functions} \label{sec:comp-bfoPDF}

In the \textit{collinear factorisation} approach, in order to resum the large $L_c$ terms, one relies on a factorisation formula~\cite{Kuraev:1985hb,AltarelliMartinelli,Nicrosini:1986sm}, similar to the standard QCD factorisation formula adopted at hadron colliders,
\begin{eqnarray}\label{eq:fact}
    \lefteqn{\dd\sigma_{e^+e^-\to X} =} \\
    & & \sum_{ij} \int {\dd}x_+ \dd x_-\, D_{i/e^+}(x_+,\mu^2,m_e^2)
    \, D_{j/e^-}(x_-,\mu^2,m_e^2) \, \dd\hat \sigma_{ij\to X}(x_+, x_-, Q^2,\mu^2)
    + \mathcal{O}\left(\frac{m_e^2}{Q^2}\right)\,, \nonumber
\end{eqnarray}
valid up to power corrections in the mass of the electron $m_e^2$ over some hard scale $Q^2$.
The term on the l.h.s.\ is the {\em particle-level} cross section 
for the process $e^+\,e^-\to X$, computed with massive electrons, whereas $\dd\hat{\sigma}_{ij\to{X}}$ appearing under integration on the r.h.s.\ is the {\em parton-level} or \emph{short-distance} cross section for the process $i\,j\to{X}$, which is free of electron mass singularities.
The collinear logarithms are resummed by means of the Parton Distribution Functions (PDF) $D_{i/e^{\pm}}(x^{\pm})$.
The partons entering the short-distance cross section are rescaled by a longitudinal momentum fraction $x_{\pm}$, and their nature can
coincide with that of the incoming particles i.e., $(i,j)=(e^+,e^-)$, or it can be different, e.g., $(i,j)=(\gamma,e^-),(e^-,e^-),\ldots$.
Finally, a factorisation scale $\mu^2$ appears both in the $D_{i/e^{\pm}}$ and in 
${\dd}\hat{\sigma}_{ij\to{X}}$. A  suitable factorisation scheme must be introduced in order to remove the collinear logarithms (or collinear divergences if massless electrons are used) present in the parton-level cross section.

At variance with hadronic PDFs, QED PDFs are entirely calculable with perturbative techniques.
One option is to resort to an iterative Monte Carlo procedure as detailed in  \secref{sec:comp-bfoPS}.
Alternatively, one can solve numerically or analytically the Dokshitzer--Gribov--Lipatov--Altarelli--Parisi (DGLAP) evolution equation \cite{Altarelli:1977zs,Gribov:1972ri,Gribov:1972rt,Dokshitzer:1977sg}, given the initial condition for the evolution at the initial scale $\mu_0^2 \simeq m_e^2$.
Analytical solutions with LL accuracy have been extensively used for numerical simulations at LEP and flavour factories. Neglecting the running of the QED coupling constant, the nonsinglet QED PDF  $D(x,\mu^2) \equiv D_{e^-/e^-}^{\rm LL}(x,\mu^2)$  (also called QED structure function in the literature) is solution of the 
DGLAP evolution equation 
\begin{equation}\label{eq:DGLAP}
    \mu^2\frac{\partial}{\partial \mu^2}D\left(x,\mu^2\right)=\frac{\alpha}{2\pi}\int_x^{1}\frac{{\rm d}z}{z}P_+(z) D\left(\frac{x}{z},\mu^2\right)\,,
\end{equation}
with initial condition $D(x,m_e^2) = \delta(1-x)$ and $P_+(z)$ the regularised $\ell\to \ell\gamma$ splitting  function. The structure function takes into account how radiation is emitted at all orders in the collinear limit and can be interpreted as the probability density of having a lepton with momentum $p'=xp$ and virtuality $\mu^2$ in a parent lepton with momentum $p$. 
In the literature, different analytical solutions have been obtained with different logarithmic accuracies: purely-soft Gribov--Lipatov~\cite{Gribov:1972rt}, hybrid additive\cite{Kuraev:1985hb,Nicrosini:1986sm,Cacciari:1992pz,Arbuzov:1999cq}, and hybrid factorised~\cite{Skrzypek:1990qs,Przybycien:1992qe,Skrzypek:1992vk} solutions. 
The typical expression of the hybrid additive solutions is of the form
\begin{align}\label{comp:LLpdf}
  D_{e^-/e^-}^{\rm LL}(x,Q^2) &=
  \frac{\exp\left[(3/4-\gamma_E)\frac{\eta}{2}\right]}{\Gamma(1+\frac{\eta}{2})}\frac{\eta}{2}(1-x)^{-1+\frac{\eta}{2}}
  -\frac{1}{4}\eta (1+x)+\mathcal{O}(\alpha^2)\,,\quad
  \eta = \frac{2 \alpha}{\pi} L_c \,,  
\end{align} 
while in the hybrid factorised solutions the Gribov--Lipatov exponentiated term is multiplied by finite-order terms. 
Such LL results are built out of an additive matching between a recursive solution up to some order in $\alpha$, and an all-order $\alpha$ solution valid in the region $x \to 1$. With $Q$ in the region of GeV we have $\eta \sim \cO(10^{-2})$. Therefore, the $(1-x)^{-1+\frac{\eta}{2}}$ factor results in a PDF that is very peaked towards $x=1$, where it diverges with an integrable singularity.
A last comment is in order: for an $e^+ e^-$ annihilation, the choice 
$\eta \to \beta = \frac{2 \alpha}{\pi} \left[ \log{(s/m_e^2)} - 1\right]$ in \eqref{comp:LLpdf} allows to correctly resum all the IR structure of the cross-section calculated with \eqref{eq:fact}~\cite{Montagna:1998sp}.

The next-to-leading logarithmic approximation for QED structure functions was first developed in~\cite{Berends:1987ab}.
Recently, there has been some additional progress to extend the accuracy to NLL.
In~\cite{Bertone:2019hks}, the electron-in-electron, positron-in-electron, and photon-in-electron PDFs have been calculated at NLL accuracy in the $\overline{\mathrm{MS}}$ factorisation and renormalisation scheme. The PDFs have been derived by solving the DGLAP equations both numerically and analytically,
by using as initial conditions for the evolution the ones derived in~\cite{Frixione:2019lga}.
In~\cite{Bertone:2022ktl}, these results have been improved in several
directions: first, with a DGLAP evolution featuring multiple fermion families
(leptons and quarks) in a variable flavour number scheme, i.e., by properly
including the respective mass thresholds; second, by taking into account an
alternative factorisation scheme, the $\Delta$ scheme~\cite{Frixione:2012wtz},
where the NLO initial condition are maximally simplified; third, by considering
two alternative renormalisation schemes, $\alpha(m_Z)$ and $G_\mu$ schemes
(where $\alpha$ is fixed).
Finally, in~\cite{Frixione:2023gmf}, the framework has been extended to include QCD and mixed QED-QCD effects in the DGLAP evolution.

NLL PDFs ready for phenomenology can be obtained with the public code
{\sc\small eMELA}~\cite{Bertone:2022ktl}. {\sc\small eMELA} is a stand-alone code, and can be linked to any external programme.
{\sc\small eMELA} can also provide PDFs with beamstrahlung effects according to
the procedure presented in~\cite{Frixione:2021zdp}.

\subsubsection{Parton shower} \label{sec:comp-bfoPS}

The parton shower procedure considered in this section is a Monte Carlo algorithm that provides an exact numerical solution to the 
DGLAP evolution equation for the nonsinglet LL QED structure function 
\eqref{eq:DGLAP}.
A regularised $\ell\to \ell\gamma$ splitting function, suitable for a Monte Carlo implementation, can 
be defined as
\begin{equation}\label{eq:regP+}
    P_+(z) = \frac{1+z^2}{1-z}\Theta(1-\varepsilon-z)-\delta(1-z)\int_0^{1-\varepsilon}{\rm d}y \frac{1+y^2}{1-y}\,,
\end{equation}
where $\varepsilon\ll1$ is an IR cutoff. In this way, \eqref{eq:DGLAP} can be rewritten as
\begin{equation}\label{eq:DGLAP_PS}
    \mu^2\frac{\partial}{\partial \mu^2}D\left(x,\mu^2\right)=\frac{\alpha}{2\pi}\int_x^{1-\varepsilon}\frac{{\rm d}z}{z}\frac{1+z^2}{1-z} D\left(\frac{x}{z},\mu^2\right) -\frac{\alpha}{2\pi}D\left(x,\mu^2\right)\int_x^{1-\varepsilon}{\rm d}z\frac{1+z^2}{1-z}\,.
\end{equation}

One can introduce the Sudakov form factor \cite{Sudakov:1954sw}
\begin{equation}\label{eq:sudakov}
\Delta(s_1,s_2) = \exp \left[-\frac{\alpha}{2 \pi} 
\int_{s_2}^{s_1} \frac{{\rm d} s'}{s'} \int_0^{1-\varepsilon} {\rm d}z\frac{1+z^2}{1-z} \right]\,.
\end{equation}
It represents the probability that the lepton evolves from virtuality $s_2$ to virtuality $s_1$ without emitting photons whose energy fraction is larger than $\varepsilon$. Then, \eqref{eq:DGLAP_PS} can be reworked in the integral form as
\begin{equation}\label{eq:DGLAP_sudakov} 
    D\big(x,\mu^2\big) = \Delta\left(\mu^2,m_\ell^2\right)D\left(x,m_\ell^2\right) + \frac{\alpha}{2\pi}\int_{m_\ell^2}^{\mu^2}\frac{{\rm d}s'}{s'}\Delta\left(\mu^2,s'\right)\int_x^{1-\varepsilon}\frac{{\rm d}z}{z}\frac{1+z^2}{1-z}D\left(\frac{x}{z},s'\right)\,.
\end{equation}
Iteratively, it becomes
\begin{multline}\label{eq:DGLAP_iterative}
    D(x,\mu^2) = \sum_{i=0}^\infty\prod_{j=1}^i\left[\frac{\alpha}{2\pi}\int_{m_\ell^2}^{s_{j-1}}\frac{{\rm d}s_j}{s_j}\Delta\left(s_{j-1},s_j\right)\int_{x/(z_1\cdot\ldots\cdot z_{j-1})}^{1-\varepsilon}\frac{{\rm d}z_j}{z_j}\frac{1+z_j^2}{1-z_j}\right]\\
    \times\Delta\left(s_{i},m_\ell^2\right)D\left(\frac{x}{z_1\cdot\ldots\cdot z_i},m_\ell^2\right)\,.
\end{multline}

The previous equation allows calculating the structure function $D(x,\mu^2)$ via a Monte Carlo method, according to an iterative procedure \cite{CarloniCalame:2000pz}. This way, a shower of photons is generated and the simulated $x$ distribution exactly follows the structure function. For leptonic processes, if the scale is set to $\mu^2=st/u$, this prescription also exponentiates the dominant contribution due to initial-final-state interference. Then, assuming that both initial-state and final-state particles emit photons, the corrected cross section at LL can be written as
\begin{align}  \label{eq:xsec_corrected}
  {\rm d}\sigma_{xx}^{(LL\geq1)}(s)=\int \Bigg(\prod_{i=1}^4 \dd x_i\,D\big(x_i,\mu^2\big)\Bigg)
  \ {\rm d}\sigma_{xx}^{(0)}(x_1x_2s)\Theta({\rm cuts})\,,
\end{align}
where $x_{1,2}$($x_{3,4}$) refer to initial(final)-state particles. For the $\gamma\gamma$ final state, the structure function is set to 1.

It is possible to exclusively generate the transverse momentum of electrons $p_\perp$ and photons, to go beyond the strictly collinear formulation of this method \cite{CarloniCalame:2001ny}. This can be done by generating the angle of the $k$-th photon according to the YFS formula \cite{Yennie:1961ad}
\begin{equation}\label{eq:phpperp}
  \cos \vartheta_k \propto - \sum_{i,j}^N \eta_i \eta_j
  \frac{1 - \beta_i \beta_j \cos\vartheta_{ij}}{(1-\beta_i \cos\vartheta_{ik})
    (1 - \beta_j \cos\vartheta_{jk})}\,.
\end{equation}
In the previous equation, $N$ is the total number of generated photons, $\beta_i$ is the speed of the $i$-th emitter, $\vartheta_{ij}$ is the angle between the $i$-th and the $j$-th particles and the symbol $\eta_i$ is equal to $1$($-1$) for incoming lepton (antilepton) or outgoing antilepton (lepton). Moreover, the diagonal terms of the YFS eikonal current account for terms of the kind $m_\ell^2/(p\cdot k)^2$ in the cross section. This makes the inclusion of finite mass corrections in the splitting functions unnecessary.

Through a matching procedure it is possible to combine the cross section \eqref{eq:xsec_corrected} with corrections up to a certain fixed order, thereby also including higher-order terms without double counting. At present, the exact NLO cross section has been matched with higher-order terms
\begin{equation}\label{eq:NLOPS}
    {\rm d}\sigma_{xx}^{(0)}+{\rm d}\sigma_{xx}^{(1)}+{\rm d}\sigma_{xx}^{(LL\ge2)}.
\end{equation}
According to \tabref{tab:logtower}, this amounts to calculating the first two rows and the first column.

In {\sc BabaYaga@NLO} \cite{Balossini:2006wc}, it is performed as follows. With no loss of generality, one can write the LL exponentiation for the emission of a single leg as
\begin{equation}\label{eq:LLxsec}
    {\rm d}\sigma_{xx}^{(LL\geq1)} = \Delta\left(Q^2,\varepsilon\right)
    \sum_{k=0}^\infty\frac{1}{k!}\cM_{xx\gamma_\varepsilon^k}^{(LL)}\,{\rm d}\Phi_k,
\end{equation}
where $\cM_{xx\gamma_\varepsilon^k}^{(LL)}$ is the squared tree-level amplitude for the emission of $k$ photons with energy fraction larger than $\varepsilon$ in LL approximation and 
$\dd\Phi_k$ is the exact $k+2$-body phase space. The expansion of \eqref{eq:LLxsec} at order $\cO(\alpha)$
\begin{align}\label{eq:xsecLLoal}
    \begin{split}
        {\rm d}\sigma_{xx}^{(LL@1)}&= \left[1-\frac{\alpha}{2\pi}\log\frac{Q^2}{m_\ell^2}\int_{0}^{1-\varepsilon}{\rm d}z\frac{1+z^2}{1-z}\right] \cM_{xx}^{(0)}\,{\rm d}\Phi_0
        +\cM_{xx\gamma_\epsilon}^{(LL)}\,{\rm d}\Phi_1\\
        &\equiv \left[1+C^{(LL@1)}(\varepsilon)\right]\cM_{xx}^{(0)}\,{\rm d}\Phi_0+
        \cM_{xx\gamma_\varepsilon}^{(LL)}\,{\rm d}\Phi_1\,,
    \end{split}
\end{align}
does not coincide with the exact 
\begin{equation}\label{eq:xsecoal}
    {\rm d}\sigma_{xx}^{(1)}=\left[1+C^{(1)}(\varepsilon)\right] \cM_{xx}^{(0)}\,{\rm d}\Phi_0+
    \cM_{xx\gamma_\varepsilon}^{(0)}\,{\rm d}\Phi_1\,.
\end{equation}
The coefficient $C^{(1)}(\varepsilon)$ contains all the virtual and soft (i.e., energy fraction less than $\varepsilon$) real squared matrix elements at $\cO(\alpha)$ and $\cM^{(0)}_{xx\gamma_\varepsilon}$ is the exact tree-level amplitude with the emission of a photon with energy larger than $\varepsilon$.
It is crucial that $C^{(LL@1)}$ have the same logarithmic structure as $C^{(1)}$ and   
$\cM_{xx\gamma_\varepsilon}^{(LL)}$ have the same singular behaviour as $\cM_{xx\gamma_\varepsilon}^{(0)}$. This makes it possible to define two finite and IR- and collinear-safe coefficients
\begin{align}\label{eq:FsvFh}\begin{split}
    F_{SV}&=1+\left(C^{(1)}-C^{(LL@1)}\right)\,,\\
    F_H&= 1+\frac{\cM_{xx\gamma_\varepsilon}^{(0)}-\cM_{xx\gamma_\varepsilon}^{(LL)}}{\cM_{xx\gamma_\varepsilon}^{(LL)}} \,.
\end{split}\end{align}
These two correction factors can be made explicit in \eqref{eq:xsecoal}, which becomes
\begin{equation}
    {\rm d}\sigma_{xx}^{(1)}=F_{SV}\big(1+C^{(LL@1)}\big) \cM_{xx}^{(0)}\,\dd\Phi_0
    +F_H \cM_{xx\gamma_\varepsilon}^{(LL)}\,\dd\Phi_1\,.
\end{equation}
Thus, the master formula \eqref{eq:LLxsec}, can be improved by writing the resummed matched cross section
\begin{equation}\label{eq:final_matching}
    {\rm d}\sigma_{xx}^{(0)}+{\rm d}\sigma_{xx}^{(1)}+{\rm d}\sigma_{xx}^{(LL\ge2)}= 
    F_{SV}\, \Delta\big(Q^2,\varepsilon\big)
    \sum_{k=0}^\infty\frac{1}{k!}\Bigg(\prod_{i=1}^kF_{H,i}\Bigg)
    \cM_{xx\gamma_\varepsilon^k}^{(LL)}\, \dd\Phi_k\,,
\end{equation}
where the $i$ subscript refers to the $i$-th emitted photon. The expansion at ${\cal O}(\alpha)$ of \eqref{eq:final_matching} exactly corresponds to \eqref{eq:xsecoal}. Moreover, the resummation of higher-order terms is conserved without any double counting.

\subsubsection{YFS resummation} \label{sec:comp-bfoYFS}

The Yennie--Frautschi--Suura theorem~\cite{Yennie:1961ad} details how logarithms associated with soft photons, both real and virtual, can be resummed to all orders, which in turn renders the entire perturbative expression IR finite order-by-order. This subsequent expression can be interpreted as a subtraction scheme, which allows us to include higher-order IR finite corrections in a systematic form. 
Another feature of the YFS theorem is how it treats the multi-photon phase space.
In contrast to the collinear approach of an electron structure function, the
YFS approach explicitly generates resolved photons, with a resolution criterion given by an energy and
angle cut-off. In so doing, the full kinematic structure of scattering events is reconstructed, which leads to a
straightforward implementation of the YFS method as both cross-section calculator and an event generator. Of course, the same properties can be achieved in the collinear picture with the supplementation of an appropriate parton shower algorithm. 
In essence, the YFS method allows us to take the cross-section expression for a generic process with the addition of infinitely many photons, both real and virtual, and rewrite it into an IR-finite expression, suitable for implementation in a Monte Carlo event generation.
For the specific case of $e^+ e^- \to x x$ considered here, where $xx$ represents any two-particle final state, the cross section is given by\footnote{For clarity and ease of understanding, the notation used in this section has been adapted from the standard YFS format used in~\cite{Krauss:2022ajk}.}
\begin{equation}
\label{yfs:xs}
    \dd\sigma_{xx}^{\mathrm{YFS}} =  \sum_{k=0}^{\infty} \frac{1}{k!}\,
    \dd\Phi_Q
    \left[\prod_{i=1}^{k}d{\Phi_i^\gamma}\right]
    \left(2\pi\right)^4
    \delta^4\left(\sum_{a=1}^2p_a
                  -\sum_{b=3}^{4} p_b
                  -\sum_{c=1}^{k} p^\gamma_c\right)
    \left\lvert \sum_{n=0}^{\infty}  \cA^{(n)}_{xx\gamma^k}\right\rvert^2\,,
\end{equation}
where $n$ denotes the number of virtual photons (for a process which is tree-level at LO this is equal to the number of loops in QED), and $\gamma^k$ denotes the emission of $k$ real photons.

If we consider the addition of one virtual photon, the amplitude will factorise, in the soft limit, as
\begin{align}\label{Eq:virtual_factorisation}
\cA_{xx}^{(1)} = \alpha \mathcal{B} A_0^0 + A_0^1\,,
\end{align}
where $\alpha$ is the QED coupling, and the remainders $A_k^n$ are the 
IR-subtracted residuals from the emission of $n$ virtual photons
($k$ is the number of real photons, which have associated IR divergences; these will be subtracted later).
Therefore, $A_0^0=\cA_{xx}^{(0)}$ is the LO matrix element.
The remainder $A_k^n$ is of 
relative size $\mathcal{O}(\alpha^{n+k/2})$ with respect to the LO amplitude.
We introduce the factor $\mathcal{B}$, which is an integrated 
off-shell eikonal encoding the universal soft-photon limit.
For a single dipole, $\mathcal{B}$ is given by
\begin{equation}
 \mathcal{B}_{ij}
=
-\frac{i}{8\pi^3}Z_iZ_j\theta_i\theta_j\int\frac{\dd^4p_\gamma}{p_\gamma^2}
\left(\frac{2p_i\theta_i-p_\gamma}{p_\gamma^2-2(p_\gamma\cdot p_i)\theta_i}
  +\frac{2p_j\theta_j+p_\gamma}{p_\gamma^2+2(p_\gamma\cdot p_j)\theta_j}\right)^2\,,
\end{equation}
where $Z_i$ and $Z_j$ are the charges of particles $i$ and $j$ in units of the positron charge, and $\theta_{i,j} = 1\,(-1)$ for final (initial) state particles.

If we consider the corrections due to up to two virtual photons, we see
\begin{align}
\cA_{xx}^{(0)}  = & A_0^0, \nonumber\\
\cA_{xx}^{(1)}  = & A_0^1 + \alpha \mathcal{B} A_0^0, \nonumber\\
\cA_{xx}^{(2)}  = & A_0^2 + \alpha \mathcal{B} A_0^1 + \frac{(\alpha \mathcal{B})^2}{2!}A_0^0\,.
\end{align} 
This generalises to any number $n$ of virtual photons, 
\begin{align}
    \cA_{xx}^{(n)} = \sum_{r=0}^{n}
    A_0^{n-r}\frac{(\alpha \mathcal{B})^r}{r!}\,,
\end{align}
and resumming all virtual photon emissions therefore gives
\begin{align}
    \sum_{n=0}^\infty
    \cA_{xx}^{(n)}  =
    \exp(\alpha \mathcal{B})\sum_{n=0}^\infty A_0^{n}\,.
\end{align}
Due to the Abelian nature of QED, this factorisation can be further generalised to produce squared matrix 
elements that include any number of additional real photon emissions, such that
\begin{align}
    \left|\sum_{k=0}^\infty
    \cA_{xx\gamma^k}^{n}\right|^2  = &
    \exp(2\alpha \mathcal{\tilde{B}})
    \left|\sum_{n}^\infty
    A_{k}^{n}\right|^2\,.
\end{align}

\noindent
To show this, we consider first the case of a single real photon. This can be expressed as
\begin{align}
    \frac{1}{2(2\pi)^3}\left|\sum_{n=0}^\infty
    A_1^{n}\right|^2
     =
    \tilde{S}(p_\gamma)
    \left|\sum_{n=0}^\infty A_0^{n}\right|^2 +
    \sum_{n=0}^\infty
    \tilde{\beta}_{1}^{n}(p_\gamma)\,.
\end{align}

In this expression, all the singularities, due to the emission of soft real photons, are contained within the eikonal, 
\begin{equation} \label{EQ::Eikonal}
\tilde{S}(p_\gamma) =  \sum_{i,j}\frac{\alpha}{4\pi^2}\,Z_iZ_j\theta_i\theta_j
           \left(\frac{p_i}{p_i\cdot p_\gamma}-\frac{p_j}{p_j\cdot p_\gamma}\right)^2,
\end{equation}
while the remaining term $\tilde{\beta}_{k}^{n}$ are the IR-finite residuals. 
Extracting all real-emission soft photon divergences through eikonal factors, 
the squared matrix element for any $k$ real emissions, 
summed over all possible virtual photon corrections, can be written as
\begin{align}\label{eq:AllVirt} 
    \lefteqn{\left(\frac{1}{2(2\pi)^3}\right)^{k}
    \left|\sum_{n=0}^\infty 
    A_{k}^{n}\right|^2} \nonumber\\ 
    \pdforweb{&}{} = &
    \;\;\;\tilde{\beta}_0 \prod_{i=1}^{k}\left[\tilde{S}(p^\gamma_i)\right]
    +\sum_{i=1}^{k}\left[\frac{\tilde{\beta}_1(p^\gamma_i)}{\tilde{S}(p^\gamma_i)}\right]
    \prod_{j=1}^{k}\left[\tilde{S}(p^\gamma_i)\right]
    +\sum_{\genfrac{}{}{0pt}{}{i,j=1}{i<j}}^{k}\left[
    \frac{\tilde{\beta}_2(p^\gamma_i,p^\gamma_j) }{\tilde{S}(p^\gamma_i)\tilde{S}(p^\gamma_j)}\right]
    \prod_{l=1}^{k}\left[\tilde{S}(p^\gamma_l)\right]
    +\dots \nonumber\\
    \pdforweb{&&{}}{&}+\tilde{\beta}_{k-1}(p^\gamma_1,\ldots,p^\gamma_{i-1},p^\gamma_{i+1},\ldots,p^\gamma_k)
         \sum_{i=1}^{k}\tilde{S}(p^\gamma_i)+\tilde{\beta}_{k}(p^\gamma_1,\ldots,p^\gamma_k)\,.
\end{align}
Within this expression, all $\tilde{\beta}_i$ are free from all IR divergences due to either real or virtual photon emissions. For convenience, we have introduced the following notation,
\begin{align}
\tilde{\beta}_{k} = 
\sum_{n=0}^\infty
\tilde{\beta}_{k}^{n}\,,
\end{align}
in which we suppress the summation over virtual contributions.
\noindent
To recombine all terms into an expression for the inclusive cross section and facilitate the cancellation of all IR singularities, it is useful to define an unresolved region $\Omega$ in which the kinematic impact of any real photon emission is unimportant and the photon itself is undetectable.
Integrating over this unresolved real emission phase space gives the integrated on-shell eikonal $\tilde{\mathcal{B}}$, defined by
\begin{equation}
2\alpha \tilde{\mathcal{B}}(\Omega) = \int \frac{\dd^3{p_\gamma}}{p_\gamma^0}\,
                             \tilde{S}(p_\gamma)
                             \big[1-\Theta(p_\gamma,\Omega)\big]\,,
\end{equation}
which contains all IR poles due to real soft photon emission.
With these corrections, we can express the YFS theorem for an arbitrary process as 
\begin{align}\label{eq:masterYFS}
    \dd\sigma^{\mathrm{YFS}} = \sum_{k=0}^\infty
        \frac{e^{Y(\Omega)}}{k!}\,
        \dd{\Phi_Q} 
        \left[\prod_{i=1}^{k}\dd{\Phi_i^\gamma}\,\tilde{S}(p^\gamma_i)\,\Theta(p^\gamma_i,\Omega)\right]
        \left(\tilde{\beta}_0
        + \sum_{j=1}^{k}\frac{\tilde{\beta}_{1}(p^\gamma_j)}{\tilde{S}(p^\gamma_j)}
        +\sum_{\genfrac{}{}{0pt}{}{j,l=1}{j<l}}^{k}
        \frac{\tilde{\beta}_{2}(p^\gamma_j,p^\gamma_l)}{\tilde{S}(p^\gamma_j)\tilde{S}(p^\gamma_l)}
        + \cdots
        \right)\,,
\end{align}
with the YFS form factor
\begin{equation}\label{EQ::FormFactor}
Y(\Omega) = 2 \alpha \sum_{i<j}  \left[
    \mathcal{B}_{ij} +
    \tilde{\mathcal{B}}_{ij}(\Omega) \right]\,.
\end{equation}
Therein, all IR singularities originating from real and virtual soft photon emission, contained in $\tilde{\mathcal{B}}$ and $\mathcal{B}$ respectively, cancel, leaving a finite remainder.

The most sophisticated variant of YFS is the Coherent Exclusive Exponentiation (CEEX)~\cite{Jadach:2000ir} currently only available in {\sc KKMCee}\pdforweb{, see \secref{sec:kkmc}}{}. Contrary to the standard YFS approach (EEX), the CEEX variant of YFS exponentiation is implemented in terms of the IR-free $\tilde{\beta}_i$ residuals constructed at the amplitude level, rather than using the spin-summed squared $\tilde{\beta}_i$. CEEX treats correctly to infinite order not only IR cancellations but also QED interferences and narrow resonances.
The YFS resummation (EEX) has been implemented in \textsc{Sherpa}~\cite{Schonherr:2008av}\pdforweb{, see \secref{sec:sherpa}}{}. The YFS approach has also been used in several earlier codes, such as \textsc{Bhwide}~\cite{Jadach:1995nk}, \textsc{Bhlumi}~\cite{Jadach:1996is} and it has been implemented in \textsc{Herwig}~\cite{Hamilton:2006xz,Bellm:2015jjp}. In a more recent development, the combination of the YFS exponentiation with collinear factorisation has been considered~\cite{Jadach:2023pka}.


\subsection{Internal hadrons} \label{sec:comp-inthad}

This section provides a brief overview on the methods used to calculate nonperturbative hadronic corrections for the processes~\eqref{intro:scan} and \eqref{intro:return} at NNLO accuracy. These corrections correspond to Feynman diagrams with insertions of Green's functions built from any number of electromagnetic currents
\begin{align}
    j_\text{em}^\mu(x) = \sum_q Q_q \bar{q}(x) \gamma^\mu q(x)
\end{align}
in pure QCD, where the sum runs over the three lightest quarks $q$ with charge $Q_q$. Correlation functions with an odd number of currents vanish as a consequence of Furry's theorem. Up to NNLO, the only contributing Green's functions are therefore the HVP tensor
\begin{align}
    \Pi_h^{\mu\nu}(q) =
    \parbox{2cm}{\includegraphics[width=2cm]{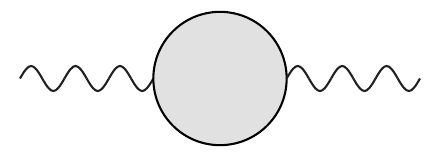}}
    = i\,e^2 \int \mathrm{d}^4 x \, e^{-i q x} 
    \langle 0 | T \{ j_\text{em}^\mu(x) j_\text{em}^\nu(0) \} |0\rangle
    = \Pi_h(q^2) (g^{\mu\nu} q^2 - q^\mu q^\nu)
\end{align}
and the HLbL tensor
\begin{align}
    \Pi_h^{\mu\nu\lambda\sigma}(q_1,q_2,q_3) &=
    \parbox{3cm}{\includegraphics[width=3cm]{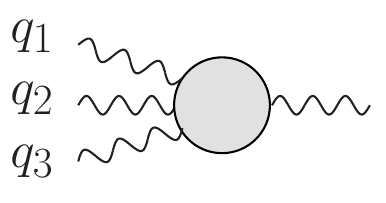}}
    \nonumber \\
    &=  i \int \mathrm{d}^4 x \, \mathrm{d}^4 y \, \mathrm{d}^4 z \,
    e^{-i(q_1 \cdot x + q_2 \cdot y + q_3 \cdot z)} 
    \langle 0 | 
        T \{ j_\text{em}^\mu(x) j_\text{em}^\nu(y)
         j_\text{em}^\lambda(z) j_\text{em}^\sigma(0) \} 
    |0\rangle \, .
\end{align}
The HLbL tensor is only relevant for the process $e^+\,e^-\to\gamma\,\gamma$ at NNLO. For our main processes it only contributes beyond NNLO. In the following, we therefore mainly focus on HVP corrections. 

At the energy scales of the experiments discussed in \pdforweb{\secref{sec:exp}}{here}, these hadronic correlation functions are nonperturbative and cannot be calculated in perturbation theory. Instead they either have to be computed in lattice gauge theory or extracted from experimental data making use of unitarity and analyticity. We only discuss the latter approach. In the case of the HVP function, the optical theorem (unitarity) gives a relation between its imaginary part and the production cross section of hadrons
\begin{align}
    \text{Im}\, \Pi_h 
    = - \frac{s}{4\pi\alpha}\,\sigma(e^+e^- \to \gamma^* \to \text{had}) \, .
\end{align}
Analyticity of the HVP function then allows for its reconstruction via the dispersion integral
\begin{align}\label{comp:hvp_disprel}
    \frac{\Pi_h^\text{ren}(q^2)}{q^2}
    = \frac{1}{\pi} \int_{4m_\pi^2}^{\infty} 
    \frac{\mathrm{d} z}{z} \frac{\text{Im} \Pi_h(z)}{(z-q^2-i0)} \, ,
\end{align}
where we introduced the renormalised (subtracted) $\Pi_h^\text{ren}(q^2)\equiv\Pi_h(q^2)-\Pi_h(0)$. 
There is a long history of specialised tools such as \texttt{alphaQED}\cite{Jegerlehner:2001ca,Jegerlehner:2006ju,Jegerlehner:2011mw} and \texttt{HVPTools}\cite{Davier:2010rnx} that compute the HVP function based on the described approach. 
The recent most precise tabulations of the full VP are provided by the KNT group
(v3.1, 2022)~\cite{Keshavarzi:2019abf}, Jegerlehner (hadr5x23 from
\texttt{alphaQED23} package, 2023)~\cite{Jegerlehner:hvp19},
and Novosibirsk (NSK) VP (v2.9, 2022)~\cite{Ignatov:2008bfz,Ignatov:hvp}.

Similarly, the HLbL tensor has been studied in detail in the context of the anomalous magnetic moment of the muon, $a_\mu^\text{HLbL}$, see~\cite{Aoyama:2020ynm} (including a corresponding prediction for the HLbL contribution based on~\cite{Melnikov:2003xd,Masjuan:2017tvw,Colangelo:2017fiz,Hoferichter:2018kwz,Gerardin:2019vio,Bijnens:2019ghy,Colangelo:2019uex,Pauk:2014rta,Danilkin:2016hnh,Jegerlehner:2017gek,Knecht:2018sci,Eichmann:2019bqf,Roig:2019reh}). In particular, a dispersive formalism was developed that, in analogy to~\eqref{comp:hvp_disprel}, aims at reconstructing the entire HLbL tensor in terms of its singularities~\cite{Colangelo:2014dfa,Colangelo:2014pva,Colangelo:2015ama}. The dominant intermediate states that contribute in such a dispersive representation are pseudoscalar poles $P=\pi^0,\eta,\eta'$~\cite{Masjuan:2017tvw,Hoferichter:2018kwz,Gerardin:2019vio,Hoferichter:2018dmo}, two-meson cuts~\cite{Colangelo:2017fiz,Colangelo:2017qdm}, and the corresponding $S$-wave rescattering, the latter encoding the effects of scalar resonance exchanges~\cite{Danilkin:2021icn}, e.g., $S=f_0(500)$, $f_0(980)$, and $a_0(980)$. Subleading corrections arise from axial-vector exchanges, $A=f_1(1285)$, $f_1(1420)$, and $a_1(1260)$, see~\cite{Hoferichter:2020lap,Zanke:2021wiq,Hoferichter:2023tgp,Hoferichter:2024fsj,Ludtke:2024ase} for their incorporation into the dispersive formalism, and tensor mesons~\cite{Ludtke:2023hvz}, where, in the case of $T=f_2(1270)$, corrections beyond a narrow-width approximation can again be captured by two-meson rescattering~\cite{Hoferichter:2019nlq,Danilkin:2019opj}.

In the application to $e^+e^-$ reactions, a dispersive approach to HLbL could be established similarly to the application for $a_\mu^\text{HLbL}$, with a few important differences. On the one hand, the dispersive frameworks for $a_\mu^\text{HLbL}$ are formulated either in four-point kinematics with fixed $t=q_2^2$~\cite{Colangelo:2015ama,Colangelo:2017fiz}, or, alternatively, directly in the soft-photon limit~\cite{Ludtke:2023hvz,Ludtke:2024ase}, whereas for $e^+e^-\to\gamma\gamma$, one requires doubly-virtual four-point kinematics with Mandelstam variables different from the photon virtualities. Simplifications compared to the case of $a_\mu^\text{HLbL}$ occur in 
the $s$-channel: first, $P$ and $S$ contributions are helicity suppressed, and thus likely irrelevant for the $e^+e^-$ channel due to the resulting scaling with $m_e$. Second, the on-shell two-photon coupling vanishes for $A$ 
on account of the Landau--Yang theorem~\cite{Landau:1948kw,Yang:1950rg}. Accordingly, the first resonantly enhanced contribution without helicity suppression in $e^+e^-\to\gamma\gamma$ is due to tensor states in the $s$-channel. Away from the resonance region, the HLbL contribution is likely dominated by $t$-channel exchanges, which are neither helicity suppressed nor forbidden by Landau--Yang, by the charged pion box, as well as short-distance contributions~\cite{Melnikov:2003xd,Colangelo:2019uex,Bijnens:2021jqo,Bijnens:2022itw}. To estimate the impact of the HLbL contribution, a simple strategy could thus rely on a narrow-width approximation for the $f_2(1270)$, with transition form factors constrained by data on $\gamma\gamma^{(*)}\to\pi\pi$~\cite{Belle:2015oin,ParticleDataGroup:2022pth} and their asymptotic behaviour~\cite{Hoferichter:2020lap}, to estimate $s$-channel effects. The $t$-channel contributions could be estimated, e.g., for $P=\pi^0$ using the known transition form factor, whereas the pion box could be evaluated with input for the pion VFF~\cite{Colangelo:2015ama}. A dispersive treatment beyond these leading contributions would need to deal with the issue of kinematic singularities in the tensor decomposition, which so far has only been addressed in the kinematic configuration of $a_\mu^\text{HLbL}$.

Once the HVP and HLbL tensors have been determined through data, their contribution to the processes can be computed perturbatively. This is trivial for diagrams where the hadronic tensors factorise from loop integrals. In the case of nonfactorisable hadronic corrections, the loop integral has to be performed semi-numerically since the hadronic tensors are only available as numerical routines. In the case of HVP corrections this is rather straightforward. After inserting the dispersive representation~\eqref{comp:hvp_disprel} into the amplitude and exchanging the order of integration, the loop integral can be performed with standard methods where $z$ simply acts as a photon mass. This yields a one-dimensional integration over $z$ that can be done numerically. This \textit{dispersive approach}\cite{Cabibbo:1961sz} has been used to calculate NNLO hadronic corrections to the muon decay~\cite{vanRitbergen:1998hn,Davydychev:2000ee}, Bhabha scattering~\cite{Actis:2007fs,Kuhn:2008zs,CarloniCalame:2011zq}, and muon--electron scattering~\cite{Fael:2019nsf,Budassi:2021twh}. An alternative approach to the dispersive technique is the \textit{hyperspherical method}~\cite{Levine:1974xh,Levine:1975jz,Fael:2018dmz}. It is based on the idea that the loop momentum routing can be chosen such that the HVP function only depends on the square of the loop momentum. In this representation the angular integration can be performed analytically. The remaining one-dimensional radial integral over the HVP function can then be computed numerically. Contrary to the dispersive approach, the integration is over space-like momenta in this case. Narrow hadronic resonances are therefore avoided, which represents a significant advantage. However, the method entails subtle analytic continuations from the Euclidean to the Minkowski region, which significantly complicates the application of the method to processes with time-like momentum transfer.


\subsection{Hadronic final states} 
\label{sec:comp-fshad}

\subsubsection{Strategies for radiative corrections to low-energy hadronic amplitudes}
\label{sec:comp-fshad-strategies}

In processes with hadronic final states, the nonperturbative nature of the strong interactions at low energies significantly complicates the description of radiative corrections compared to purely leptonic processes. On the one hand, the strong dynamics itself requires nonperturbative techniques, on the other hand the additional interplay with QED corrections constitutes a highly nontrivial problem. In principle, the QED interaction can still be treated perturbatively. At each order in QED perturbation theory, nonperturbative matrix elements show up, which can be defined in pure QCD. One complication is the fact that the parameters of pure QCD (i.e., strong interaction in the limit of vanishing electromagnetic coupling) need to be expressed in terms of the physical parameters in a scheme- and scale-dependent matching~\cite{Gasser:2003hk,Gasser:2010wz,Carrasco:2015xwa}. Since the masses of the hadronic states depend on electromagnetic effects, QED corrections also lead to shifted thresholds. In practice, one often chooses a different approach than a strict separation of QED and QCD effects, e.g., the LO HVP contribution to $g-2$ is conventionally defined to be the one-particle-irreducible hadronic two-point function including photonic corrections~\cite{Aoyama:2020ynm}. The separation by topologies leads to a fully consistent treatment of higher-order insertions of HVP corrections, but does no longer follow a strict power counting in $\alpha$.

For the description of the nonperturbative strong dynamics and the interaction of the hadronic states with photons, different techniques and approximations with varying levels of rigour exist. Lattice QCD provides an approach based on first principles. The restriction of lattice simulations to Euclidean space severely impacts the applicability to Minkowski-space processes~\cite{Luscher:1986pf,Lellouch:2000pv}, but recent years have seen promising conceptual developments.
At very low energies, the interaction of the pseudoscalar Goldstone bosons with photons and leptons is described by ChPT~\cite{Gasser:1983yg,Gasser:1984gg,Urech:1994hd,Neufeld:1995mu,Knecht:1999ag}. While this effective-field-theory framework consistently describes structure-dependent radiative corrections and has been applied to a variety of low-energy processes~\cite{Cirigliano:2001er,Cirigliano:2001mk,Gasser:2002am,Cirigliano:2002ng,Cirigliano:2007ga,Colangelo:2008sm,Hoferichter:2009ez,Hoferichter:2009gn,Stoffer:2013sfa}, higher-order contributions involve poorly known nonperturbative low-energy constants. Furthermore, the energies of interest are often beyond the range of applicability of ChPT. Dispersion theory provides an approach to the strong dynamics based on the fundamental principles of unitarity, analyticity, and crossing, establishing model-independent relations between different measurable processes. While for the leading hadronic processes in \eqref{intro:scan} dispersion theory is well established~\cite{Colangelo:2018mtw,Hoferichter:2019mqg}, dispersive treatments of the corresponding radiative corrections are only partially available or under development~\cite{Monnard:2021pvm,Colangelo:2022prz,Colangelo:2022lzg,Hoferichter:2023bjm}, as will be discussed in more detail in \secref{sec:comp-eepipi}, \secref{sec:comp-eepipigamma}, and \secref{sec:comp-eepipipi}. Often, a combination of dispersive techniques with input from ChPT and, if available, lattice QCD provides the ideal description of radiative corrections in hadronic processes by taking into account as many theoretical and experimental constraints as possible~\cite{Seng:2021boy,Seng:2022wcw}. 

In practice, for many processes the more rigorous approaches mentioned above are either not applicable or not yet available. In this case, one typically resorts to models of the interaction of hadrons with photons. A hadronic model is a phenomenological approach that is not fundamentally derived from QCD and comes with intrinsic uncertainties that cannot be systematically reduced and can be assessed only by comparison with more rigorous approaches. For $e^+e^-\to\pi^+\pi^-$, existing Monte Carlo tools are all based on different levels of hadronic modelling, which we categorise in the following.
\begin{description}
	\item[``sQED''] Scalar QED: this is the $U(1)$ gauge theory describing the renormalisable interaction of (pseudo-)scalars with photons. Its use is motivated by the fact that sQED corresponds to the subset of renormalisable interactions of LO ChPT. In contrast to ChPT, sQED assumes pions to be pointlike. Note that LO ChPT includes further nonrenormalisable interactions, which at the loop level lead to UV divergences that necessitate the introduction of NLO ChPT counterterms. For loop processes including photons and leptons, these are the counterterms of~\cite{Urech:1994hd,Neufeld:1995mu,Knecht:1999ag}. sQED avoids the chiral counterterms by neglecting the effects of pion structure.
	\item[``F$\times$sQED''] Scalar QED multiplied by form factors for external virtualities~\cite{Tracz:2018}: this prescription preserves the nice properties of scalar QED, such as renormalisability and the cancellation of IR divergences, while keeping the dominant form-factor effects, but only the virtualities corresponding to the $e^+e^-$ and $\pi^+\pi^-$ invariant masses can be taken into account in this way.
	\item[``GVMD''] Generalised vector-meson dominance~\cite{Ignatov:2022iou}: the photon--pion coupling is modified by (a sum of) vector-meson Breit--Wigner propagators. This approach allows one to include form-factor effects within loops. However, the form factor does not fulfill analyticity constraints and the use of a constant vector-meson width introduces unphysical subthreshold imaginary parts.
	\item[``FsQED''] Form-factor sQED~\cite{Colangelo:2014dfa,Colangelo:2015ama,Colangelo:2022lzg}: in this approach, form factors are kept everywhere by isolating pion poles in dispersion theory. This allows one to account for form-factor effects inside loop integrals in a well-defined manner as long as the required matrix elements are sufficiently simple so that pion poles are uniquely defined.
	\item[``full''] Full hadronic matrix elements: the inclusion of a more complete description of the hadronic matrix elements as explained above is not yet available in existing Monte Carlo tools. However, as we will explain below, we expect even FsQED to be insufficient for more complicated matrix elements, such as $\pi^+\pi^-\to3\gamma^*$. This is on the one hand due to the fact that the definition of pion poles becomes ambiguous as it depend on the choice of dispersion relations, on the other hand FsQED does not account for other potentially important intermediate states beyond pion poles.
\end{description}

\subsubsection[Structure-dependent radiative corrections in \texorpdfstring{$e^+e^-\to\pi^+\pi^-$}{ee->pipi}]{Structure-dependent radiative corrections in $\boldsymbol{e^+e^-\to\pi^+\pi^-}$}
\label{sec:comp-eepipi}

\begin{figure}[t]
	\centering
	\ampamp{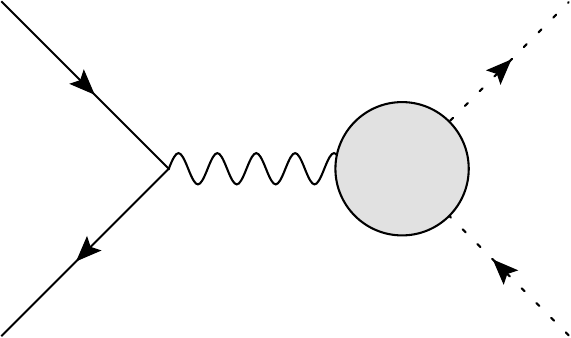}{figs/had/LO.pdf}
	\caption{LO contribution to the $e^+e^-\to\pi^+\pi^-$ cross section.}
	\label{fig:eepipi-LO}
\end{figure}

\begin{figure}[t]
	\centering
    \begin{tabular}{lcc}
	ISC: &
		\ampamp{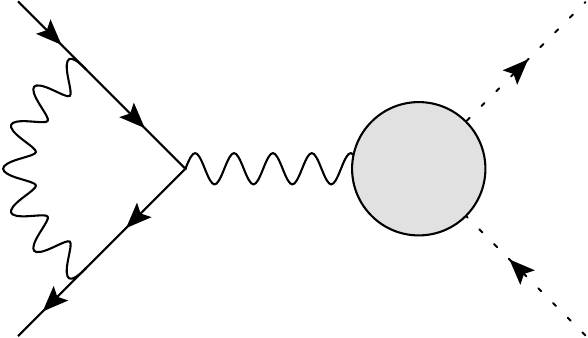}{figs/had/LO}
		&
		\ampamp{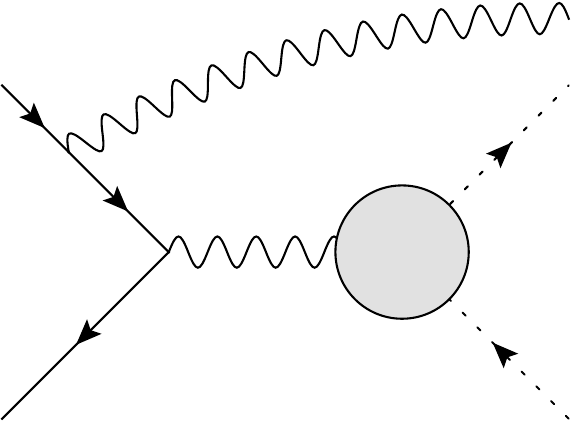}{figs/had/NLO-ISR}
		\\[10pt]
	FSC: &
		\ampamp{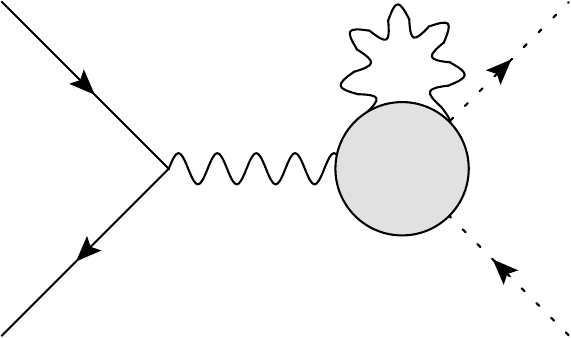}{figs/had/LO}
		&
		\ampamp{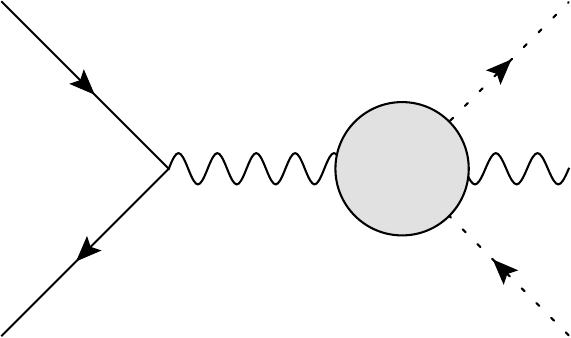}{figs/had/NLO-FSR}
		\\[10pt]
	mixed: &
		\ampamp{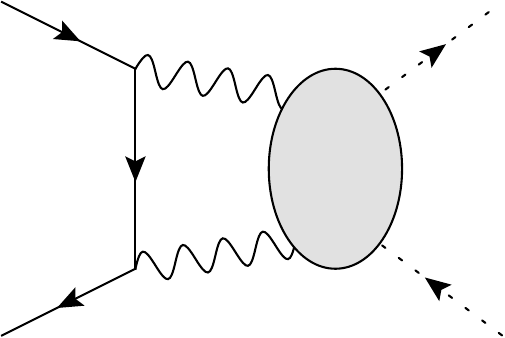}{figs/had/LO}
		&
		\ampamp{figs/had/NLO-ISR}{figs/had/NLO-FSR}
		\\[10pt]
	VP: &
		\ampamp{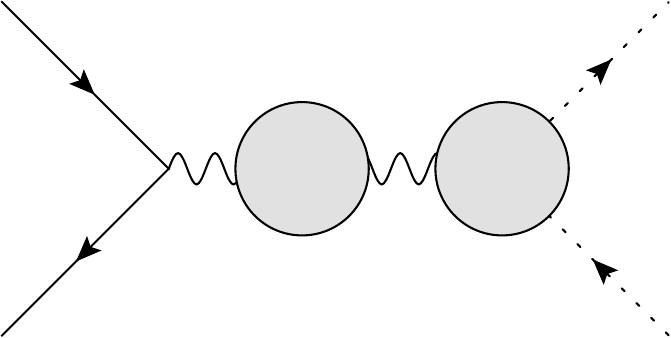}{figs/had/LO}
    \end{tabular}
	\caption{Representative NLO contributions to the $e^+e^-\to\pi^+\pi^-(\gamma)$ cross section: the first line shows ISC, consisting of initial-state virtual corrections and ISR. The second line shows FSC, consisting of virtual and real photon corrections to the pion VFF. The third line shows mixed corrections and the last additional VP.}
	\label{fig:eepipi-NLO}
\end{figure}

The photon-inclusive process $e^+e^-\to\pi^+\pi^-(\gamma)$ is used to measure the pion electromagnetic form factor in $e^+e^-$ experiments in scan mode. At leading order, the cross section shown in \figref{fig:eepipi-LO} is directly proportional to the squared modulus of the pion vector form factor (VFF) in pure QCD, defined by the current matrix element
\begin{equation}
\label{VFF_def}
  \langle 0|j_\mu(0)|\pi^+(p_+)\pi^-(p_-)\rangle = (p_+-p_-)_\mu F_\pi(s)\,,\qquad s=(p_++p_-)^2\,.  
\end{equation}
Comprehensive dispersive analyses of the form factor are available~\cite{Ananthanarayan:2013zua,Ananthanarayan:2018nyx,Colangelo:2018mtw,Colangelo:2020lcg,Colangelo:2022prz}. The representation of~\cite{Colangelo:2018mtw} can be used in the energy range up to $1\,\GeV$ and reads
\begin{equation}
	\label{eq:VFF}
	F_\pi(s) = \Omega_1^1(s) \times G_\omega(s) \times G_\mathrm{in}^N(s) \, ,
\end{equation}
where
\begin{equation}
	\Omega_1^1(s) = \exp\left\{ \frac{s}{\pi} \int_{4\mpi^2}^\infty ds^\prime \frac{\delta_1^1(s^\prime)}{s^\prime(s^\prime-s)} \right\}
\end{equation}
is the Omn\`es function with the elastic $\pi\pi$-scattering $P$-wave phase shift $\delta_1^1(s)$ as input~\cite{Ananthanarayan:2000ht,Caprini:2011ky}. The second factor in~\eqref{eq:VFF} describes the resonantly enhanced isospin-breaking $\rho$--$\omega$ interference effect
\begin{equation}
	G_\omega(s) = 1 + \frac{s}{\pi} \int_{9\mpi^2}^\infty ds^\prime \frac{\Im\,g_\omega(s^\prime)}{s^\prime(s^\prime-s)} \left( \frac{1 - \frac{9\mpi^2}{s^\prime}}{1 - \frac{9\mpi^2}{m_\omega^2}} \right)^4 \, , \quad g_\omega(s) = 1 + \epsilon_\omega \frac{s}{(m_\omega - \frac{i}{2} \Gamma_\omega)^2 - s} \, ,
\end{equation}
and additional inelastic contributions are parameterised by a conformal polynomial $G_\mathrm{in}^N(s)$ with a cut starting at the $\pi^0\omega$ threshold. The dispersive representation~\eqref{eq:VFF} respects the constraints of unitarity and analyticity and it depends on only a few free parameters. Although there are systematic differences between the different cross-section measurements, the dispersive representation can be fit with a good $\chi^2$ to almost all major experimental low-energy data sets, see~\cite{Stoffer:2023gba} for an overview.

\begin{figure}[t]
	\centering
	\includegraphics[width=0.9\textwidth]{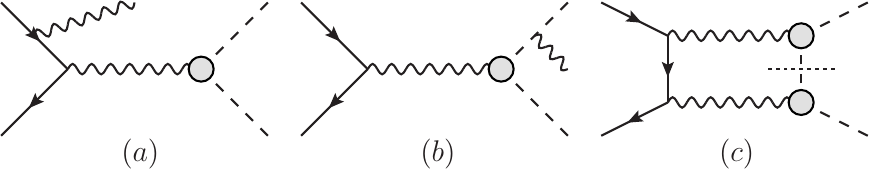}
	\caption{Diagrams for the forward--backward asymmetry $A_\text{FB}$ in $e^+e^-\to\pi^+\pi^-$, taken from~\cite{Colangelo:2022lzg}. The short-dashed line in $(c)$ indicates that only the pion-pole contribution is kept, not the general pion Compton tensor.}
	\label{fig:diagrams_AFB_eepipi}
\end{figure}

At NLO, within a strict fixed-order counting in $\alpha$ the radiative corrections can be split into the different gauge-invariant contributions introduced in \secref{sec:comp-fo}, as shown in \figref{fig:eepipi-NLO}. All ISC scaling as $\sim q_e^4 q_h^2$, where $q_h$ denotes a generic coupling of the photon to hadronic states, can be treated as described in \secref{sec:comp-fo} and pose no additional conceptual difficulty. The FSC, however, scaling as $\sim q_e^2 q_h^4$ and shown in the second line of \figref{fig:eepipi-NLO}, include two additional nonperturbative matrix elements: the amplitude $\gamma^*\to\pi^+\pi^-\gamma$ for the real corrections, related by crossing to the pion Compton tensor, as well as virtual corrections to the VFF itself. So far, these corrections were treated in sQED, which leads to the inclusive cross section due to FSC
\begin{equation}
	\sigma(e^+e^-\to\pi^+\pi^-(\gamma)) = \biggl[ 1 + \frac{\alpha}{\pi} \eta(s) \biggr] \sigma(e^+e^-\to\pi^+\pi^-) \, ,
 \label{eq:2piFSC}
\end{equation}
where the correction factor $\eta(s)$ determined in sQED can be found in~\cite{Schwinger:2019zjk,Gluza:2002ui,Czyz:2004rj,Bystritskiy:2005ib}. The reliability of the sQED determination of the factor $\eta(s)$ is currently under investigation, by making use of dispersion theory for the hadronic matrix elements entering the FSC. The pion Compton amplitude was analysed dispersively in~\cite{Garcia-Martin:2010kyn,Hoferichter:2011wk,Moussallam:2013una,Colangelo:2015ama,Danilkin:2018qfn,Hoferichter:2019nlq,Danilkin:2019opj}. Preliminary results for the virtual corrections to the VFF are available in~\cite{Monnard:2021pvm}, where no large effects beyond the sQED approximation were observed, which can be explained at least partially by the dominance of the IR-enhanced effects. The correction factor $\eta(s)$ is also used to capture the dominant radiative effects in $e^+e^-\to K^+K^-(\gamma)$, see~\cite{Stamen:2022uqh}, where, in addition, the comparison to the Sommerfeld--Gamow--Sakharov factor~\cite{sommerfeld1921atombau,Gamow:1928zz,Sakharov:1948plh}  
\begin{equation}
\label{Gamow}
	Z(s) = \frac{\pi\alpha}{\sigma_K(s)}\frac{1+\alpha^2/\big(4\sigma_K^2(s)\big)}{1-\exp\big(-\pi\alpha/\sigma_K(s)\big)},\qquad \sigma_K=\sqrt{1-\frac{4m_K^2}{s}}\, ,
\end{equation}
was studied, resumming threshold-enhanced higher-order effects in $\alpha$. However, already for the $K^+K^-$ channel the non-Coulomb corrections contained in $\eta(s)$ prove more important. 

\begin{figure}[t]
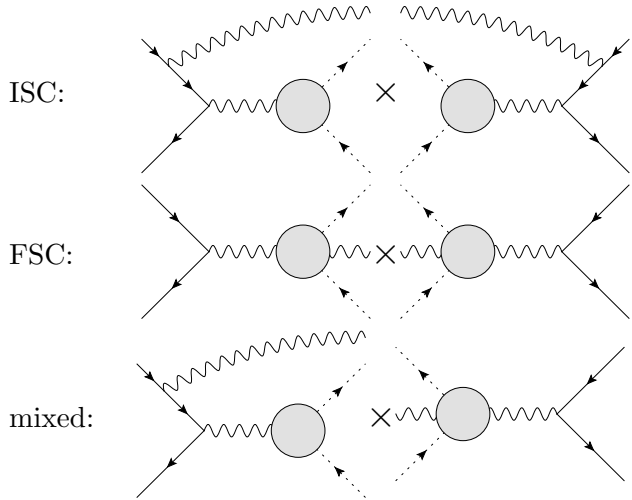

	\centering
    \begin{tabular}{lcc}
		ISC: &
		\ampamp{figs/had/NLO-ISR}{figs/had/NLO-ISR}
		\\[10pt]
		FSC: &
		\ampamp{figs/had/NLO-FSR}{figs/had/NLO-FSR}
		\\[10pt]
		mixed: &
		\ampamp{figs/had/NLO-ISR}{figs/had/NLO-FSR}
    \end{tabular}
	\caption{LO contributions to the $e^+e^-\to\pi^+\pi^-\gamma$ cross section.}
	\label{fig:eepipigamma-LO}
\end{figure}

The third line in \figref{fig:eepipi-NLO} shows the mixed corrections $\sim q_e^3 q_h^3$, where there is an additional photon coupling to both initial and final states. It is important to note that this contribution comes with an odd number of photon couplings to the hadronic states, hence it is $C$-odd and only contributes to the differential cross section and the charge or forward--backward asymmetry, whereas these mixed corrections cancel in the total cross section. Due to one additional photon coupling to the hadronic final state, the mixed corrections to $e^+e^-\to\pi^+\pi^-$ only depend on two hadronic matrix elements: the pion VFF itself as well as the pion Compton tensor. The charge asymmetry was studied in~\cite{Ignatov:2022iou} within the GVMD model and in~\cite{Colangelo:2022lzg} within the FsQED approximation, i.e., taking into account only the dispersively defined pion-pole contribution to the pion Compton tensor, see \figref{fig:diagrams_AFB_eepipi}. Both analyses came to similar conclusions and found relatively large deviations from the point-like sQED approximation and a much better agreement with measurements of the asymmetry. Generalisations of these treatments should be possible by including effects in the Compton tensor beyond the pion pole, in particular two-pion rescattering in $S$- and $D$-waves as determined in~\cite{Colangelo:2015ama,Hoferichter:2019nlq,Danilkin:2019opj}.

Finally, the last line in  \figref{fig:eepipi-NLO} involves the VP corrections $\sim q_e^2q_h^2 \Pi^{(1)}$. Conventionally, this correction is included in what is reported by $e^+e^-$ experiments as the form factor, while it is excluded from the so-called ``bare cross section,'' which only includes FSC.

\begin{figure}[t]
	\centering
		\minidiagSize{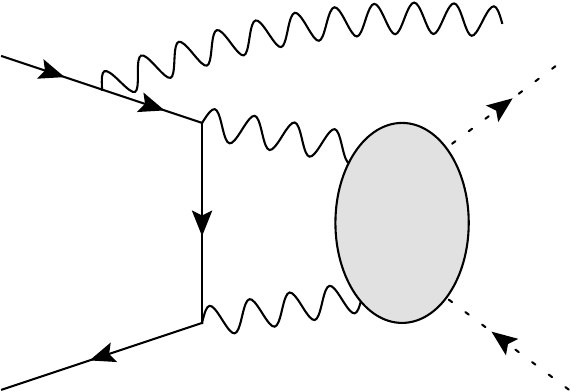}{3cm} \quad
		\minidiagSize{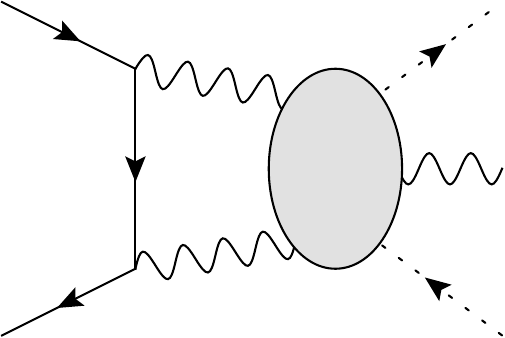}{3cm} \quad
		\minidiagSize{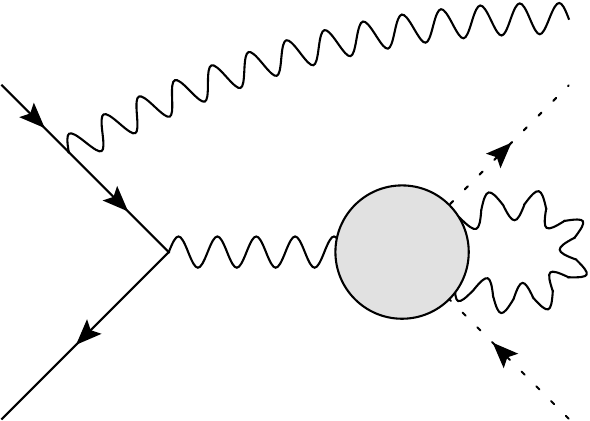}{3cm} \quad
		\minidiagSize{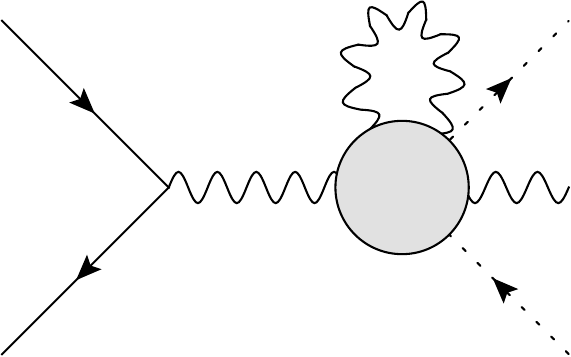}{3cm}
	\caption{Representative virtual NLO contributions to the $e^+e^-\to\pi^+\pi^-\gamma$ amplitude, omitting ISC. The ISR photon can be attached anywhere at the lepton line.}
	\label{fig:eepipigamma-NLO-Amp}
\vspace{30pt}
	\centering
		\ampamp{figs/had/NLO-FSR-FSV}{figs/had/NLO-FSR}
		\qquad
		\ampamp{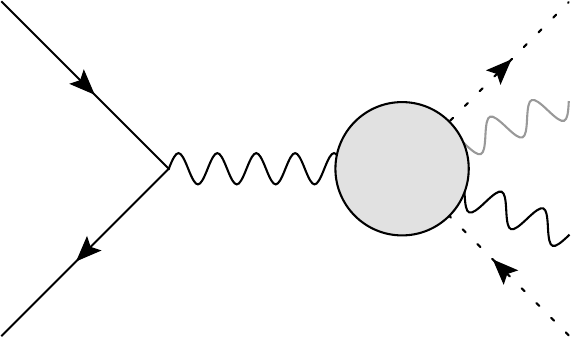}{figs/had/NLO-FSR-FSR}
	\caption{FSC NLO contributions to the $e^+e^-\to\pi^+\pi^-\gamma(\gamma)$ cross section. The grey photon is soft and the grey blobs denote the hadronic matrix elements.}
	\label{fig:eepipigamma-NLO-FSC}
\end{figure}

\subsubsection[Structure-dependent radiative corrections in \texorpdfstring{$e^+e^-\to\pi^+\pi^-\gamma$}{ee->pipi gamma}]{Structure-dependent radiative corrections in $\boldsymbol{e^+e^-\to\pi^+\pi^-\gamma}$}
\label{sec:comp-eepipigamma}

Radiative-return experiments measure the process $e^+e^-\to\pi^+\pi^-\gamma$ with a hard photon in the final state. As in \eqref{comp:loR}, the LO cross section consists of ISC, FSC, and a mixed contribution, shown in \figref{fig:eepipigamma-LO}. The first line shows the ISC $\sim q_e^4 q_h^2$, which can be determined in standard QED and only depends on the VFF.
The FSC $\sim q_e^2 q_h^4$ shown in the second line of \figref{fig:eepipigamma-LO} only depends on the singly-virtual pion Compton tensor and in principle could be determined using a dispersive description of the hadronic sub-amplitude~\cite{Moussallam:2013una,Colangelo:2015ama,Hoferichter:2019nlq,Danilkin:2019opj}.
The mixed contribution $\sim q_e^3 q_h^3$ shown in the third line of \figref{fig:eepipigamma-LO} depends on both the pion VFF and the Compton tensor. As before, it is $C$-odd and contributes to the charge asymmetry of the process.

\begin{figure}[ht]
	\centering
    \begin{tabular}{lcc}
	2PE-ISR: &
		\ampamp{figs/had/ISR-NLO-IFV}{figs/had/NLO-ISR}
		&
		\ampamp{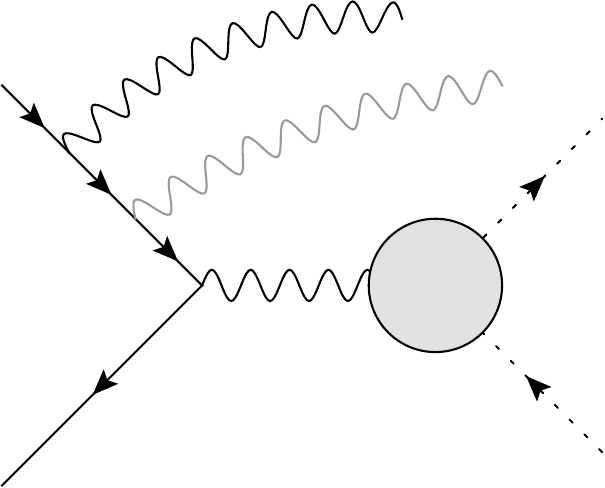}{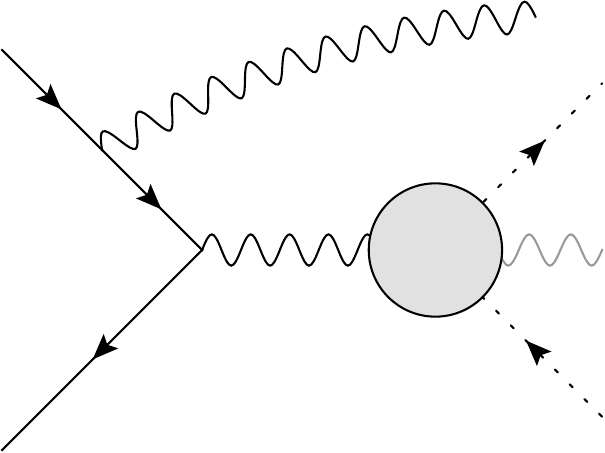}
		\\[10pt]
	2PE-FSR: &
		\ampamp{figs/had/NLO-IFV-FSR}{figs/had/NLO-FSR}
		&
		\ampamp{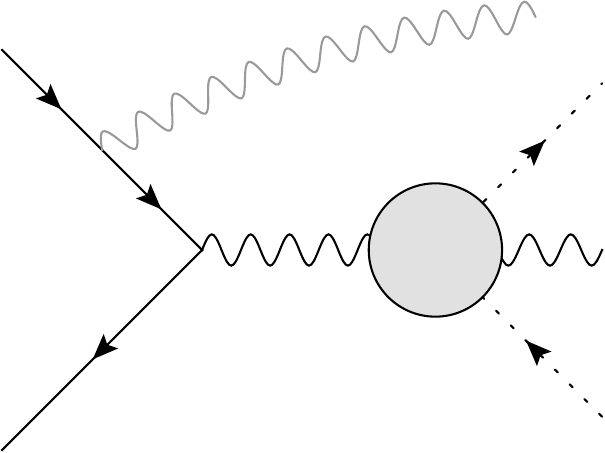}{figs/had/NLO-FSR-FSR}
		\\[10pt]
        &
		\ampamp{figs/had/ISR-NLO-FSV}{figs/had/NLO-FSR}
		\\[-1cm]
	1PE: & &
		\ampamp{figs/had/NLO-FSR-ISR-2}{figs/had/NLO-FSR-FSR}
		\\[-1cm]
        &
		\ampamp{figs/had/NLO-FSR-FSV}{figs/had/NLO-ISR}
    \end{tabular}
	\caption{Representative $C$-odd mixed NLO contributions to the $e^+e^-\to\pi^+\pi^-\gamma(\gamma)$ cross section. The grey photon is soft, the grey blobs denote the hadronic matrix elements, and the ISR photons can be attached anywhere at the lepton line.}
	\label{fig:eepipigamma-NLO-C-odd}
\end{figure}

The radiative corrections at NLO can again be split into ISC, FSC, and mixed contributions. By definition, the ISC part only depends on the pion VFF as hadronic sub-amplitude and we do not discuss it further. In the following, we also ignore VP corrections, which do not introduce new hadronic sub-amplitudes compared to the LO contributions. The virtual corrections to the radiative amplitude are shown in \figref{fig:eepipigamma-NLO-Amp}, where we omit the two ISC diagrams. We see that the radiative corrections depend on the pion Compton tensor, which appears in the first diagram (\mbox{2PE-ISR}), the five-point amplitude $\gamma^*\gamma^*\to\pi^+\pi^-\gamma$, which is part of the second diagram (\mbox{2PE-FSR}), as well as on the virtual corrections to the pion VFF and the pion Compton tensor. At present, all these corrections have been studied within sQED, multiplied by external form factors~\cite{Czyz:2003ue,Tracz:2018,Campanario:2019mjh}.

As previously discussed, dispersive approaches are available for the doubly-virtual pion Compton tensor~\cite{Colangelo:2015ama,Hoferichter:2019nlq,Danilkin:2019opj}, which consist of the dispersive pion pole and potentially vector-meson left-hand cuts, together with a unitarisation of the two-pion rescattering in the $S$- and $D$-waves. While in the case of $e^+e^-\to\pi^+\pi^-$, \cite{Colangelo:2022lzg} includes the pure pion pole in the loop diagram shown in \figref{fig:diagrams_AFB_eepipi}, a similar study for the radiative process is not yet available. The rescattering contribution has been studied neither in $e^+e^-\to\pi^+\pi^-$ nor in the radiative process. However, at least at low energies below $1\,\GeV$ the pion pole is known to be a decent approximation of the full Compton amplitude~\cite{Aoyama:2020ynm}.
The radiative corrections to the pion VFF studied dispersively in~\cite{Monnard:2021pvm} could be included both in $e^+e^-\to\pi^+\pi^-$ as well as in the radiative process.

At present, no dispersive study exists for radiative corrections to the Compton tensor. The complicated five-point amplitude $\gamma^*\gamma^*\gamma\to\pi^+\pi^-$ appears in a new dispersive approach to HLbL, proposed in~\cite{Ludtke:2023hvz}. There, the gauge-invariant Bardeen--Tung--Tarrach tensor decomposition into structures free of kinematic singularities was performed, a necessary starting point for a dispersive analysis, which is work in progress in the context of HLbL. There, it is needed only in the next-to-next-to-soft approximation for the real photon~\cite{Ludtke:2023hvz}, whereas in the present context the amplitude is required with a hard real photon.

The NLO cross section is obtained from the three-body phase-space integral over the product of the LO and NLO amplitudes, together with the four-body phase-space integral over the square of the doubly-real-emission amplitude. We follow the classification of~\cite{Czyz:2003ue,Tracz:2018,Campanario:2019mjh} of the different contributions to the cross section into IR-safe categories, neglecting again pure ISC and VP contributions.

\figref{fig:eepipigamma-NLO-FSC} shows pure FSC $\sim q_e^2 q_h^6$, where the soft real photon is shown in grey. Going beyond a model estimate might be difficult: these corrections require virtual-photon and soft-real corrections to the pion Compton tensor, which have not yet been studied dispersively. However, the pure FSC are generally expected to be small~\cite{Czyz:2003ue}. If the conclusions of the preliminary dispersive studies of FSC for the nonradiative process~\cite{Monnard:2021pvm} can be transferred, they are well approximated by sQED estimates.

The mixed corrections can be split up further. In \figref{fig:eepipigamma-NLO-C-odd}, we show the $C$-odd contributions, which can be divided into three gauge-invariant and IR-finite sets. The first one, scaling as $\sim q_e^5 q_h^3$, involves a 2PE-ISR diagram in the virtual corrections. The hadronic matrix elements in this class are the pion VFF and the pion Compton tensor. The second class in \figref{fig:eepipigamma-NLO-C-odd}, scaling as $\sim q_e^3 q_h^5$, involves a virtual correction with a 2PE-FSR diagram. The required hadronic sub-amplitudes are the pion Compton tensor, as well as the complicated five-point amplitude $\gamma^*\gamma^*\to\pi^+\pi^-\gamma$. The last class in \figref{fig:eepipigamma-NLO-C-odd} also scales as $\sim q_e^3 q_h^5$ and contains virtual corrections to the pion VFF and the pion Compton tensor, together with real corrections that cancel the IR divergences of the virtual corrections. Although these different corrections partially involve hadronic matrix elements that are difficult to compute beyond model approximations, the overall $C$-odd contributions will cancel if the experimental cuts are symmetric under change conjugation. Measurements of the charge asymmetry in the radiative process could be used to constrain the hadronic sub-processes~\cite{Czyz:2003ue}.

We now turn to $C$-even mixed corrections. The class of diagrams shown in \figref{fig:eepipigamma-NLO-VFF} corresponds to virtual and soft real corrections to the pion VFF, overall scaling as $\sim q_e^4 q_h^4$. The hard real photon is emitted from the initial state. For these corrections, the dispersive analysis of~\cite{Monnard:2021pvm} should be applied.

\begin{figure}[t]
	\centering
		\ampamp{figs/had/ISR-NLO-FSV}{figs/had/NLO-ISR}
		\qquad
		\ampamp{figs/had/NLO-FSR-ISR-2}{figs/had/NLO-FSR-ISR-2}
	\caption{Representative $C$-even mixed NLO contributions to the $e^+e^-\to\pi^+\pi^-\gamma(\gamma)$ cross section, corresponding to virtual and soft real corrections to the pion VFF with an additional initial-state hard photon.}
	\label{fig:eepipigamma-NLO-VFF}
\vspace{20pt}
	\centering
    \begin{tabular}{lcc}
	2PE-ISR: &
		\ampamp{figs/had/ISR-NLO-IFV}{figs/had/NLO-FSR}
		&
		\ampamp{figs/had/NLO-ISR-ISR}{figs/had/NLO-FSR-FSR}
		\\[10pt]
	2PE-FSR: &
 		\ampamp{figs/had/NLO-IFV-FSR}{figs/had/NLO-ISR}
		&
		\ampamp{figs/had/NLO-FSR-ISR-1}{figs/had/NLO-FSR-ISR-2}
    \end{tabular}
	\caption{Representative $C$-even mixed NLO contributions to the $e^+e^-\to\pi^+\pi^-\gamma(\gamma)$ cross section, where the hard photon interferes between initial and final states.}
	\label{fig:eepipigamma-NLO}
\end{figure}

This finally leaves us with the class of diagrams shown in \figref{fig:eepipigamma-NLO}, which form an IR-finite subset scaling as $\sim q_e^4 q_h^4$. In this class, the hard real photon interferes between initial and final states. While these corrections were previously assumed to be negligible~\cite{Czyz:2003ue}, they were later computed within the sQED approximation~\cite{Campanario:2019mjh}, multiplied by external form factors. The studies of the charge asymmetry in $e^+e^-\to\pi^+\pi^-$ of~\cite{Ignatov:2022iou,Colangelo:2022lzg} however indicates that larger corrections beyond the point-like approximation are possible~\cite{Abbiendi:2022liz}. In the case of the 2PE-ISR corrections, a study similar to~\cite{Ignatov:2022iou,Colangelo:2022lzg} should be performed. At least at low energies, the pion Compton tensor can be well approximated by a dispersively defined pion pole.

In the case of the 2PE-FSR correction, the situation is more challenging. It is unclear if even a dispersively defined pion-pole approximation is sufficient, since the two-pion state is in a $P$-wave. Final-state rescattering effects giving rise to the $\rho$-resonance call for the inclusion of a unitarisation of the pole contribution. Furthermore, due to the hard real photon additional internal resonance enhancement is possible. At present, these effects have not been estimated in a model-independent framework.

\subsubsection[\texorpdfstring{$e^+e^-\to\pi^+\pi^-\pi^0$}{ee->pi pi pi}]{$\boldsymbol{e^+e^-\to\pi^+\pi^-\pi^0}$}
\label{sec:comp-eepipipi}

In analogy to~\eqref{VFF_def},  the general matrix element for $\gamma^*(q)\rightarrow \pi^+(p_+)\pi^-(p_-)\pi^0(p_0)$  is defined by
\begin{equation}
    \langle0|j_\mu(0)|\pi^+(p_+)\pi^-(p_-)\pi^0(p_0)\rangle = -\epsilon_{\mu\nu\alpha\beta}p_+^\nu p_-^\alpha p_0^\beta \mathcal{F}(s,t,u;q^2)\,,
\end{equation}
with $q=p_++p_-+p_0$, $s=(p_++p_-)^2$, $t=(p_-+p_0)^2$, $u=(p_++p_0)^2$, subject to the constraint $s+t+u=3m_\pi^2+q^2$.  In contrast to the pion electromagnetic form factor, the function $\mathcal{F}(s,t,u;q^2)$ does not only depend on the virtual-photon momentum, but also on the invariant masses of all two-pion subsystems; expressed in terms of resonances, in addition to the isoscalar vector three-pion resonances such as $\omega(782)$, $\phi(1020)$, \ldots, also two-pion resonances of isospin 1 and odd angular momentum play an important role, most significantly the $\rho(770)$.
The total $e^+e^-\to3\pi$ cross section is obtained according to
\begin{equation}
\sigma_{e^+ e^- \to 3\pi}(q^2) = \alpha^2\int_{s_\text{min}}^{s_\text{max}} \dd s \int_{t_\text{min}}^{t_\text{max}} \dd t \,
\frac{s[\kappa(s,q^2)]^2(1-z_s^2)}{768 \, \pi \, q^6}  \, |\mathcal{F}(s,t,u;q^2)|^2\,, 
\end{equation}
with the relevant kinematic quantities
\begin{align}
        z_s&=\cos\theta_s =\frac{t-u}{\kappa(s,q^2)}\,, \qquad
        \kappa(s,q^2)=\sigma_\pi(s)\lambda^{1/2}(q^2,m_\pi^2,s)\,,\notag\\
     \lambda(x,y,z)&=x^2+y^2+z^2-2(xy+yz+xz)\,,\qquad  \sigma_\pi(s)=\sqrt{1-\frac{4\mpi^2}{s}}\,,  
\end{align}
and the boundaries of the Dalitz-plot integration given by
\begin{align}
s_\text{min} &= 4 m_\pi^2, \qquad\qquad \,s_\text{max} = \big(\sqrt{q^2}-m_\pi \big)^2\,, \notag \\ 
t_\text{min/max}&= (E_-^*+E_0^*)^2-\bigg( \sqrt{E_-^{*2}-m_\pi^2} \pm  \sqrt{E_0^{*2}-m_\pi^2} \bigg)^2\,,
\end{align}
and
\begin{equation}
E_-^*=\frac{\sqrt{s}}{2}\,,\qquad E_0^*=\frac{q^2-s-m_\pi^2}{2\sqrt{s}}\,.
\end{equation}
The function $\mathcal{F}(s,t,u;q^2)$ can be decomposed into simpler ones using 
a reconstruction theorem when neglecting discontinuities in the two-pion invariant masses of angular momentum 3 and higher, leading to~\cite{Niecknig:2012sj,Hoferichter:2012pm,Hoferichter:2014vra,Hoferichter:2019mqg}
\begin{equation}
 \mathcal{F}(s,t,u;q^2) = \mathcal{F}(s,q^2)+\mathcal{F}(t,q^2)+\mathcal{F}(u,q^2)\,,
\end{equation}
which corresponds to a symmetrised partial-wave expansion stopping beyond $P$-waves.  The neglected $F$-wave contributions are irrelevant before production of the $\rho_3(1690)$ becomes feasible, which happens around $\sqrt{q^2} \approx 1.83\,\GeV$~\cite{Niecknig:2012sj,Hoferichter:2017ftn,Hoferichter:2019mqg}.
The $s$-dependence of the functions $\mathcal{F}(s,q^2)$ can be calculated dispersively in the Khuri--Treiman formalism~\cite{Khuri:1960zz}, which takes two-pion $P$-wave rescattering in all subchannels into account consistently, using the corresponding phase shift.  They require a $q^2$-dependent subtraction or normalisation function that incorporates information on all $3\pi$ resonances, which can be modelled in a way consistent with analyticity and fit to experimental data~\cite{Hoferichter:2014vra,Hoferichter:2019mqg,Hoferichter:2023bjm}.  Its value at $q^2=0$ is constrained by the chiral anomaly~\cite{Adler:1971nq,Terentev:1971cso,Aviv:1971hq}, including quark-mass corrections~\cite{Bijnens:1989ff,Hoferichter:2012pm,Niehus:2021iin}.

\begin{figure}[t]
    \centering
    \includegraphics[width=0.8\textwidth]{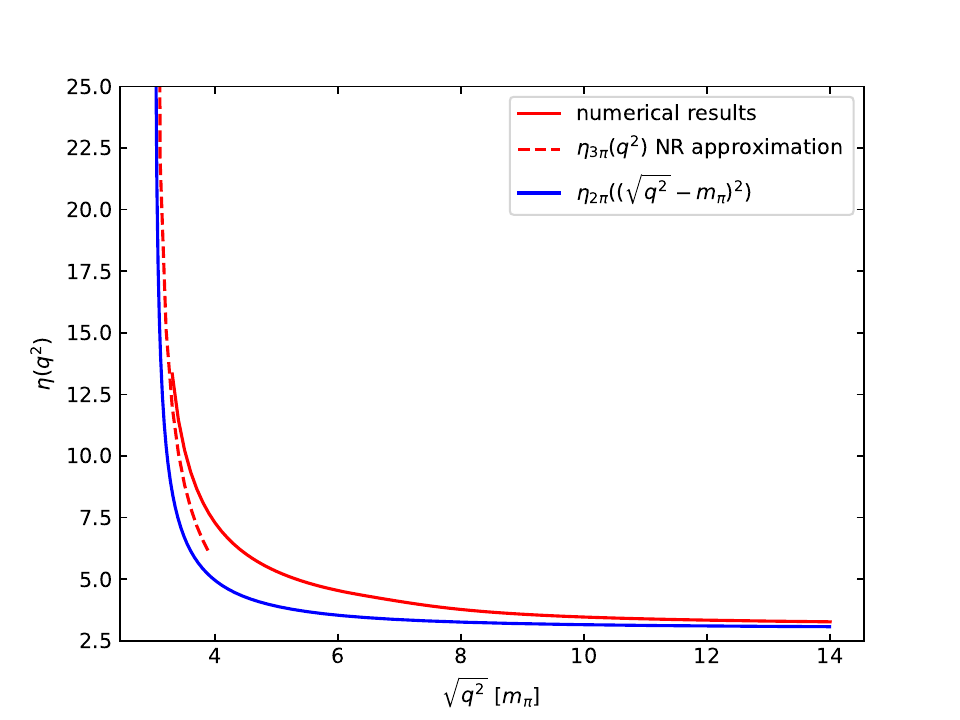}
    \caption{IR-enhanced radiative corrections for $e^+e^-\to 3\pi$ incorporated in $\eta_{3\pi}$ (red solid). The numerical result is compared to a nonrelativistic approximation (red dashed)
    and the $2\pi$ analogue $\eta_{2\pi}\equiv \eta$ shifted to the $3\pi$ threshold (blue solid). Figure adapted from~\cite{Hoferichter:2023bjm}.}
    \label{fig:eta3pi}
\end{figure}

The assessment of radiative corrections to the $\gamma^*\to3\pi$ matrix element is complicated by the fact that it is not accessible in the framework of sQED: it is given, at LO in the chiral expansion, by the Wess--Zumino--Witten anomaly~\cite{Wess:1971yu,Witten:1983tw}, which, due to its derivative structure, does not constitute a renormalisable theory, but requires counterterms to render loop corrections UV finite.  ChPT calculations~\cite{Ametller:2001yk} or other approaches to radiative corrections near threshold~\cite{Ahmedov:2002tg,Bakmaev:2005sg} furthermore miss the resonance dynamics of the $3\pi$ final state and are therefore phenomenologically insufficient in the energy region relevant for HVP studies.

While, as a consequence, an unambiguous determination of structure-dependent radiative corrections in the $3\pi$ channel would be a formidable task, the IR-enhanced effects have been studied in~\cite{Hoferichter:2023bjm}, owing to the observation that these constitute the by far dominant contributions to the radiative corrections in the $2\pi$ channel~\cite{Moussallam:2013una}.  Due to the dominance of pion--pion $P$-waves in the partial-wave decomposition of the $3\pi$ final state, it was observed that the sQED correction factor $\eta(s)$, cf.~\eqref{eq:2piFSC}, can similarly be applied to the $\pi^+\pi^-$ subsystem in $3\pi$, whose invariant mass is, however, $s$, not $q^2$.  
The cross section including IR-enhanced radiative corrections can therefore be written as
\begin{align}
&\sigma_{e^+ e^- \to 3\pi(\gamma)}(q^2) \notag\\&= \alpha^2\int_{s_\text{min}}^{s_\text{max}} \dd s \int_{t_\text{min}}^{t_\text{max}} \dd t \,
\frac{s[\kappa(s,q^2)]^2(1-z_s^2)}{768 \, \pi \, q^6}  \, |\mathcal{F}(s,q^2)+\mathcal{F}(t,q^2)+\mathcal{F}(u,q^2)|^2 \Big(1+\frac{\alpha}{\pi}\eta(s)\Big)\, , 
\end{align}
and an effective analogue correction factor $\eta_{3\pi}(q^2)$ be defined as
\begin{equation}
\label{eta3pi}
1+\frac{\alpha}{\pi}\eta_{3\pi}(q^2)=\frac{\sigma_{e^+e^-\to 3\pi(\gamma)} (q^2)}{\sigma^{(0)}_{e^+e^-\to 3\pi} (q^2)}\, ,
\end{equation}
which has to be calculated numerically.  Near threshold $q^2=9m_\pi^2$, $\eta_{3\pi}(q^2)$ shows a Coulomb-pole-like divergence $\propto (q^2-9m_\pi^2)^{-1/2}$ inherited from the divergence in $\eta(s) \propto (s-4m_\pi^2)^{-1/2}$, whose coefficient can be calculated analytically from a nonrelativistic expansion~\cite{Hoferichter:2023bjm}. The resulting correction factor is reproduced in \figref{fig:eta3pi}, compared to its nonrelativistic approximation and the (shifted) $\eta(s)$ correction.

\section{Generators and integrators}\label{sec:gen}

In this section we will briefly review each of the seven codes that are used in this comparison, in alphabetical order.
We focus mainly on the physics effects included in each generator,
using the language defined in~\secref{sec:comp}.
This will allow us to study the relative impact of different contributions \pdforweb{in~\secref{sec:mcc}}{later}.

\setlength{\parindent}{11.4pt}
The seven codes discussed during Phase I are \afkqed, {\sc{Babayaga@NLO}}, {\sc{KKMCee}}, MCGPJ, \mcmule, \phokhara, and \Sherpa. These codes have been developed by groups and authors that are not necessarily part of this community effort. While the responsibility of the codes clearly lies with the original authors, we provide list of responsible contact persons for each code on the website at
\begin{quote}
\url{https://radiomontecarlow2.gitlab.io/code-responsible.html}
\end{quote}
This list can serve as a first point of contact for potential users of the codes.
Other codes may be added in Phase II.

    \subsection{\afkqed} \label{sec:afkqed}

\afkqed{} is a Monte Carlo event generator that simulates radiative processes, i.e., those with a hard photon
\begin{equation}
    e^{+}e^{-} \to X\gamma,
\end{equation}
where $X$ in the final state can represent a $\mu^+\mu^-$ pair, or a number of hadronic processes, including those with pions, kaons and protons.

The simulation of most hadronic processes is based on the EVA event generator~\cite{Binner:1999bt,Czyz:2000wh}, which was originally used to generate $2\pi\gamma$ and $4\pi\gamma$ final states. The code was developed in Fortran mainly by V.~Druzhinin at Novosibirsk, and further final states were implemented in a modular way. \afkqed{} was used by the BaBar collaboration in 2009 for acceptance determination~\cite{BaBar:2012bdw}. In later analyses, the BaBar collaboration used \phokhara{} for this purpose. 

The zipped file \texttt{AfkQed.tar.gz} in the repository
\begin{quote}
    \url{https://gitlab.com/radiomontecarlow2/monte-carlo-results/-/tree/root/codes/afkqed}
\end{quote}
contains all modules necessary for execution, including the stand-alone test programme \texttt{Afkcrs.F} in Fortran, the \texttt{GfiAfkQed.cc} code written in \texttt{C++} which was used within the BaBar software environment, and the stand-alone code \texttt{afkrun.cpp}, which was used for the purpose of this report.

For most hadronic channels, including $\pi^+\pi^-\gamma$, the simulation is based on the approach developed in~\cite{Czyz:2000wh}. For the hadronic processes $p^+p^-\gamma$, $3\pi^+3\pi^-\gamma$, $2\pi^+2\pi^-2\pi^0\gamma$, the Bonneau--Martin formula is used~\cite{Bonneau:1971mk}. For the $\pi^+\pi^-\gamma$ process, the VFF is implemented through the F$\times$sQED approach described in \secref{sec:comp-fshad-strategies}. For the $\mu^+\mu^-\gamma$ process, the calculation of the hard photon matrix element is based on the Born cross section formulae in~\cite{Arbuzov:1997pj}, and the simulation includes ISC, FSC, and mixed corrections.

In \afkqed, the LO hard photon can be generated at large angle, within the detector acceptance range. At higher orders, additional ISR photons are generated first, through the structure function method described in \secref{sec:comp-bfoPDF}. Such photons are assumed to be collinear to the $e^+$ or $e^-$ beams, so they do not have the angular distribution of \eqref{eq:phpperp}, and photon jets radiated from each beam are resummed as a single photon along that beam. After one or two collinear ISR photons have been simulated, the final state, including a photon with an angular distribution, is generated in the boosted system. Additional FSR photons are generated with the \Photos{} package~\cite{Barberio:1993qi}.
In all cases, the photon generated at a large angle is assumed to be hard and is not complemented by  soft photon emission corrections.
For the $\mu^+\mu^-\gamma$ process, HVP is included, where in the original code it could be chosen to include leptonic contributions only (from~\cite{Arbuzov:1997pj}), or both hadronic ($\rho+\omega+\phi$) and leptonic. 
By default, the current version on the repository employs the Novosibirsk (NSK) VP~\cite{Ignatov:2008bfz,Ignatov:hvp} applied in the Dyson resummed form.

\afkqed{} events are built in a factorised way, meaning that the modules displayed in \tabref{tab:afkqed_nlo} for additional ISR or FSR, or for HVP, can be switched on or off at the user's discretion.

\begin{table}[tb]
\renewcommand{\arraystretch}{1.5}
\centering
\begin{tabular}{ll||c|c}
 & & $e^+ e^- \to \mu^+ \mu^- \gamma$ & $e^+ e^- \to \pi^+ \pi^- \gamma$ \\
\hline\hline
\multirow{2}{*}{ISC} & LO & Exact matrix elements~\cite{Arbuzov:1997pj} & EVA~\cite{Czyz:2000wh} \\ \cline{2-4} 
 & NLO & \multicolumn{2}{c}{Collinear structures~\cite{Caffo:1994dm,Caffo:1997yy}} \\
\hline
\multirow{3}{*}{FSC} & \multirow{2}{*}{LO} & Exact matrix elements~\cite{Arbuzov:1997pj} & \multirow{2}{*}{No FSR at LO}\\
 & & including ISR-FSR interference& \\ \cline{2-4} 
 & NLO & \multicolumn{2}{c}{\Photos~\cite{Barberio:1993qi}} \\
\hline
\multicolumn{2}{c||}{\multirow{3}{*}{HVP}} & None, leptonic only~\cite{Arbuzov:1997pj}, & \multirow{2}{*}{F$\times$sQED} \\
 & & leptonic~\cite{Arbuzov:1997pj} + hadronic, & \multirow{2}{*}{Customisable VFF}\\
  & & or NSK VP~\cite{Ignatov:2008bfz,Ignatov:hvp} & \\

\end{tabular}
\caption{
Overview of the modules implemented in \afkqed{} for $e^+e^-\to X^+ X^- \gamma$ processes.
}
\label{tab:afkqed_nlo}
\end{table}

For the simulation of different scenarios, the following parameters in the centre-of-mass frame can be specified: the centre-of-mass energy, the minimum energy and polar angle of the LO hard photon, and --- in case photons from collinear structures are switched on --- the minimum squared invariant mass of observed particles in the final state (i.e., the system composed of hadrons or muons and of the photon emitted to large angle).

\afkqed{} is not suitable for use  in cases when the LO photon at a large angle is not treated as the hardest photon after selection cuts, because this photon is not supplemented by soft and virtual corrections. In the KLOE-like small-angle scenario (see \secref{sec:KLOEsasc}), the collinear structures for additional ISR will almost always be the hardest photons in all events, whereas the LO photon at a large angle will be a correction. In the KLOE-like large-angle scenario (see \secref{sec:KLOElasc}), if \Photos{} is switched on, the additional FSR might be detected as the main ISR photon, and for this reason we do not include FSR corrections from \Photos{} in the KLOE-like scenarios. In the BESIII-like scenario, the issue of the full treatment of the LO photon is addressed by applying a soft photon cut of $400\,$MeV at the generation level, together with the experimental requirement that only one hard photon be detected at large angle. This will force the LO photon to be considered as the main radiative photon. Finally, in the $B$ scenario there is a requirement for a photon energy of at least $3\,$GeV at the generator and experimental photon selection levels. We also require the invariant mass of all observed final-state particles to be greater than 8\,GeV. Together, these requirements allow \afkqed{} to work well for these scenarios.

    \subsection{BabaYaga@NLO} \label{sec:babayaga}

{\sc BabaYaga@NLO} is a fully differential Monte Carlo event generator which was developed for high-precision simulations of QED processes at flavour factories up to $\sqrt{s}\simeq 10\,\GeV$~\cite{CarloniCalame:2000pz, CarloniCalame:2001ny, CarloniCalame:2003yt, Balossini:2006wc, Balossini:2008xr, Barze:2010pf,Budassi:2024whw}. It has been used by most of the experimental collaborations for luminosity determination and other physics studies.
It is based on the matching of exact ${\cal O}(\alpha)$ corrections with a QED parton shower algorithm. The {\sc BabaYaga@NLO} code allows the generation of fully exclusive events with exact kinematics and phase space. Via the parton shower, the generator is also able to exactly reconstruct all the generated photon momenta, conserving both the total four-momentum and the mass shell conditions of all the particles. A detailed description of the parton shower algorithm as implemented in {\sc BabaYaga@NLO} can be found in \secref{sec:comp-bfoPS}. 
In addition to resumming all leading logarithmic terms of the form $\alpha^n L_c^n$ for $n \geq 2$, the matching procedure takes into account the next-to-leading logarithmic terms of the form $\alpha^n L_c^{n-1}$ for $n = 2$, as well as the leading mass terms proportional to $m_i^2/(p_i\cdot q)$.
The resummation is extended to the ISC, FSC, mixed, and VPC contributions for all channels.

\tabref{tab:by_processes} shows a list of all the processes one can calculate using {\sc BabaYaga@NLO}. All the SM processes are calculated at NLO with a consistent matching of a QED parton shower, without double counting, as specified by the NLOPS tag in \tabref{tab:by_processes}. Pion pair production is implemented according to three of the possible alternative approaches discussed in \secref{sec:comp-fshad}, namely F$\times$sQED, GVMD, and FsQED~\cite{Budassi:2024whw}.

Different pion VFF parametrisations are included for comparative studies.
Any other parametrisation can be added to the code, either as an analytical function or as a numerical table.
In addition to the SM processes, it is possible to study the production of a vector boson according to the dark matter models discussed in~\cite{Boehm:2003hm,Arkani-Hamed:2008hhe,Batell:2009yf}. This calculation has been performed at LOPS~\cite{Barze:2010pf}. 

The accuracy of the code is estimated to be at the $0.1\%$ level for calculations at NLOPS for the $2\to2$ processes with typical event selections at flavour factories. It is important to note that for radiative events, i.e., $2\to2+\gamma$ processes the accuracy of {\sc BabaYaga@NLO} is LL, estimated at the $\mathcal{O}(1\%)$ level.

\begin{table}[t]
\renewcommand{\arraystretch}{1.5}
\centering
\begin{tabular}{r@{\hspace{0.6cm}}c@{\hspace{0.6cm}}c@{\hspace{0.4cm}}c@{\hspace{0.2cm}}c}

       & $e^+e^-\to e^+e^-$ & $e^+e^-\to \mu^+\mu^-$ & $e^+e^-\to\gamma\gamma$ & $e^+e^-\to \pi^+\pi^-$ \\
    \pdforweb{\Xcline{2-5}{1 pt}}{\hline}
Order    & NLOPS            & NLOPS            & NLOPS                  & NLOPS                   \\
Accuracy & ${\cal O}(0.1\%)$ &  ${\cal O}(0.1\%)$   & ${\cal O}(0.1\%)$      & ${\cal O}(0.1\%)$       \\
\end{tabular}
\caption{All the processes that are calculated by {\sc BabaYaga@NLO}, the order of their calculation, and their estimated accuracy. NLOPS means the matching of the full NLO calculation with a PS algorithm, as described in \secref{sec:comp-bfoPS}.}
\label{tab:by_processes}
\end{table}

{\sc BabaYaga@NLO} can generate both weighted and unweighted events. In the former case, each event comes with a different weight which has to be carried throughout the whole detector simulation and analysis. This procedure allows for a very fast generation and a better Monte Carlo error convergence. In the case of unweighted generation, all events come with the same weight. This means that they are distributed according to the cross section. However, the generation of unweighted events can be very slow due to the unweighting procedure, which will be inefficient if there are large fluctuations in weights.

To calculate one-loop amplitudes, for $\mu^+\mu^-$ and $\pi^+\pi^-$ production, {\sc BabaYaga@NLO} is interfaced to Collier~\cite{Denner:2016kdg} (and optionally to LoopTools~\cite{Hahn:1998yk,Hahn:2010zi}), and mass effects are fully included. The mass of the external particles is instead neglected in the non-IR parts $\delta_{\rm virt}^{\rm non-IR}$ for the $e^+e^-\to e^+e^-$ and $e^+e^-\to\gamma\gamma$ processes. For all implemented processes, {\sc BabaYaga@NLO} keeps the full dependence of the masses in the kinematics and in the real photon radiation processes.

Infrared divergences are treated by giving a vanishingly small mass to real and virtual photons. Moreover, the real photon contribution is computed using the phase space slicing technique. This amounts to imposing an arbitrary cutoff $\varepsilon$ on the photon energy and then separating the two different phase space regions where the photon energy is smaller or larger than the cutoff. The phenomenological results obtained with {\sc BabaYaga@NLO} do not depend on the choice of the infrared separator $\varepsilon$.

In {\sc BabaYaga@NLO}, the hadronic contribution to the vacuum polarisation can be included by using various routines, for example by KNT~\cite{Keshavarzi:2019abf}, Jegerlehner~\cite{Jegerlehner:2017gek, Jegerlehner:hvp19} or NSK~\cite{Ignatov:hvp}. 
    \subsection{KKMC} \label{sec:kkmc}
{\sc KKMCee} is a Monte Carlo event generator applicable for electron--positron annihilation processes, accounting for multiple photon emission:

\begin{equation}
e^-e^+ \rightarrow f \bar{f} + n\gamma, \qquad f = \mu, \tau, \nu, u, d, s, c, b, \qquad n = 0,1, \dots \infty 
\end{equation} 
{\sc KKMCee} allows the generation of fully exclusive events with exact kinematics and phase space, which is crucial for realistic data analysis. The Monte Carlo integration over the phase-space is executed through the \texttt{FOAM} algorithm~\cite{Jadach_2000}.  
The generator can produce weighted and unweighted events depending on the user's needs.   {\sc KKMCee} is extensively used in data analysis for most existing electron colliders, such as BES and Belle, and in research for future electron colliders like FCCee, CLIC, and ILC.

Effects due to photon emission from incoming beams and outgoing fermions
are calculated in QED up to second order in $\alpha_\text{QED}$, including all interference effects,
within Coherent Exclusive Exponentiation (CEEX)~\cite{Jadach:2000ir}, 
which is based on Yennie--Frautschi--Suura exponentiation~\cite{Yennie:1961ad}, see \secref{sec:comp-bfoYFS}.
CEEX not only treats infrared cancellations and the description of soft photons correctly to infinite order, but also includes the correct description of QED interferences and narrow resonances.
Contrary to the standard YFS approach, in which higher-order corrections are effected at the level of the amplitude squared, the CEEX scheme is devised in terms of spin amplitudes and can therefore account for ISR-FSR interferences.
A detailed description of CEEX and the older EEX scheme can be found in~\cite{Jadach:2000ir,Jadach:2002vd}.
In addition to higher order QED corrections the weak corrections are provided by \texttt{Dizet} library~\cite{BARDIN1990303},
 and polarised $\tau$ decays are included using the \texttt{TAUOLA} programme~\cite{Jadach:1990mz}.

The current version of the code, written in \texttt{C++}, was released in 2022~\cite{Jadach_2023} and it is publicly available at
 \begin{quote}
    \url{https://github.com/KrakowHEPSoft/KKMCee}
\end{quote}

    \subsection{MCGPJ} \label{sec:mcgpj}

The Monte Carlo Generator with Photon Jets (MCGPJ)~\cite{Arbuzov:2005pt}
was developed more than 20 years ago by
the Dubna--Novosibirsk collaboration, based on~\cite{Arbuzov:1997pj,Arbuzov:1997je}. 
The main purpose of the
generator is to simulate processes of electron--positron annihilation
to two particles for the energy scan experiments,
specifically for the CMD-2 experiment at the VEPP-2M electron--positron collider.
This includes Bhabha scattering, production of two charged pions,
kaons, lepton pairs, and $3\gamma$.  The programme is based
on the exact ${\cal O}(\alpha)$ differential amplitude supplemented by
logarithmically enhanced higher order contributions. The exact NLO amplitude
includes emission from the pions and kaons in the pointlike (sQED)
approximation. The LL effects are taken into account using the
structure function approach in the collinear approximation, where
multi-photon jet objects are emitted along the electrons' direction
of motion. For the structure function, MCGPJ uses the jet energy emission function $D(z)$~\cite{Kuraev:1985hb,Arbuzov:1997pj}, which is an extension of \eqref{comp:LLpdf} 
to $\mathcal{O}(\alpha^3)$ and to $\mathcal{O}(\eta^2)$ in the exponent.
The original paper~\cite{Kuraev:1985hb} also includes part of the NLL corrections due to
virtual and real $e^+e^-$ pair production.
In MCGPJ, the jet emission is performed only from the light electron and positron lines,
including final states in the Bhabha process, while for heavier final-state particles no final-state resummation is applied.

Unlike in many other generators, the matching of the resummation to higher-order 
corrections is achieved additively in MCGPJ. 
Event generation is performed independently using the exact NLO amplitude or
by simultaneous jet emissions from all electron and positron lines.
An energy cut $\Delta=\Delta\epsilon/\epsilon$ is applied to the real photon which 
is generated with NLO accuracy, and isolation criteria are applied to electrons and positrons, meaning the NLO-accurate photon emission does not occur within a narrow cone of width $\theta_0$ around electrons/positrons. 
Most of the photon radiation comes from
the narrow collinear region with angles around $\sim 1/\gamma \sim \sqrt{1-\beta^2}$
(which justifies the use of the collinear jet approximation in most cases),
and the auxiliary parameter $\theta_0$ is taken  $\sim 1/\sqrt{\gamma}$
to make it outside of the collinear region.
The corresponding one-photon contribution, integrated over the volume defined by
$\Delta$ and $\theta_0$, is subtracted from jets when only one jet is above 
the $\Delta\epsilon$ cut energy. The
corresponding soft region of all jets is also matched with the one-photon
soft and virtual corrections. 
While the final result does not depend on $\Delta$ or $\theta_0$, a physically-motivated 
selection of these auxiliary parameters helps to suppress negative weighted events.

The theoretical precision for the integrated cross section is estimated to be better than 0.2\%. Until now, MCGPJ was the only generator capable of simulating two-pion events with such precision for scan experiments.

The collinear approximation can limit the precision of the differential
cross section prediction in tails, especially in kinematic regions where events with two resolved photon
emissions (which are not included in the NLO amplitude) can pass experimental cuts.
To improve this situation, the original version of MCGPJ was modified
to take into account the angular distribution of photon emissions.
The angle distribution of photon jets was generated according to the one-photon approximation
\begin{align} \label{eq:mcgpj1}
f(c=\cos(\theta),x=\omega/E) \sim 
\frac{1}{1-\beta c} - \frac{1-x}{1+(1-x)^{2}} \frac{1-\beta^2}{(1-\beta c)^2}\, ,
\end{align}
where $\theta$ is the angle of the jet relative to the radiating particle and
$x$ is the emitted energy ratio. This modification was implemented only for
the Bhabha process and led to the ability to run the event
generation in partially weighted mode, with a significant presence of normalised negative weighted events.

The latest version of MCGPJ also includes the contribution of the two-photon exchange diagram calculated in the above F$\times$sQED approximation,
taken into account either using the GVMD model or in the dispersive
formalism~\cite{Ignatov:2022iou,Colangelo:2022lzg}.
The correction is pre-tabulated in $\sqrt{s}$ and the
$\theta_\text{avg}$ of final-state particles.
The VPC are included according to
the Novosibirsk (NSK) compilation~\cite{Ignatov:2008bfz,Ignatov:hvp,WGRadCor:2010bjp}.

\begin{table}[tb]
\renewcommand{\arraystretch}{1.5}
\centering
\begin{tabular}{l||c|c|c}
  &  $e^+ e^- \to e^+ e^-(\gamma)$  & $e^+ e^- \to \mu^+ \mu^-(\gamma)$  &
  $e^+ e^- \to \pi^+ \pi^-(\gamma)$ 
  \\ \hline\hline
  \multirow{2}{*}{NLO} & \multicolumn{3}{c}{exact amplitude} \\
                       &      &  & F$\times$sQED \\ \hline
  \multirow{2}{*}{next LL orders} & SF with angles &\multicolumn{2}{c}{collinear structures} \\
  & ISC+FSC & ISC & ISC \\ \hline
  VP & \multicolumn{2}{c|}{Novosibirsk VP table} & VFF \\ \hline
  2PE & & & F$\times$sQED, GVMD  \\
      & & & or dispersive  \\ \hline 
\end{tabular}
\caption{Overview of the corrections to $e^+e^-\to X^+ X^-$ that are available in MCGPJ}
\label{tab:mcgpj_processes}
\end{table}

    \subsection{McMule} \label{sec:mcmule}
In its current version, the Monte Carlo framework \mcmule{}~\cite{Banerjee:2020rww}
\begin{quote}
    \url{https://mule-tools.gitlab.io}
\end{quote}
is a parton-level integrator that performs fully differential fixed-order QED calculations to high precision. The available processes are $ee \to XX$ with $X\in\{e,\mu,\pi\}$ and others such as $\ell p \to \ell p$~\cite{Engel:2023arz} and $\mu\,e\to\mu\,e$~\cite{Broggio:2022htr}, all at NNLO accuracy. Further, \mcmule{} also supports electroweak and polarisation effects for selected processes such as $ee\to\mu\mu$~\cite{Kollatzsch:2022bqa}.
$2 \to 3$ QED processes, i.e., those with an additional photon in the final state, are implemented at NLO accuracy.

\mcmule{} builds upon modern methods developed for higher-order QCD calculations, adapting them to the case of QED with massive fermions. 

On the one hand, this translates into a simplification, as the presence of fermion masses makes the handling of infrared divergences solvable to all orders in $\alpha$. In particular, infrared singularities are regularised in $d=4-2\epsilon$ dimensions, and dealt with in phase-space integrations with the FKS$^\ell$ method~\cite{Engel:2019nfw}, an adaption of FKS subtraction~\cite{Frixione:1995ms,Frederix:2009yq} to any loop order in QED. The extension of the subtraction scheme to all orders is made possible by the absence of collinear singularities in QED, allowing for the exponentiation of soft singularities, as in YFS~\cite{Yennie:1961ad} (see \secref{sec:comp-bfoYFS}).
The FKS$^\ell$ method introduces an unphysical parameter as a bookkeeper of contributions with different real-photon multiplicities, and the independence of the final result of said parameter is used as a strong consistency check. 
The method is exact and does not require splitting photon radiation into a soft and a hard part, nor does it introduce a photon mass as a regulator, since dimensional regularisation is instead used. If all squared matrix elements are known, the numerical integration of FKS$^\ell$-arranged squared matrix elements, combined with a measurement function to define IR-safe observable(s), results in a fully-differential Monte Carlo code.

On the other hand, while the presence of non-vanishing fermion masses simplifies the infrared structure, the additional~(though typically small) scales can cause serious complications. This is particularly relevant for the evaluation of loop amplitudes and the numerical stability of phase-space integrations. 

For one-loop amplitudes \mcmule{} relies on external libraries such as OpenLoops~\cite{Buccioni:2017yxi,Buccioni:2019sur} and Collier~\cite{Denner:2016kdg}. Two-loop matrix elements are the main bottleneck since no general approach exists, especially because of the higher number of scales. This issue can be mediated using massification~\cite{Penin:2005eh, Becher:2007cu, Engel:2018fsb,Bonciani:2021okt}, if the corresponding two-loop matrix element with vanishing fermion masses is known, and if some external fermions have small masses compared to all other scales in the process, such as the common case $m^2_e \ll Q^2$. Using this small-mass expansion, all terms that are not polynomially suppressed in $m_e^2/Q^2$, i.e.,~the logarithmically enhanced ones as well as the constant terms, can be recovered. 

In particular, the massive result can be related to the massless one via the factorisation formula
\begin{equation}
\label{eq:massification}
\cM_n(m_e) = \biggl( \prod_j Z(m_e) \biggr) \times S \times \cM_n(m_e=0) + \cO(m_e) \,,
\end{equation}
where the product is over all external fermion legs with a small mass $m_e$, and $n$ refers to any non-radiative final state such as $n=mm=\mu^+\mu^-$. The factorisation is a consequence of collinear and soft degrees of freedom factorising at leading power, as shown in soft-collinear effective theory~(SCET)~\cite{Bauer:2000yr, Bauer:2001yt, Beneke:2002ph}. Fermions with small masses correspond to highly-energetic particles in the external states, thus defining a collinear sector in SCET and contributing one power of the massification constant $Z$. This process-independent factor does not depend on any hard scale and, apart from a trivial factorised $m_e$ dependence, is a constant now known up to three loop~\cite{Ulrich:2023mfs}. The soft part $S$ is process dependent, starts at NNLO, and is obtained from fermionic corrections only~\cite{Becher:2007cu}. However, it is often advantageous to compute the fermionic corrections of two-loop squared matrix elements semi-numerically as it includes the HVP data. This has the added advantage of eliminating $S$ in~\eqref{eq:massification}, thus rendering massification completely universal.

The comparison of massified calculations with their equivalent full-mass calculation has allowed the quantification of the massification error. At the level of the differential NNLO cross section, massification introduces an error of the order $\alpha^2 \times 10^{-3}$~\cite{Broggio:2022htr}. It was also verified, for the case of the muon decay, that the massified result at NNLO gives a very good approximation to the result with exact $m_e$ dependence~\cite{Engel:2019nfw}. In general, applying massification corresponds at NNLO to a parametric error of order $(\alpha/\pi)^2 \, m_e^2/Q^2$, potentially multiplied by a $L_c^1$.

Non-vanishing fermion masses, in addition to problematic evaluations of loop amplitudes, cause instabilities in the numerical integration, in particular, at NNLO, for the real-virtual  contributions. OpenLoops is remarkably stable in the bulk of the phase space, and even in very soft and collinear regions. However, for simultaneously extremely soft and collinear kinematics the numerical stability in double precision, with on-the-fly support in quadruple, is not sufficient. In order to overcome this problem, \mcmule{} applies next-to-soft~(NTS) stabilisation, i.e.~replaces the real-virtual squared matrix elements by numerically adequate squared matrix elements in the problematic soft regions of the phase space~\cite{Banerjee:2021mty}. While the leading power eikonal approximation is in general not sufficiently accurate, an expansion in the soft photon energy, $E_\gamma$, 
up to the next-to-leading power proves to be accurate enough. 

The LBK theorem~\cite{Low:1958sn, Burnett:1967km} can provide the NTS limit at tree level, however for the real-virtual contribution the one-loop NTS limit has to be used. The real-virtual squared matrix element can be written as~\cite{Engel:2021ccn}
\begin{equation}
\label{eq:ntsstab}
\cM_{n+\gamma}^{(1)} = 
\bigl( {\cal E} + {\cal D}\bigr) \, \cM_n^{(1)} + {\cal S}^{(1)} \, \cM_n^{(0)}+ \cO(E_\gamma^0) \,,
\end{equation}
in terms of the non-radiative one-loop squared matrix element, where ${\cal E}$ is the eikonal factor, that scales as $1/E_\gamma^2$. In addition, we must include the LBK operator ${\cal D}$, and the one-loop contribution ${\cal S}^{(1)}$, which takes into account soft virtual corrections. These contributions scale as $1/E_\gamma$. Recent results~\cite{Engel:2023ifn, Engel:2023rxp} have shown that the LBK theorem holds at any loop order in QED, since the soft function ${\cal S}^{(1)}$ is one-loop exact, as well as for any number of photon emissions.

From the comparison of results obtained with NTS stabilisation with the same obtained with OpenLoops in full quadruple precision~(at the cost of at least ten times longer running times), the error due to the former could be quantified, at the level of the differential NNLO cross section, to $\alpha^2 \times 10^{-2}$~\cite{Broggio:2022htr}.

Similar instabilities in the numerical integration due to the presence of non-vanishing fermion masses are found in the case of tree-level contributions with the emission of one or more photons, for example real corrections at NLO and double-real corrections at NNLO. Since the mass of the fermion is often small compared to its energy, radiative amplitudes exhibit narrow peaks: remnants of the collinear singularities (which have been regularised by the masses). A reliable integration, less hampered by such peaks, is achieved in \mcmule{} by a partitioning and tuning of the phase space~\cite{Engel:2022kde} to directly match the collinearity with a variable of the adaptive integration algorithm.

In \mcmule{}, NLO corrections to a radiative process $ee\to X\,X\,\gamma$ are obtained as a subset of the NNLO corrections of the non-radiative process $ee\to X\,X$. As described in \secref{sec:comp-fo}, 
this amounts to simply dropping the pure virtual terms (assuming an IR-safe observable). The availability of a non-radiative process at N$^n$LO corresponds to the availability of the related radiative process at N$^{n-1}$LO.
Hence, we only describe the implemented corrections for non-radiative processes in \tabref{tab:mcmulenlonnlo}.

\begin{table}[tb]
\renewcommand{\arraystretch}{1.5}
\centering
\scalebox{0.975}{
\begin{tabular}{ll||c|c|c}
                       &                       &  $e^+ e^- \to e^+ e^-$ \cite{Banerjee:2021mty,Banerjee:2021qvi} & $e^+ e^- \to \mu^+ \mu^-$ \cite{Broggio:2022htr,Kollatzsch:2022bqa}  &  $e^+ e^- \to \pi^+ \pi^-$ \cite{Banerjee:2020rww} \\ \hline\hline
\multirow{3}{*}{ISC}   & NLO                & full mass dependence                                                            & full mass dependence                                                                 & full mass dependence                                               \\ \cline{2-5} 
                       & \multirow{2}{*}{NNLO}& massified                                                                       & full mass dependence                                                                 & full mass dependence                                               \\
                       &                   & OpenLoops+NTS                                                                   & OpenLoops+NTS                                                                        & OpenLoops+NTS                                                      \\ \hline
\multirow{3}{*}{FSC}   & NLO                   & full mass dependence                                                            & full mass dependence                                                                 &                                                                    \\ \cline{2-5} 
                       & \multirow{2}{*}{NNLO} & massified                                                                       & full mass dependence                                                                 &                                                                    \\
                       &                       & OpenLoops+NTS                                                                   & OpenLoops+NTS                                                                        &                                                                    \\ \hline
\multirow{3}{*}{mixed} & NLO                   & full mass dependence                                                            & full mass dependence                                                                 &                                                                    \\ \cline{2-5} 
                       & \multirow{2}{*}{NNLO} & massified                                                                       & massified                                                                            &                                                                    \\
                       &                       & OpenLoops+NTS                                                                   & OpenLoops+NTS                                                                        &                                                                    \\ \hline
                    \multirow{2}{*}{VP}    &                       & HVP with \texttt{alphaQED}                                                           & \multicolumn{1}{c|}{HVP with \texttt{alphaQED}}                                                                 & HVP with \texttt{alphaQED}                                                               \\
                       &                       &         or NSK                                                                        & \multicolumn{1}{c|}{or NSK}                                                                                     & or NSK and VFF         
\end{tabular}
}
\caption{
Overview of the corrections to $e^+e^-\to X^+ X^-$ that are available in \mcmule{} at NLO and NNLO. 
At LO and NLO all implementations have full mass dependence. Radiative processes $e^+e^-\to X^+ X^- \gamma$ at NLO are deduced from NNLO results of $e^+e^-\to X^+ X^-$.
}
\label{tab:mcmulenlonnlo}
\end{table}

\mcmule{} currently does not support any effects related to the hadronic structure of the pion in the final state beyond one-photon exchange contributions. As indicated in Table~\ref{tab:mcmulenlonnlo}, only ISC and VPC are included together with a customisable pion form-factor vertex. Work to implement the remaining contributions is currently in progress.

With \mcmule{} it is possible to separately compute contributions involving the leptonic and hadronic VP, including at NNLO. The most recent version of the Fortran library \texttt{alphaQED}~\cite{Jegerlehner:2001ca, Jegerlehner:2006ju, Jegerlehner:2011mw, Jegerlehner:hvp19}, \texttt{alphaQEDc23}, or the NSK VP~\cite{Ignatov:2008bfz,Ignatov:hvp} are employed for the evaluation of the HVP.
The leptonic VP can even be split into separate leptons. However, typically \mcmule{} combines the full leptonic VP parts with the photonic corrections. 

Fermionic corrections are treated differently according to how the VP is inserted in the diagrams. At NNLO, if $\Pi(Q^2)$ factorises from the rest of the amplitude, the correction reduces to quantities that have already been computed at NLO. Non-factorisable contributions need instead a special treatment, which corresponds to a dispersive~\cite{Cabibbo:1961sz} or hyperspherical~\cite{Levine:1974xh,Levine:1975jz,Fael:2018dmz} approach in \mcmule{}. In the dispersive approach, the VP in the original integrand is replaced by a massive photon propagator, where the photon mass is the dispersion parameter. This integral is then computed with existing one-loop tools: in \mcmule{}, Collier is used due to its stability in the presence of large cancellations for large values of the dispersion parameter. However, for extremely large values of the latter, a sufficiently precise evaluation of the kernel becomes even more difficult. This necessitates expanding the amplitude, typically using the method of regions~\cite{Beneke:1997zp}.
    \subsection{Phokhara} \label{sec:phokhara}

The Monte Carlo event generator \phokhara{} was conceived
to provide theoretical predictions for
cross section measurements
at fixed low-energy meson factories. 
In particular, the current version of the generator \phokhara{} 10.0 focuses on the radiative return processes $e^+ e^- \to \mu^+ \mu^- \gamma$~\eqref{intro:returnMu} and $e^+ e^- \to \pi^+ \pi^- \gamma$~\eqref{intro:returnPi} at NLO.
These theoretical predictions include virtual and soft photon corrections to one-photon emission
events and the emission of two real hard photons, 
accounting for the complete NLO corrections. 

\phokhara{} can be downloaded from
\begin{quote}     
\url{https://looptreeduality.csic.es/phokhara/}.
\end{quote} 

In addition to these flagship processes, \phokhara{} also contains a variety of hadronic production channels. 
We summarise in Table~\ref{tab:phokharalonlo} all physical processes which are available in \phokhara{}, as well as some features of the main channels.
\begin{table}[t]
\renewcommand{\arraystretch}{1.5}
\centering
\begin{tabular}{c||c|c|c|c}
 $e^+e^-\to$ &  Order  & VP & VFF  & Extras  \\ \hline\hline
$\mu^+\mu^-$ & LO & \texttt{alphaQED}, & & Narrow resonances  \\ \cline{1-2} 
 \multirow{2}{*}{$\mu^+\mu^-\gamma$}  & NLO with full & from \cite{Hagiwara:2003da,Hagiwara:2006jt} & - & of $J/\psi$ and $\psi(2S)$ \\ 
 & mass dependence & or NSK  & &\\ \cline{1-5}
 $\pi^+\pi^-$ & LO & \texttt{alphaQED}, & F$\times$sQED & Narrow resonances \\ \cline{1-2}
\multirow{2}{*}{$\pi^+\pi^-\gamma$}  & NLO with full & from \cite{Hagiwara:2003da,Hagiwara:2006jt} & choice of & of $J/\psi$ and $\psi(2S)$\\ 
 & mass dependence & or NSK & 3 VFF & Radiative $\phi$ decays \\ \hline\hline

\multirow{2}{*}{$X$} & \multicolumn{4}{c}{$X \in $ 2$\pi^0\pi^+\pi^-$, $2\pi^+2\pi^-$, $p\bar{p}$, $n\bar{n}$, $K^+K^-$, $K^0\bar{K}^0$, $\pi^+\pi^-\pi^0$, $\Lambda(\to\pi^-p)\bar{\Lambda}(\to\pi^+\bar{p})$,}\\
& \multicolumn{4}{c}{$\eta\pi^+\pi^-$, $\pi^0\gamma$, $\eta\gamma$, $\eta'\gamma$, $\chi_{c1}\to J/\psi(\to\mu^+\mu^-)\gamma$, $\chi_{c2}\to J/\psi(\to\mu^+\mu^-)\gamma$} \\

\end{tabular}
\caption{%
Overview of the corrections to 
$e^+e^-\to X$, 
$e^+e^-\to X^+ X^-$,
$e^+e^-\to X^+ X^-+\gamma$,
that are available in \phokhara{} at LO and NLO. 
}
\label{tab:phokharalonlo}
\end{table}

In the following, we briefly discuss the history and main features of the generator.
The first versions of \phokhara{} 
were based on EVA \cite{Binner:1999bt}, a LO Monte Carlo generator for the pion radiative-return process~\eqref{intro:returnPi}, extended to include the muon process~\eqref{intro:returnMu}, as well as the emission of an extra hard photon in the final state from the incoming electrons. 
In \phokhara{}, ISC and FSC are treated separately as independent contributions using the gauge-invariant splitting method described in \secref{sec:comp-fo}. By switching them on or off, an accurate determination of the contributions to the physical
region under consideration in the experiment can be explored. When looking at these physical
observables within \phokhara{}, one can identify interesting properties
of the physical process, which are useful during all phases of an experiment.

The treatment of hadrons in the final state, in particular pions, 
is carried out under the approximation of the pions being point-like particles described by sQED 
(see~\figref{fig:comp-WP}). This approach, not being rigorous, is augmented by the use of a non-perturbative form factor. 
For further details, we refer the reader to~\cite[Figure~2.10]{Punzi2023}. 

The decomposition between perturbative and non-perturbative contributions 
to the pion production cross section~\eqref{intro:returnPi} motivated the addition 
of initial-state radiative corrections as a first attempt in \phokhara{} 
to simulate emission of photons at large-angle
($\theta\gg m_{e}^{2}/s$)~\cite{Rodrigo:2001cc,Rodrigo:2001jr,Rodrigo:2001kf}, and small-angle ($\theta\ll m_{e}^{2}/s$)~\cite{Kuhn:2002xg} regions.
To further improve theoretical predictions, for final-state radiation off the pions, \phokhara{} adopted the F$\times$sQED approach (see \secref{sec:comp-fshad-strategies}). 
The accuracy of this approach is under scrutiny. 
Using these methods, \phokhara{} includes
the emission of one photon from the initial state and one photon from the final state, requiring
one of the photons to be hard~\cite{Czyz:2003ue,Czyz:2004rj}. 
The relevant Feynman diagrams can be seen 
in the right column of \figref{fig:eepipigamma-NLO-C-odd} and~\figref{fig:eepipigamma-NLO-VFF}.

Regarding other important channels for the evaluation of the hadronic vacuum polarisation, \phokhara{} considers the
hadronic channels of $\pi^{+}\pi^{-}\pi^{0}$, kaon pairs $K^+K^-$
and $K^0\bar{K}^{0}$~\cite{Czyz:2002np,Czyz:2005as}, 
nucleon pairs $p\bar{p}$ and $n\bar{n}$, and the radiative
$\phi$ decay contributions to the reaction $e^{+}e^{-}\to\pi^{+}\pi^{-}\gamma$
relevant when running at $\phi$-factory energy (see \secref{sec:exp-frascati}).

The version of \phokhara{} which is employed for the Monte Carlo  comparisons \pdforweb{in \secref{sec:mcc}}{here} considers
the complete set of Feynman diagrams that contribute to the NLO theoretical
prediction of the scattering processes~\eqref{intro:returnMu} and~\eqref{intro:returnPi}.
On top of considering ISC
and FSC virtual and real corrections, this version includes, for the
first time, the gauge-invariant group of diagrams
containing two virtual photons, referred to as the penta-box contribution
(see~\figref{fig:comp-eemmy2PE}). The presence of these diagrams gives a complete NLO theoretical prediction. 

In the calculation of the muon production cross section~\eqref{intro:returnMu}, the evaluation of Feynman integrals
appearing at intermediate steps of the calculation is achieved by the use of publicly automated software~\cite{Ellis:2007qk,Fleischer:2011zz,vanHameren:2010cp,vanOldenborgh:1990yc}.
A more detailed discussion on the QED calculation and the implementation
of these contributions in \phokhara{} can be found in~\cite{Campanario:2013uea}. 
From the Feynman diagrams needed at NLO, one can recognise a decomposition in terms of the 
couplings between the photon and leptons (analogously to~\eqref{comp:olA}),
\begin{align}
\mathcal{A}_{mm\gamma\gamma}^{\left(0\right)}\left(q_{e}q_{m}\right)=\ &\mathcal{A}_{mm\gamma\gamma}^{\left(0\right)}\left(q_{e}^{3}q_{m}\right)+\mathcal{A}_{mm\gamma\gamma}^{\left(0\right)}\left(q_{e}^{2}q_{m}^{2}\right)+\mathcal{A}_{mm\gamma\gamma}^{\left(0\right)}\left(q_{e}q_{m}^{3}\right)\,,\\
\mathcal{A}_{mm\gamma}^{\left(1\right)}\left(q_{e}q_{m}\right)= \ &
\mathcal{A}_{mm\gamma}^{\left(1\right)}\left(q_{e}^{4}q_{m}\right)+\mathcal{A}_{mm\gamma}^{\left(1\right)}\left(q_{e}^{2}q_{m}^{3}\right)+\mathcal{A}_{\text{PB};mm\gamma}^{\left(1\right)}\left(q_{e}^{3}q_{m}^{2}\right)+\mathcal{A}_{mm\gamma}^{\left(1\right)}\left(q_{e}^{2}q_{m}\Pi^{\left(1\right)}\right)
\notag\\
    &+\left(q_{e}\pdforweb{\leftrightarrow}{\to} q_{m}\right)\,,
\end{align}
where $\mathcal{A}_{mm\gamma\gamma}^{\left(0\right)}\left(q_{e}q_{m}\right)$
accounts for the real radiation of two photons,
and $\mathcal{A}_{mm\gamma}^{\left(1\right)}\left(q_{e}q_{m}\right)$ 
is understood as the virtual correction to emission of a photon from a leptonic line.
This decomposition can be elucidated as the following gauge invariant pieces, 
with $f,F\in\{e,\mu\}$ and $f\ne F$, 
\begin{itemize}
\item 
$f^+f^-\to\gamma^*\to F^+F^- + \gamma $,
\item
$f^+f^-\to\gamma \gamma^*\to F^+F^- $,
\item 
$f^+f^-\to F^+F^- +\gamma$ (only diagrams containing two virtual photons),
\item
Insertion of VP. 
\end{itemize}
ISC and FSC real corrections in \eqref{intro:returnMu}, needed at NLO, that are accounted by $\mathcal{A}_{mm\gamma\gamma}^{\left(0\right)}\left(q_{e}q_{m}\right)$
with the emission of two hard photons are calculated in terms of helicity amplitudes.
This is done, in order to overcome numerical instabilities.  

For the process~\eqref{intro:returnPi} the organisation of virtual and real corrections
is less evident within \phokhara{} framework, since the 
organisation of this calculation has not been documented. 
However, a similar approach to the muon channel was carried out.
Additionally, because \phokhara{} makes extensive use of scalar QED, it always looks for 
consistency physical checks at various steps of the calculation. 
In particular, the study of the infrared structure of virtual and real corrections determines if the code needs to switch to running in quadruple precision.

Finally, for both~\eqref{intro:returnMu} and \eqref{intro:returnPi}, VP diagrams are not included, and their contribution is instead taken into account by an overall factor, included at NNLO. To calculate it, three routines are available, as summarised in~\tabref{tab:phokharalonlo}. Furthermore, for the two main processes and some hadronic final states, \phokhara{} has the option to simulate narrow resonances and their decay into kaon, pion and muon pairs, also listed in the aforementioned table. This allows the generator to produce a description of the experimental data at the peaks of the resonances, as explored in~\cite{Czyz:2010hj}.

In the modified version of \phokhara{} 10.0, available in the GitLab repository associated with this manuscript, the generator is now able to incorporate fully customised cuts at the generation level, as well as producing and storing histograms of differential cross sections versus user defined variables. Finally, the four-momenta and weights of all events can be stored in a \verb".csv" file for further analysis.

\subsection{Sherpa} \label{sec:sherpa}
\Sherpa{} is a general-purpose Monte Carlo event generator 
which was originally developed to model high-multiplicity processes 
at the LHC. Since its inception, the generator has been under 
active development and recent developments include the addition of 
dedicated modules for the simulation of processes at lepton 
colliders, in addition to many other 
physics improvements. \Sherpa{} includes two inbuilt matrix-element 
generators, \Amegic{} \cite{Krauss:2001iv} and \Comix{} \cite{Gleisberg:2008fv,Hoche:2014kca}.
These allow the automated generation of tree-level matrix 
elements in the complete Standard Model.

A number of QED radiation methods are implemented in \Sherpa.
Primarily, there are two implementations of the YFS 
resummation: the {\sc Photons} module for final-state radiation 
in decays \cite{Schonherr:2008av,Krauss:2018djz,Flower:2022iew}, and the {\sc YFS} module 
for full QED radiation from the initial and final states 
in lepton--lepton collisions \cite{Krauss:2022ajk}. These 
two modules have been stringently tested and are in very  
good agreement with each other and with \Photos{} \cite{Gutschow:2020cug}.
To produce the results \pdforweb{in \secref{sec:mcc}}{presented here}, the YFS method has been
used to resum all QED IR divergences to all orders (see \secref{sec:comp-bfoYFS} for a detailed description).

This resummation method can be enhanced further by including higher-order corrections. 
Two main approaches exist for incorporating these higher-order effects. 
The first approach, called exclusive exponentiation (EEX), adheres to the framework outlined in the original YFS paper. 
This approach constructs $\tilde{\beta}_{ij}$ through analytic differential distributions derived from the 
corresponding Feynman diagrams. These expressions are typically expressed in terms of products of four-vectors; 
at lower orders, this frequently involves Mandelstam variables or similar quantities.

A key advantage of the EEX method is its relatively straightforward implementation, as well as the ease with which one can verify the behaviour of these expressions in specific limits, such as soft or collinear limits. Additionally, constructing terms independently for ISR and FSR contributions simplifies the automation and implementation of ISR effects, particularly for lepton colliders (such as $e^+e^-$ or $\mu^+\mu^-$). However, due to the potential complexity of the final states, automating FSR contributions is challenging and generally requires handling each case individually.
The EEX corrections we have included were computed in~\cite{Berends:1980yz,Berends:1986fz,Kuraev:1985hb,Burgers:1985qg,Passarino:1978jh,Berends:1987ab,Berends:1983mi,Blumlein:2011mi,Blumlein:2020jrf}, and explicit expressions for the infrared-subtracted terms can be found in~\cite{Jadach:2000ir}. We summarise these
corrections in \tabref{TAB::YFS::BETS}.

The code is available to download at
\begin{quote}
    \url{https://gitlab.com/sherpa-team/sherpa}
\end{quote}

\begin{table}
\begin{center}
\setlength{\tabcolsep}{11pt}
\renewcommand{\arraystretch}{1.5}
 \begin{tabular}{c|| c c||c}
  Order  & ISR Corrections &
  FSR Corrections & Reference \\
  \hline
    $\tilde{\beta}^{0}_{0}$
            & Born
            &  Born &  \\
    $\tilde{\beta}_{1}^{1} + \tilde{\beta}_{0}^{1}$ & $\mathcal{O}(\alpha,\alpha L)$ & $\mathcal{O}(\alpha,\alpha L)$ & Eq. (13) \\
   $ \tilde{\beta}_{2}^{2} + \tilde{\beta}_{1}^{2} + \tilde{\beta}_{0}^{2}$ & $\mathcal{O}(\alpha^2 L^2)$ & $\mathcal{O}(\alpha^2 L)$ &  Eq. (18,19) \\
    $\tilde{\beta}_{3}^{3} + \tilde{\beta}_{2}^{3} + \tilde{\beta}_{1}^{3} $& $\mathcal{O}(\alpha^3 L^3)$ & --- &  Eq. (26) \\
\hline
   \end{tabular}
  {\caption{
    The explicit beta terms that have been implemented in \Sherpa's 
    EEX. The reference column provides the equation number from~\cite{Jadach:2000ir} where the explicit form of the corrections are given. }
  \label{TAB::YFS::BETS}}
\end{center}
\end{table}
\section{Monte Carlo comparisons}\label{sec:mcc}

In this section we present results obtained with the Monte Carlo codes described in \secref{sec:gen}. To this end, we define a set of scenarios, i.e., simplified setups consisting of acceptance cuts, reminiscent of the experiments described in \secref{sec:exp}. The purpose is not to precisely match the experimental analyses carried out. Rather, we want to provide simplified but still realistic phenomenological scenarios that can be used to validate the codes and assess the importance of the various contributions to differential cross sections. As we will see, the impact and importance of these contributions is strongly dependent on the observables. Thus, generic statements are difficult to make and should be treated with care. 

The source codes that have been used to obtain these results can be found at
\begin{quote}
    \url{https://radiomontecarlow2.gitlab.io/monte-carlo-results/}
\end{quote}
Furthermore, these results can also be used to benchmark future theoretical developments. It is foreseen that the repository is updated if new theoretical computations become available. 

Throughout this section we use the same colour coding of the seven Monte Carlo codes, as shown in \figref{fig:KLA-mxx}. However, for most scenarios not all codes have provided results. Some codes are specialised either to scan $2\to{2}$ or to radiative $2\to{3}$ processes. Furthermore, sometimes particular final states or VPC have not (yet) been implemented. 

The errors that are indicated in the plots are only the statistical Monte Carlo errors. We obtain rough indications of the expected theoretical errors by comparing differences in various approaches and approximations. However, a reliable estimate of theoretical error is beyond the scope of the present article. In order to disentangle the impact of input parameters from other differences, in this section we use a standard implementation of the HVP and pion form factor, as discussed in \secref{sec:input}.

In the process of preparing these results we have also carried out numerous technical validations, by comparing identical results produced by different codes.

\subsection{Input values and observables} \label{sec:input}

We consider the processes 
\begin{subequations}\label{eq:processes}
\begin{align}
    e^+\,e^- &\to X^+(p_+)\,X^-(p_-) \label{eq:process}  \\
    e^+\,e^- &\to X^+(p_+)\,X^-(p_-)\,\gamma(p_\gamma) \label{eq:processRad}
\end{align}
\end{subequations}
with  $X\in\{e,\mu,\pi\}$ and momenta $p_\pm=(E_\pm, \vec{p}_\pm)$. The polar and azimuthal angles are denoted by $\theta_\pm$ and $\phi_\pm$ and we sometimes use the notation $\vp_\pm\equiv|\vec{p}_\pm|$.  The processes \eqref{eq:process} are considered in a CMD-like scenario. The radiative processes \eqref{eq:processRad} are looked at for scenarios related to KLOE, BES~III, and B~factories. In some of these (tagged cases), the photon is explicitly detected, allowing for cuts on 
$p_\gamma=(E_\gamma,\vec{p}_\gamma)$.  Of course, beyond LO the radiative processes will have more than one photon in the final state. Cuts on $E_\gamma$ below indicate there is at least one photon with
energy above this cut. If there are more photons in the final state, we assume they can always be separated.  The photon $\gamma(p_\gamma)$ is the hardest photon passing the cut. 
In other scenarios (untagged cases), the photon is indirectly inferred from missing momentum. In this case we define $\vec{p}_{\widetilde\gamma}\equiv -(\vec{p}_{+}+\vec{p}_{-})$.

Apart from the components of the various momenta we also consider the following quantities:
\begin{subequations} 
\begin{align}
   &\mbox{invariant mass}&  &M_{XX}\equiv \sqrt{(p_++p_-)^2}& &\mbox{or}& 
     & M_{XX\gamma}\equiv \sqrt{(p_++p_-+p_\gamma)^2}  \\
    &\mbox{(polar angle) acollinearity}& &\xi\equiv|\theta^+ + \theta^- - \pi|& \\
     &\mbox{average polar angle}  & & \theta_\text{avg}\equiv (\theta^- - \theta^+ + \pi)/2 & \\
    &\mbox{azimuthal angle acollinearity} &  &\big| |\phi^+ - \phi^-| - \pi\big|& 
\end{align}
\end{subequations}

We use on-shell coupling and masses 
\begin{align}\label{eq:inputval}
  \begin{split}
    \begin{aligned}
      \alpha&=1/137.03599908, &
      m_e&=0.510998950\, \MeV,  &
      m_\mu&=105.658375\, \MeV, \\
      m_\tau&=1776.86\, \MeV,  &
      m_\pi&= 139.57039\,\MeV  \,  .
\end{aligned}
\end{split}
\end{align}

\subsubsection{Vacuum polarisation} \label{sec:mccVP}

\begin{figure}[t]
    \centering
    \includegraphics[width=0.49\textwidth]{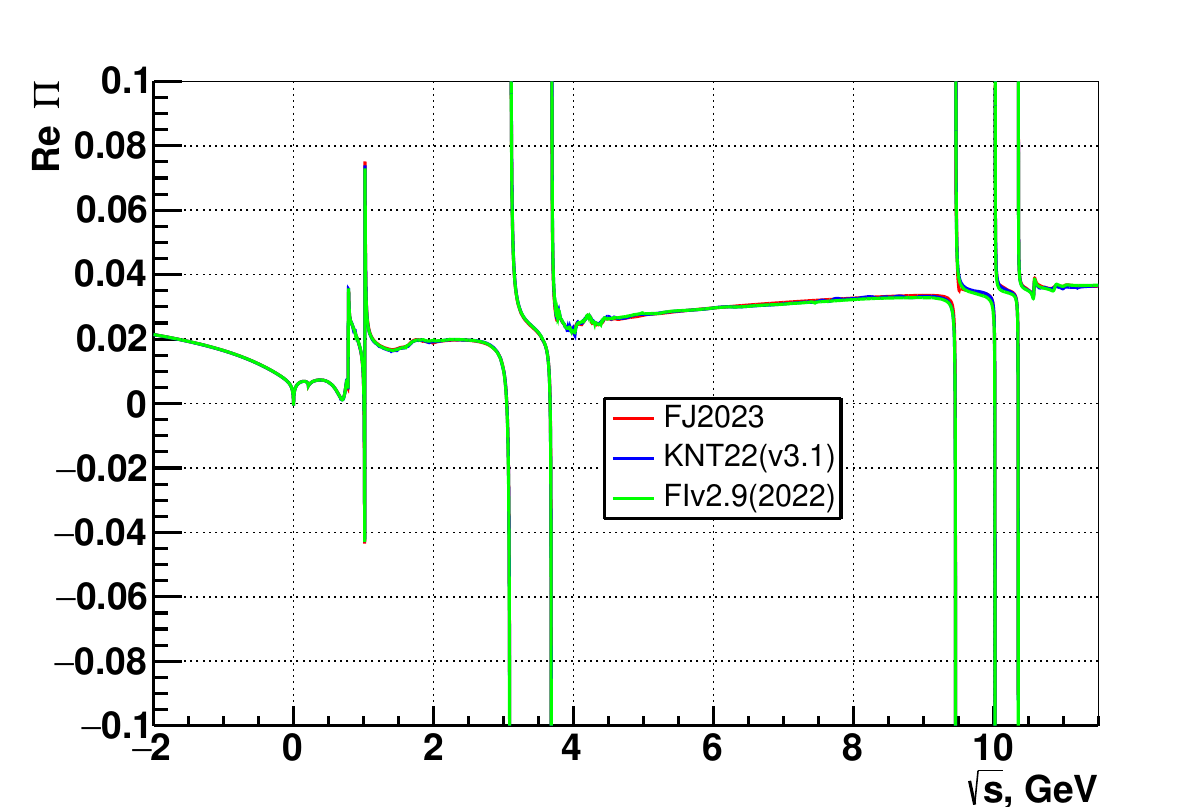}
    \includegraphics[width=0.49\textwidth]{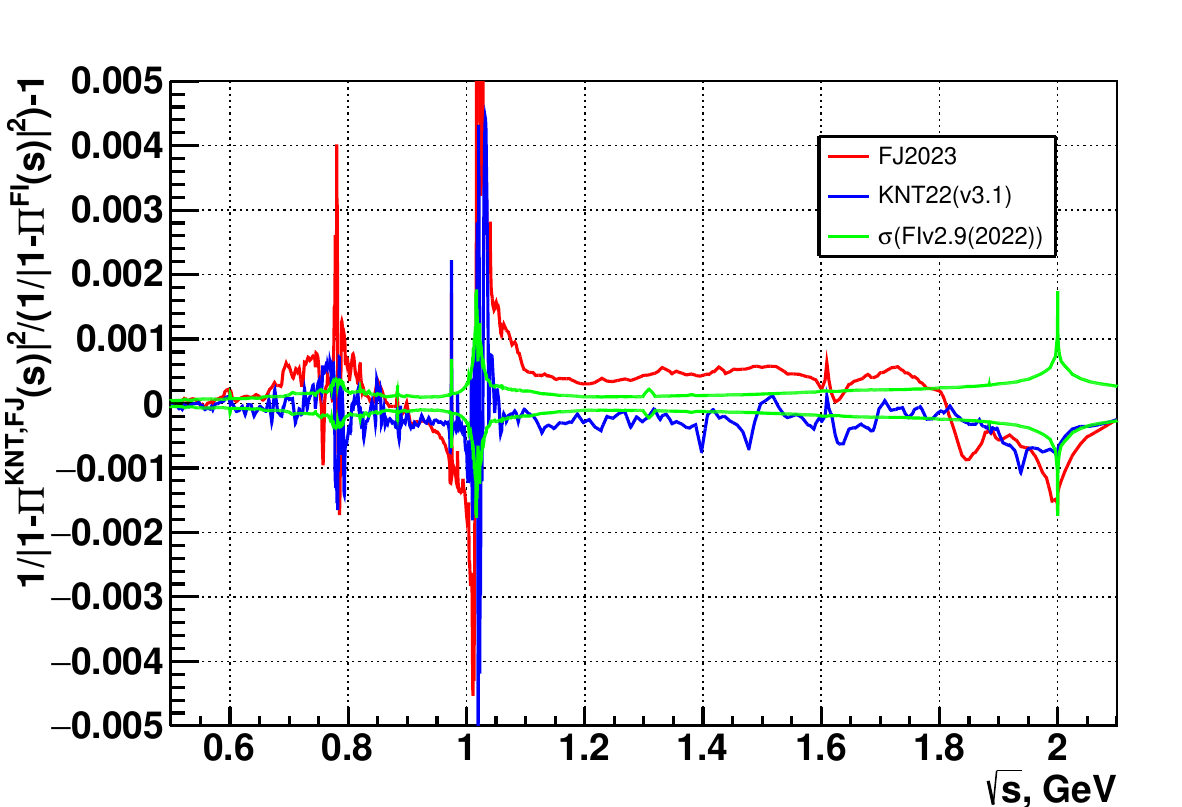}
    \caption{The comparison of the VP function from different packages. The
      left plot corresponds to $\Re\,\Pi(s)$, where the red line is
      parameterisation from Jegerlehner (2023), the blue line -- KNT
      v3.1, the green line -- NSK VP v2.9. The normalised difference of
      $1/|1-\Pi(s)|^2$ relative to the NSK VP is shown on the right plot, where green lines indicate the
      uncertainty of the NSK VP evaluation.
    }
    \label{fig:vpcomp}

    \centering
    \includegraphics[width=0.49\textwidth]{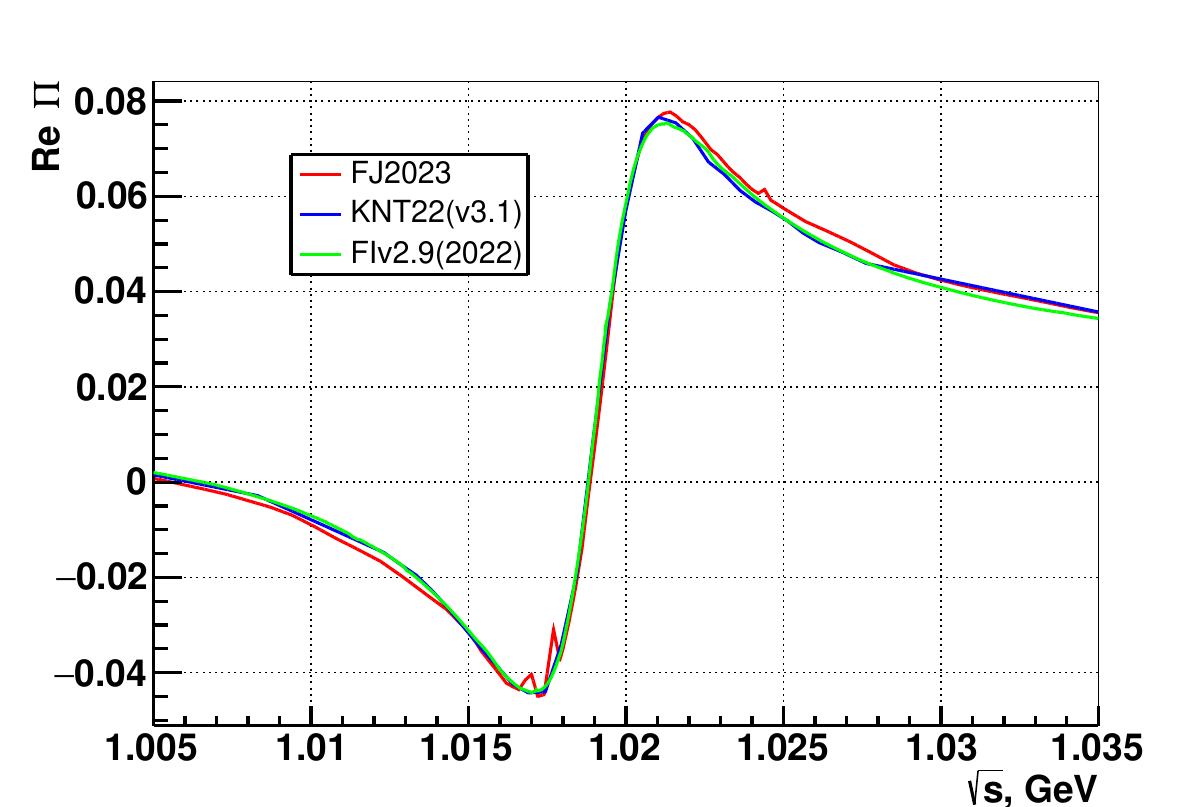}
    \includegraphics[width=0.49\textwidth]{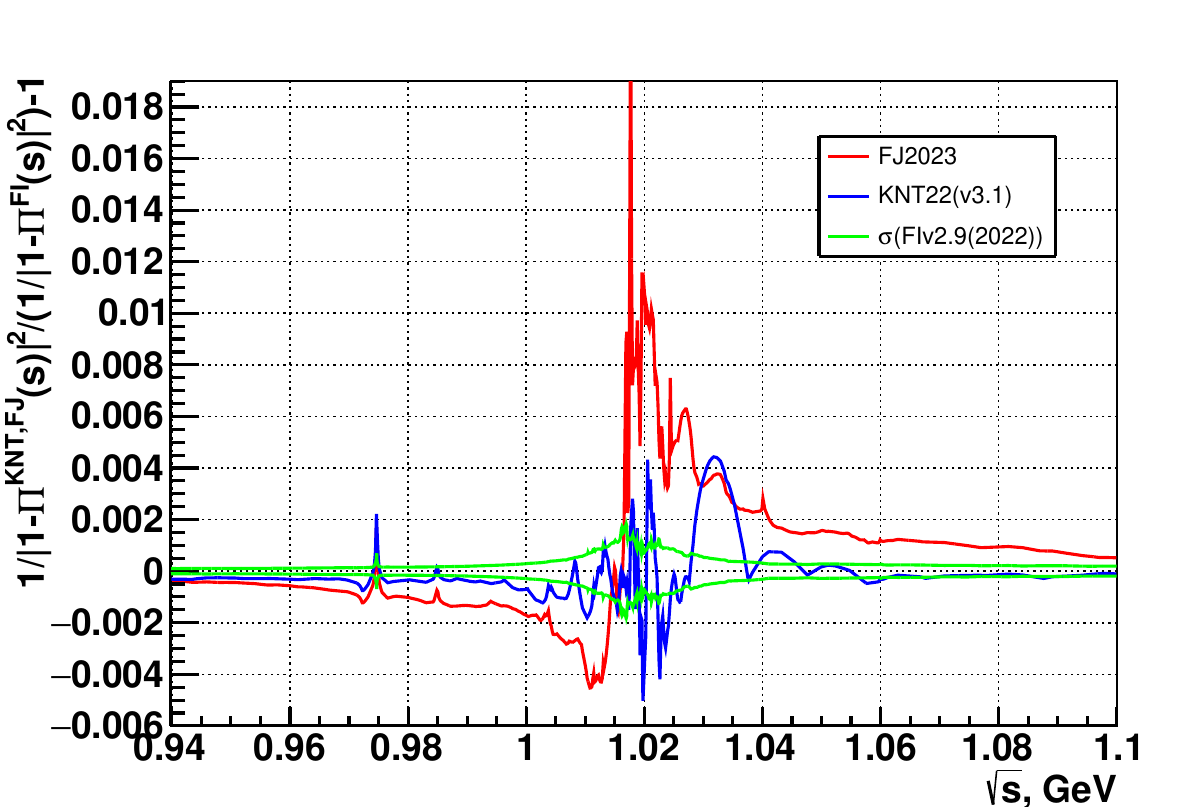}
    \caption{The comparison of the VP function from different packages at the
      $\phi$ resonance. The left plot corresponds to $\Re\,\Pi(s)$, and 
      right plot the normalised difference of $1/|1-\Pi(s)|^2$ relative to the NSK VP.}
    \label{fig:vpcompPhi}
\end{figure}

\begin{figure}[bth]
    \centering
    \includegraphics[width=0.49\textwidth]{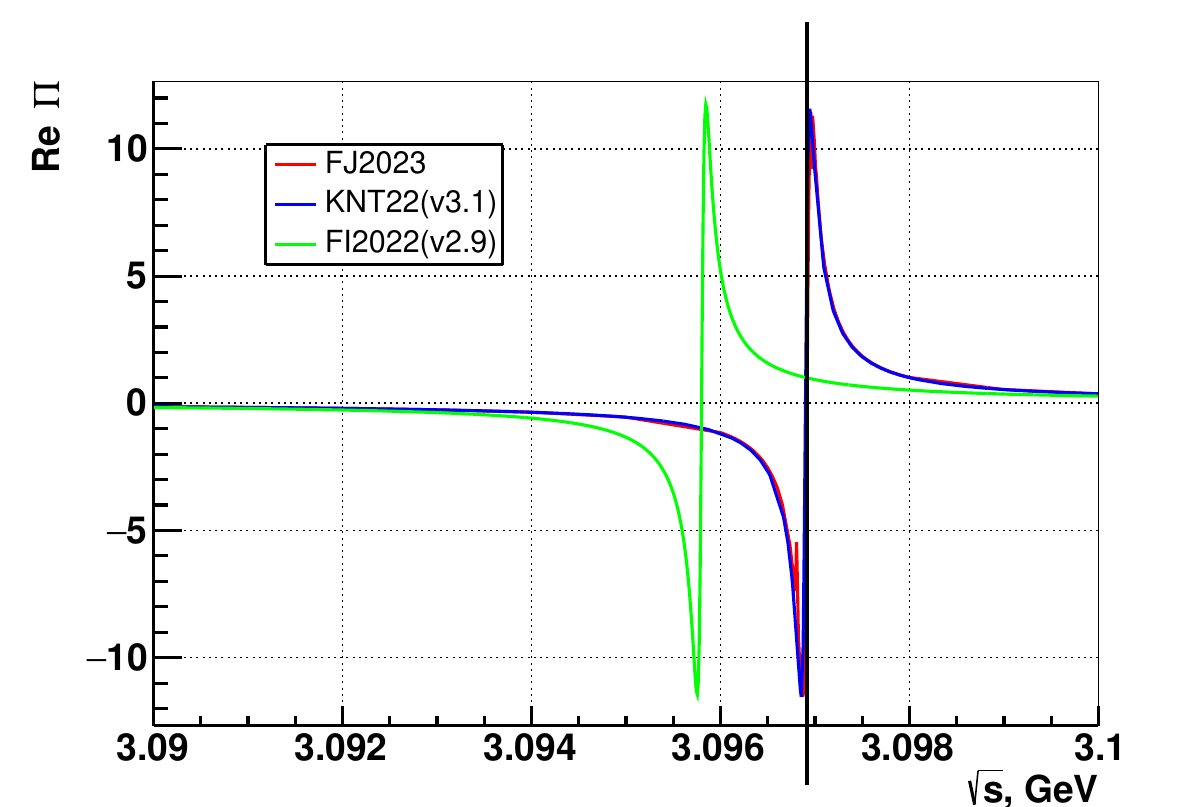}
    \caption{The  behaviour of  $\Re\,\Pi(s)$ around the $J/\psi$
      resonance. The green line corresponds to the proper evaluation using
      bare resonance parameters, while the blue and red lines were obtained
      using dressed PDG parameters.}
    \label{fig:vpcompJpsi}
\end{figure}

The VP is responsible for
the running of the electromagnetic coupling $\alpha(q^2)=\alpha/(1-\Delta\alpha(q^2))$, and it
is an essential ingredient of any radiative corrections calculation. Generalising \eqref{comp:hvp_disprel} to also include leptonic final states $\Pi^\text{ren}(s)=\Pi_\ell^\text{ren}(s) + \Pi_h^\text{ren}(s)$, the VP can be determined from $e^+e^-$ annihilation cross section using
the dispersion relation based on analyticity and unitarity
\begin{eqnarray}
\Pi^\text{ren}(s)= \frac{s}{4\pi^2\alpha}\bigl[\mathrm{PV}\!\!\!
\int\limits_{4 m_{l}^2}^{\infty}\frac{\sigma_{e^+e^- \to \gamma^* \to X}^{\text{bare}}
(s')ds'}{s-s'}- \mathrm{i}\pi\sigma_{e^+e^- \to \gamma^* \to X}^{\text{bare}}(s)\bigr],
\label{eq:pv-int}
\end{eqnarray}
where the total cross section corresponds to the production of all final states $X$ (including leptons and hadrons) via the one-photon process.  The
intermediate photon is undressed from the VP contribution itself.
In this case, the VP function represents the one-particle irreducible diagram.
The Dyson resummation of the VP corresponds to the leading effect of the VPC on the total (say $e^+e^- \to \mu^+\mu^-$) cross section
\begin{align}\label{eq:mccVPresum}
 \dd\sigma_{mm}(q_e^2\,q_m^2\,\Pi)&=\frac{\dd\sigma_{mm}(q_e^2\,q_m^2)}{|1-\Pi^\text{ren}(s)|^{2}} \,.
\end{align} 
It should be noted that the neglect of the imaginary part of $\Pi(s)$ results in a 1.6\%
systematic error of the $e^+e^- \to \mu^+\mu^-$ cross section on the peak of the $\phi$ resonance.

While the leptonic part of the VP can be computed in QED as discussed in \secref{sec:comp-fo}, the hadronic part can be determined with sufficient precision only by using experimental $e^+e^-\to \text{hadrons}$ data at this
moment. This requires proper combination and merging of all available
experimental $e^+e^-$ datasets, taking into account possible correlations in the systematic uncertainties between different channels. In this subsection we point out some differences between the recent most precise tabulations of the full VP, provided by the KNT group (v3.1, 2022)~\cite{Keshavarzi:2019abf}, Jegerlehner (hadr5x23 from \texttt{alphaQED23} package, 2023)~\cite{Jegerlehner:hvp19},
and Novosibirsk (NSK) VP (v2.9, 2022)~\cite{Ignatov:2008bfz,Ignatov:hvp}.

With the current knowledge of the $e^+e^-$ cross sections the precision of the 
$|1-\Pi^\text{ren}(s)|^2$ normalisation factor
is better than $0.05\%$ at c.m. energies below 5\,GeV, except at narrow resonances.
There the precision is somehow degraded and,
for example, the latest combination of experimental data gives
0.2\% statistical accuracy at the peak of the $\phi$ resonance.
The comparison of different VP compilations is shown in \figref{fig:vpcomp}.
A good consistency is seen outside the $\omega$ and $\phi$ resonances.
The \texttt{alphaQED} version prior to the 2023 release used dressed instead of bare
$\phi$ resonance parameters. This resulted in an
additional $\sim2.5\%$ systematic bias of the cross section at these
c.m. energies. The latest 2023 release has switched to bare resonance
treatment, but it still deviates from KNT and NSK compilations at the level of $-0.4$\% to +1\% at the left and right side 
of the $\phi$ peak, as shown in \figref{fig:vpcompPhi}.
The KNT v3.1 does not have a sufficient number of tabulated points at the fast changing
$\phi$ interference, which gives an additional ~0.5\% error after using linear
interpolation between the tabulated points (as seen at $\sqrt{s}=1.03\,\text{GeV}$ in \figref{fig:vpcompPhi}).
Both KNT and Jegerlehner's compilations are using dressed PDG mass parameters for
resonances like $J/\psi$ and $\psi'$, which gives a much larger effect in theses cases compared to the $\phi$ resonance. 
For example, the shift from dressed to bare $J/\psi$ mass
is about 1.1\,MeV, which is much larger than the resonance width itself.
The significant changes of the bare full and leptonic widths need to be taken into account as well~\cite{Ignatov:2008bfz,Anashin:2010ogq,Anashin:2011ku}.
The consistent definition of the bare resonance parameters and usage of the
resummed VP function leads to the physically observed
$\sigma^\text{bare}/|1-\Pi^\text{ren}(s)|^{2}$ cross section on resonances, without convergence problems of the Dyson series which occur when using a real running coupling definition. 
Improper usage of the dressed resonance parameters in the dispersive
integral~\eqref{eq:pv-int} will give unreliable VP energy dependence
in the vicinity of them as shown in \figref{fig:vpcompJpsi}.
Using the full VP from such packages is not applicable to energies around narrow resonances. 
Note that for the KNT compilation, when used close to a very narrow resonance, it was recommended to switch off its particular contribution to the running coupling and instead add the corresponding narrow resonance contribution by hand.

All packages mentioned above
include at least the NLO term in the leptonic part of the VP.  This
is more than sufficient compared to the precision level of the hadronic part. The NSK tabulation of the leptonic 
$\Pi_\ell(s)$ was calculated numerically through the integral from the total $e^+e^-\to \ell^+\ell^-\gamma$ cross
section. This includes the LO FSR correction~\cite{Bystritskiy:2005ib} and the additional
enhancement on the threshold from the Coulomb final-state interaction.
(the $\Pi_\ell(s)$ by the dispersive integral from such cross
sections without the Sommerfeld--Gamow--Sakharov factor gives numerically
identical results to the one-two-loop analytical formula of the VP in the paper~\cite{Laporta:2024aok}). Similar methods have been used to describe contributions of low-virtuality photons to jets at LHC~\cite{Denner:2019zfp}.

Some Monte Carlo generators still use old implementations of the VP function,
which implies additional systematic uncertainties.
Comparisons of older VP versions were given in~\cite{WGRadCor:2010bjp}.
For example, the original version of KKMC was not supposed to be used at
low energies $\sqrt{s}<2$\,GeV. The usage of the VP implemented in this package leads up to 10\%
systematic variations of the total $e^+e^-\to\mu^+\mu^-$ cross section in this case.
The original \afkqed{} generator uses a simplified evaluation of the VP with
precision up to 0.4\% of the cross section below $\sqrt{s}<2$\,GeV (except $\phi$-resonance, where it degrades up to 1.8\% variation).

In order to disentangle effects from different versions of VP from other differences, we use~\cite{Ignatov:hvp} as default for the hadronic VP for the results presented in this section. However, the procedure of resummation of VPC -- or the lack thereof -- differs between various codes. These differences are typically beyond NLO, but can be numerically significant, in particular near resonances. 
The precise treatment of the VPC in the codes is described in \secref{sec:gen} and in the following comparisons we will highlight their impact for the observables under consideration. 

\subsubsection{Pion form factor}

For the pion VFF $F_\pi(q^2)$ we adopt a harmonised parameterisation for all generators.
This avoids introducing discrepancies due to different choices of the VFF, which would be already visible at LO.
The chosen parameterisation, based on the vector meson dominance (VMD) model~\cite{Sakurai:1972wk}, is inspired by the fit functions which are usually used by the experiments to fit $F_\pi(q^2)$, for instance in~\cite{BaBar:2012bdw,BESIII:2015equ,CMD-3:2023alj}.
In particular, we consider the sum of four $\rho$-resonances with the inclusion of the $\rho$--$\omega$ and $\rho$--$\phi$ interference.
The explicit parameterisation reads
\begin{align}
\begin{aligned}
F_\pi(q^2) &= \frac{\textrm{BW}_\rho^\textrm{GS}(q^2) \left[1+ (q^2/m_\omega^2) \, c_\omega\, \textrm{BW}_\omega(q^2) + (q^2/m_\phi^2) \, c_\phi\, \textrm{BW}_\phi(q^2) \right]}{1+c_{\rho'}+c_{\rho''}+c_{\rho'''}} \\[2pt]
&+  \frac{c_{\rho'}\, \textrm{BW}_{\rho'}^\textrm{GS}(q^2) 
+ c_{\rho''}\, \textrm{BW}_{\rho''}^\textrm{GS}(q^2)
+ c_{\rho'''}\, \textrm{BW}_{\rho'''}^\textrm{GS}(q^2)}{1+c_{\rho'}+c_{\rho''}+c_{\rho'''}}
\end{aligned}
\end{align}
The amplitude of each resonance is a complex number, i.e., $c_v = |c_v|e^{i\varphi_v}$.
The narrow $\omega$- and $\phi$-resonances are described by a Breit--Wigner~(BW) function with a constant width
\begin{align}
\textrm{BW}_v(q^2) = \frac{m_v^2}{m_v^2 - q^2 - i m_v \Gamma_v} \qquad v = \omega,\phi
\end{align}
The broad $\rho$ resonances are described by a Gounaris--Sakurai~(GS)~\cite{Gounaris:1968mw} function
\begin{align}
\textrm{BW}_v^\textrm{GS}(q^2) = \frac{m_v^2 + d(m_v)\,m_v\,\Gamma_v}{m_v^2 - q^2 + f(q^2,m_v,\Gamma_v) - i\, m_v\, \Gamma(q^2,m_v,\Gamma_v)} \qquad v = \rho,\rho',\rho'',\rho'''
\end{align}
where
\begin{align}
& \mathclap{\Gamma(q^2,m_v,\Gamma_v) = \Gamma_v \frac{m_v}{\sqrt{q^2}} \left[ \frac{p_\pi(q^2)}{p_\pi(m_v^2)} \right]^3 \qquad p_\pi(q^2) = \frac12 \sqrt{q^2 - 4m^2_\pi}
} \\[2pt]
& \mathclap{d(m_v) = \frac{3}{\pi} \frac{m_\pi^2}{p_\pi^2(m_v^2)} \log{\frac{m_v + 2p_\pi(m_v^2)}{2m_\pi}} + \frac{m_v}{2\pi p_\pi(m_v^2)} - \frac{m_v m_\pi^2}{\pi p_\pi^3(m_v^2)}} \\[2pt]
& \mathclap{f(q^2,m_v,\Gamma_v) = \frac{\Gamma_v m_v^2}{p_\pi^3(m_v^2)} \bigg[ p^2_\pi(q^2) \left[h(q^2) - h(m_v^2)\right] + p^2_\pi(m_v^2) \left(m_v^2-q^2\right) \,\frac{\textrm{d}h}{\textrm{d}q^2}\Big\vert_{q^2=m_v^2} \bigg]} \\[2pt]
& \mathclap{h(q^2) = \frac{2}{\pi} \frac{p_\pi(q^2)}{\sqrt{q^2}} \log{\frac{\sqrt{q^2}+2p_\pi(q^2)}{2m_\pi}} \qquad
\frac{\textrm{d}h}{\textrm{d}q^2} = \frac{h(q^2)}{8} \left[\frac{1}{p^2_\pi(q^2)} - \frac{4}{q^2} \right] + \frac{1}{2\pi q^2}}
\end{align}

\begin{figure}[t]
    \centering
    \includegraphics[width=0.75\textwidth]{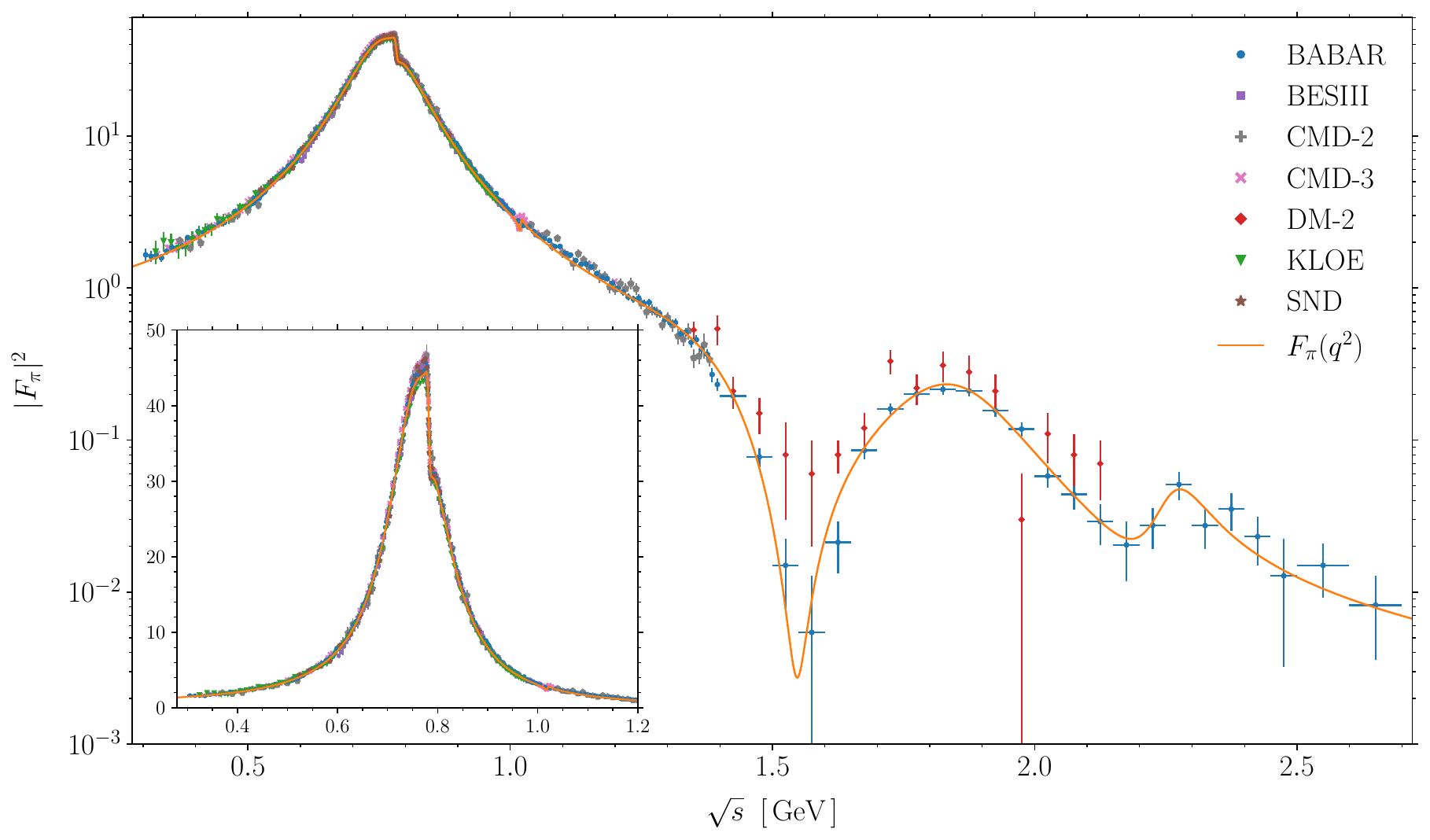}
    \caption{Comparison between our toy model VFF and experimental data.}
    \label{fig:ffpi}
\end{figure}
\begin{table}[h!]
\centering
\begin{tabular}{c|cccccc}
 & $\rho$ & $\rho'$ & $\rho''$ & $\rho'''$ & $\omega$ & $\phi$ \\[1pt]
\hline
$m_v$ (MeV) & 774.56 & 1485.9 & 1866.8 & 2264.5 & 782.48 & 1019.47 \\
$\Gamma_v$ (MeV) & 148.32 & 373.60 & 303.34 & 113.27 & 8.55 & 4.25 \\
$|c_v|$ & - & 0.14104 & 0.0614 & 0.0047 & 0.00158 & 0.00045 \\
$\varphi_v$ (rad) & - & 3.7797 & 1.429 & 0.921 & 0.075 & 2.888 \\

\end{tabular}
\caption{Input parameters for our model of pion VFF $F_\pi(q^2)$.}
\label{tab:ffpi}
\end{table}

The parameter values are inspired by the $F_\pi(q^2)$ measurement of the BaBar, BES~III, CMD-2, DM-2, KLOE, and SND experiments.
The CMD-3 data are used only to fix the parameters describing the $\phi\to\pi^+\pi^-$ resonance. 
The chosen numerical values are listed in \tabref{tab:ffpi}, while a comparison between our toy model VFF and experimental data is shown in \figref{fig:ffpi}.
The parameter values are also chosen to fulfil the dispersive sum rule
\begin{align}
\frac1\pi \int^\infty_{4m_\pi^2} \textrm{d}s' \frac{\Im F_\pi(s')}{s'} = 1 
\end{align}
and the unitary condition $\Im F_\pi(q^2<4m_\pi^2) = 0$ with a 
permille accuracy.

We do not claim that this expression for $F_\pi(q^2)$ is a proper combination of all experimental data. 
It is simply a fixed parameterisation inspired by real data and mainly serves the purpose of allowing generator comparisons without impact from VFF variations. 
Thus, we do not analyse the discrepancy with experimental data or give an error of the VFF parameters. 
Other parametrisations are possible, such as \eqref{eq:VFF}, but for the Monte Carlo comparison we adopt the well-known sum of Gounaris--Sakurai functions for simplicity.
Although the VFF parameterisation is fixed, each generator implements it following one of the approaches described in \secref{sec:comp-fshad}, namely factorised sQED, GVMD, and FsQED.
While all approaches are equivalent at LO, they can give different results at higher orders.
These differences contribute to the theoretical error of the generators.

\subsection{CMD-like scenario} \label{sec:CMDsc}

For the CMD-like scenario we consider the processes $e^+\,e^-\to X^+(p_+)\,X^-(p_-)$ with
$X\in\{e,\mu,\pi\}$ at $\sqrt{s}=0.7\,\GeV$ and apply the following kinematic selection cuts:
\begin{subequations}
\begin{align} \label{eq:cmd-cuts}
&1\,\rad\le\theta_\text{avg}\le\pi-1\,\rad \, ,\\
&\vp_\pm > 0.45\cdot \sqrt{s}/2 \, , \\
&\big| |\phi^+ - \phi^-| - \pi\big| < 0.15\,\rad \, ,\\
&\xi\equiv|\theta^+ + \theta^- - \pi| < 0.25\,\rad \, .
\end{align}
\end{subequations}
In the following subsections we give an example for all three cases $X\in\{e,\mu,\pi\}$.

\subsubsection{Muon final state} \label{sec:CMDscM}

\begin{figure}[t]
    \centering
    \includegraphics[width=0.9\textwidth]{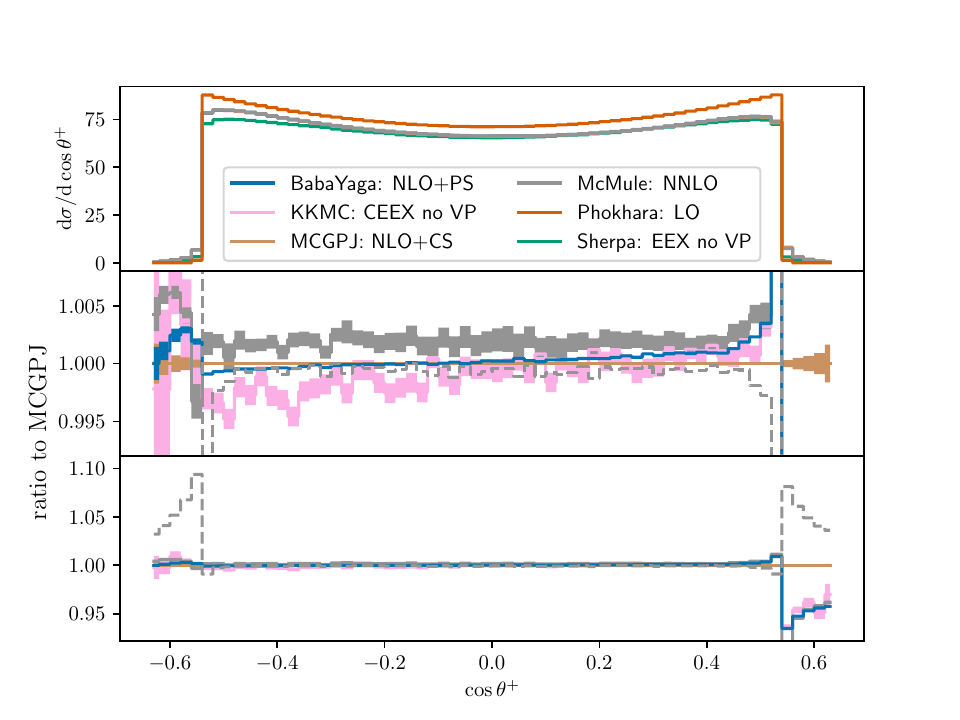}
    \caption{Distribution of the $\mu^+$ angle in the CMD-like scenario for $e^+\,e^-\to\mu^+\,\mu^-$. The dashed \mcmule{} line shows the fixed-order NLO result.}
    \label{fig:CMD-cth}
\end{figure}

There are several codes that provide results for $e^+\,e^-\to\mu^+\,\mu^-$ and we depict their results for the 
$\cos\theta^+$ distributions in the top panel of \figref{fig:CMD-cth}. Since \phokhara{} is not intended for $2~\to~2$ processes, we use its LO result to illustrate the effect of higher-order corrections. If mixed corrections are neglected, i.e., for LO and the \Sherpa{} result, the distribution is symmetric w.r.t.\ $\cos\theta^+=0$. The middle panel shows the ratio of the more specialised codes to the MCGPJ result. KKMC and \Sherpa{} do not include VPC, neither leptonic nor hadronic. However, at this particular energy it happens that the impact of VPC at NLO is minimal. This is due to the near vanishing of $\Pi(\sqrt{s}=0.7\,\GeV)$, see \figref{fig:vpcomp}. Hence, a comparison of KKMC with codes including VPC is still meaningful. In the bulk of the distribution where the LO result does not vanish, there is an agreement within less than 0.5\% of all specialised codes. The difference between the grey dashed line (\mcmule{} NLO) and the grey band (\mcmule{} NNLO) indicates the impact of full NNLO corrections. At NNLO, VPC are not smaller any longer than photonic corrections. In fact, purely photonic NNLO corrections are negative for most values of $\cos\theta^+$. Only by adding NNLO terms with a VP insertion (which amounts to an effect of about 0.5\%) the result with a positive total NNLO effect shown as the grey band is obtained. The full NNLO corrections amount to a few permille in the bulk, but are much larger at the boundaries. This is visible in the lower panel (a zoom out of the middle panel), with the NLO far off the parton-shower improved results at the edges of the distribution. However, the NNLO calculation (which in fact is only accurate at NLO in this region) reproduces the bulk of the parton shower. This is an indication that the parton shower is dominated by one additional emission. 

\subsubsection{Pion final state} \label{sec:CMDscP}

\begin{figure}[bt!]
    \centering
    \includegraphics[width=0.9\textwidth]{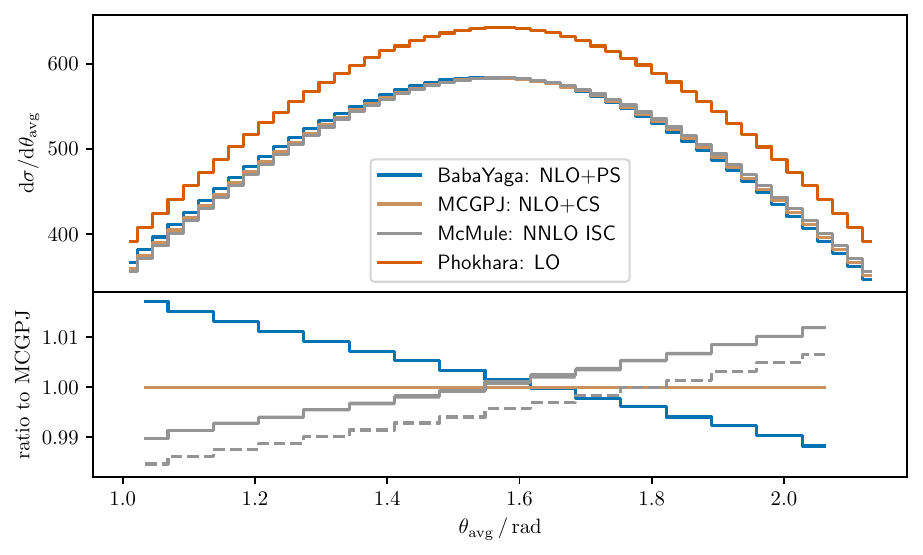}
    \caption{Distribution of $\theta_\text{avg}$ in the CMD-like scenario for $e^+\,e^-\to\pi^+\,\pi^-$. The dashed \mcmule{} line shows the NLO result (with ISC only).}
    \label{fig:CMDpp-thav}
\bigskip
    \centering
    \includegraphics[width=0.9\textwidth]{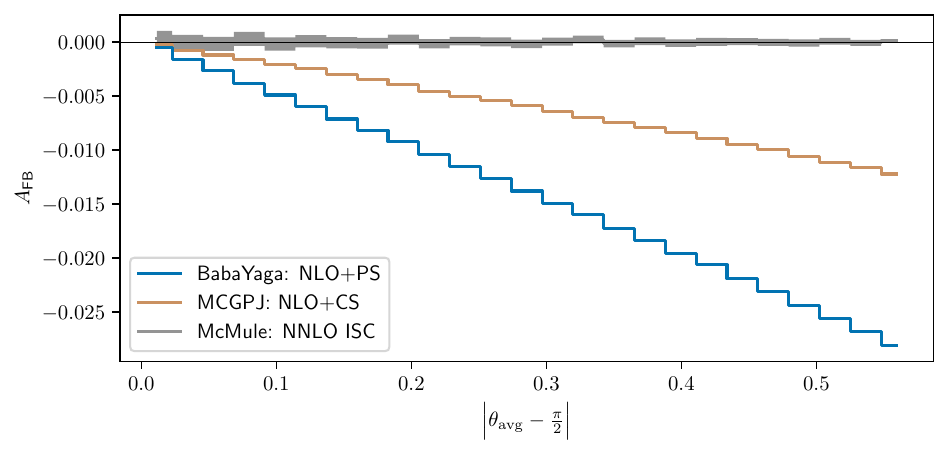}
    \caption{The asymmetry w.r.t.\ $\theta_\text{avg}$ in the CMD-like scenario for $e^+\,e^-\to\pi^+\,\pi^-$. }
    \label{fig:CMDpp-asym}
\end{figure}

The reliability of theoretical predictions for $e^+\,e^-\to\pi^+\,\pi^-$ is unfortunately not at the same level as for $e^+\,e^-\to\mu^+\,\mu^-$. This is related to the difficulties of treating radiation off pions, as described in \secref{sec:comp-fshad}. The two codes that provide a full NLO result combined with additional radiation -- MCGPJ and {\sc BabaYaga@NLO} --  differ by more than 1\%. This is shown in \figref{fig:CMDpp-thav} where we take $\theta_\text{avg}$ as an example. Again, we misuse \phokhara{} to provide a LO distribution to indicate the size of higher-order corrections. \mcmule{} includes NNLO corrections, but only ISC. Hence, its $\theta_\text{avg}$ distribution is symmetric w.r.t.\ $\theta_\text{avg}=\pi/2$. This is particularly evident in \figref{fig:CMDpp-asym} that shows the forward-backward asymmetry w.r.t.\ $\theta_\text{avg}$
\begin{align}\label{eq:CMDasym}
A_\text{FB}&= 
\frac{\frac{\dd\sigma}{\dd\theta_\text{avg}}\big(\theta_\text{avg}>\frac{\pi}{2}\big)
-\frac{\dd\sigma}{\dd\theta_\text{avg}}\big(\theta_\text{avg}<\frac{\pi}{2}\big)}{\frac{\dd\sigma}{\dd\theta_\text{avg}}\big(\theta_\text{avg}>\frac{\pi}{2}\big)+\frac{\dd\sigma}{\dd\theta_\text{avg}}\big(\theta_\text{avg}<\frac{\pi}{2}\big)}\, .
\end{align}
Both full NLO codes produce an asymmetry, as expected, but they differ substantially. This is related to the different treatment of 2PE contributions. MCGPJ uses GVMD, whereas the {\sc BabaYaga@NLO} results displayed here are obtained with F$\times$sQED. As discussed in \secref{sec:comp-eepipi}, the asymmetry has been studied extensively, also in the FsQED approach. While this leads to a reasonable agreement with GVMD, there are notable differences w.r.t.\ sQED. A recent comparison can also be found in~\cite{Budassi:2024whw}. We have to conclude that in the current codes for fully differential distributions of the process $e^+\,e^-\to\pi^+\,\pi^-$, there are similar aspects as for the asymmetry. Hence, different options of treating the 2PE contributions lead to a difference above the percent level. As we will see, these problems get more severe in the case of processes with additional photon radiation.

\subsubsection{Electron final state} \label{sec:CMDscE}

\begin{figure}[b!]
    \centering
    \includegraphics[width=0.9\textwidth]{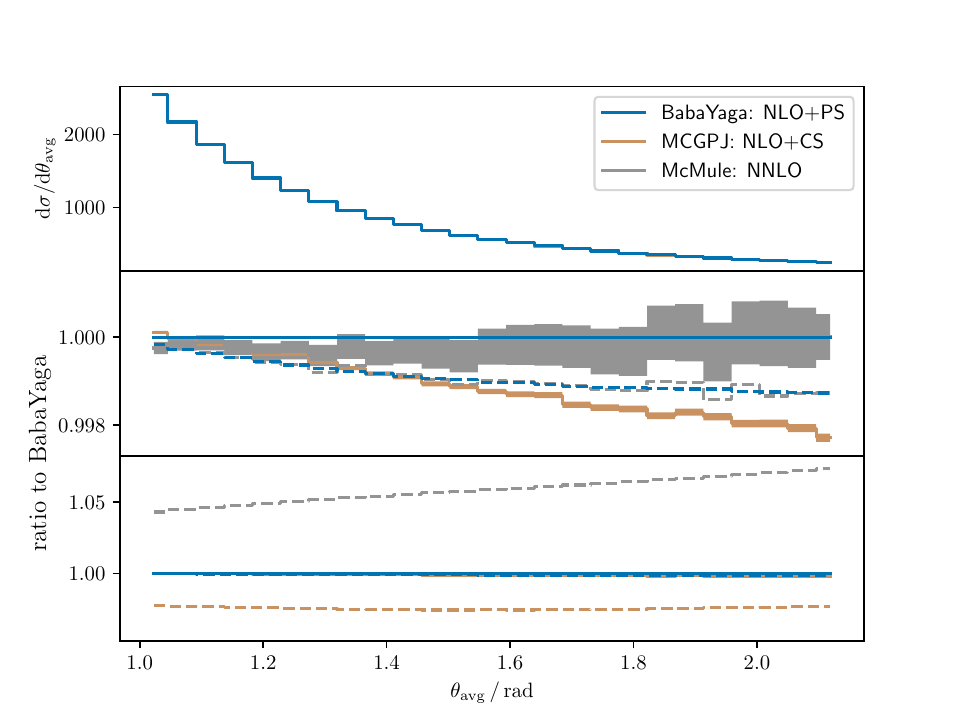}
    \caption{Distribution of $\theta_\text{avg}$ in the CMD-like scenario for $e^+\,e^-\to e^+\,e^-$. In the lower panel we show the LO result (dashed \mcmule{} line) and the NLO+CS result without VP (dashed MCGPJ line). In the middle panel the dashed lines show the full NLO result. }
    \label{fig:CMDee-thav}
\end{figure}

The situation for Bhabha scattering $e^+\,e^-\to e^+\,e^-$ is similar to muon pair production, but there are fewer codes that have implemented this process. Still, we have a full NNLO calculation of \mcmule{} and two complete NLO calculations with additional radiation from MCGPJ and {\sc BabaYaga@NLO}.

\begin{figure}[tb]
    \centering
    \includegraphics[width=0.9\textwidth]{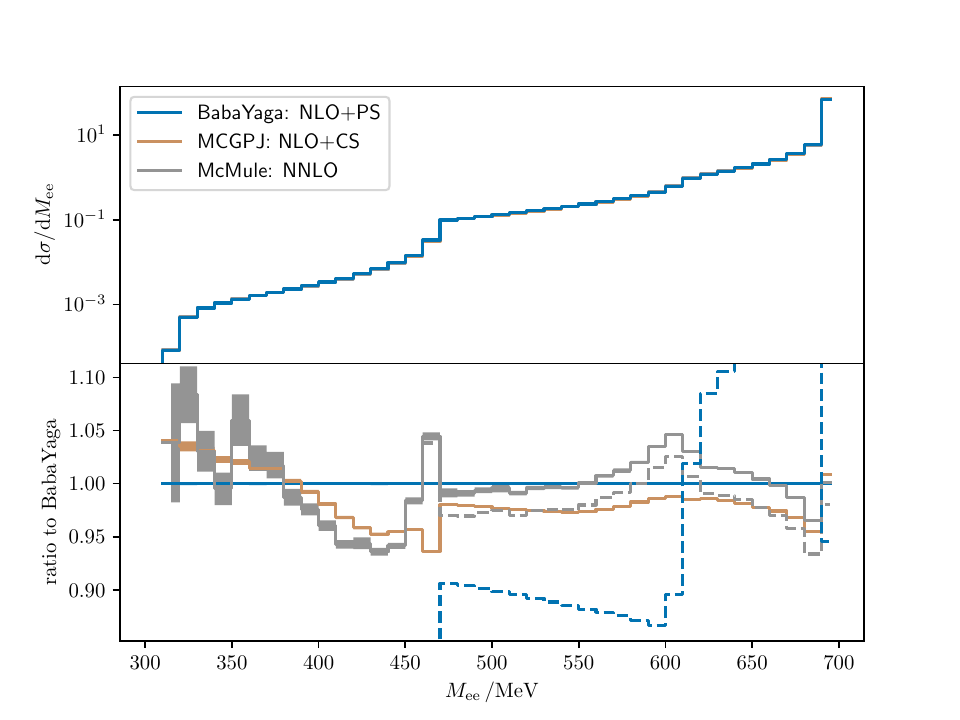}
    \caption{Distribution of the invariant mass of $M_{ee}$ in the CMD-like scenario for $e^+\,e^-\to e^+\,e^-$. In the lower panel we also show the NLO result (dashed {\sc BabaYaga@NLO} line) as well as the NNLO result without VP contributions (dotted \mcmule{} line).}
    \label{fig:CMDee-mee}
\end{figure}

We first consider an observable that is non-vanishing in the full range, namely $\theta_\text{avg}$. The results of \figref{fig:CMDee-thav} show an excellent agreement between the three codes. From the lower panel we can infer that NLO corrections are of the order of 5\% whereas VPC at NLO amount to $1-2$\%. Thus, in contrast to $e^+\,e^-\to \mu^+\,\mu^-$, VPC are not negligible in this case. The $t$-channel VPC are not affected by the accidental suppression mentioned in \secref{sec:CMDscM}. In the middle panel we zoom in to an accuracy of a few permille. First, there is perfect agreement between the NLO results of {\sc BabaYaga@NLO} and \mcmule{}, depicted as dashed lines. The full NNLO corrections of \mcmule{} are of the order of 0.1\% and are almost perfectly reproduced by the additional parton-shower emission in {\sc BabaYaga@NLO}. The MCGPJ result also agrees within 0.2\%. However, compared to the NLO result, it has a different sign. MCGPJ includes the interference between ISR and FSR only at NLO, whereas {\sc BabaYaga@NLO} resums them.
Still, we can conclude that for well-behaved observables, non-vanishing at LO, the specialised codes for Bhabha scattering produce results with an error well below 0.5\%.

The situation is more complicated for observables with a restricted range at LO. As an example we consider the invariant mass of the two final-state electrons, $M_{ee}$. At tree level, the distribution is a delta peak at $M_{ee}=\sqrt{s}$. Higher-order corrections produce a tail. Thus, the NLO computation produces a LO result for $M_{ee} < \sqrt{s}$. This is illustrated in \figref{fig:CMDee-mee}, where the bottom panel shows that the NLO result (dashed blue line) deviates up to 10\% for moderately large values of $M_{ee}$ w.r.t.\ the parton-shower improved results. For smaller values of $M_{ee}$, the NLO result is completely unreliable. However, the NNLO computation reproduces the parton shower for the whole range of $M_{ee}$ within a few percent. The difference between the additional radiation of {\sc BabaYaga@NLO} and MCGPJ is of similar size. A theoretical description of $M_{ee}$ over the whole range at the percent level would require to combine full NNLO corrections with a parton shower. The need of a complete NNLO computation is also confirmed by the impact of NNLO VPC (indicated by the grey dashed line) that are of the order of 2\% in the tail.

\subsection{KLOE-like small-angle scenario} \label{sec:KLOEsasc}

The KLOE-like small-angle scenario considers the radiative processes $e^+\,e^-\to X^+\,X^-\,\gamma$ where, however, the photon is not directly detected. The angle $\theta_{\widetilde\gamma}$ associated with the 'untagged' photon momentum $\vec{p}_{\widetilde\gamma}\equiv -(\vec{p}_{+}+\vec{p}_{-})$ is assumed to be small w.r.t.\ the beam. The energy is set to $\sqrt{s}=1.02\,\GeV$.

More precisely, the cuts we apply are
\begin{subequations}
\begin{align} \label{eq:kloesa-cuts}
& \theta_{\widetilde\gamma}\le 15^\circ \quad \mbox{or} \quad \theta_{\widetilde\gamma}>165^\circ \, ,\\
& 0.35\,\GeV^2 \le M_{XX}^2 \le 0.95\,\GeV^2\, , \\
& 50^\circ\le\theta^\pm\le 130^\circ\, , \\
& |\vp^z_\pm| > 90 \, \MeV \quad \mbox{or} \quad \vp^\perp_\pm > 160 \, \MeV \, ,
\end{align}
\end{subequations}
where $\vp^z_\pm$ and $\vp^\perp_\pm$ denote the $z$ and transverse components of the charged final-state particles. In the following subsections we give an example for the cases $X\in\{\mu,\pi\}$ .

\subsubsection{Muon final state} \label{sec:KLOESAscM}

\begin{figure}[tb]
    \centering
    \includegraphics[width=0.9\textwidth]{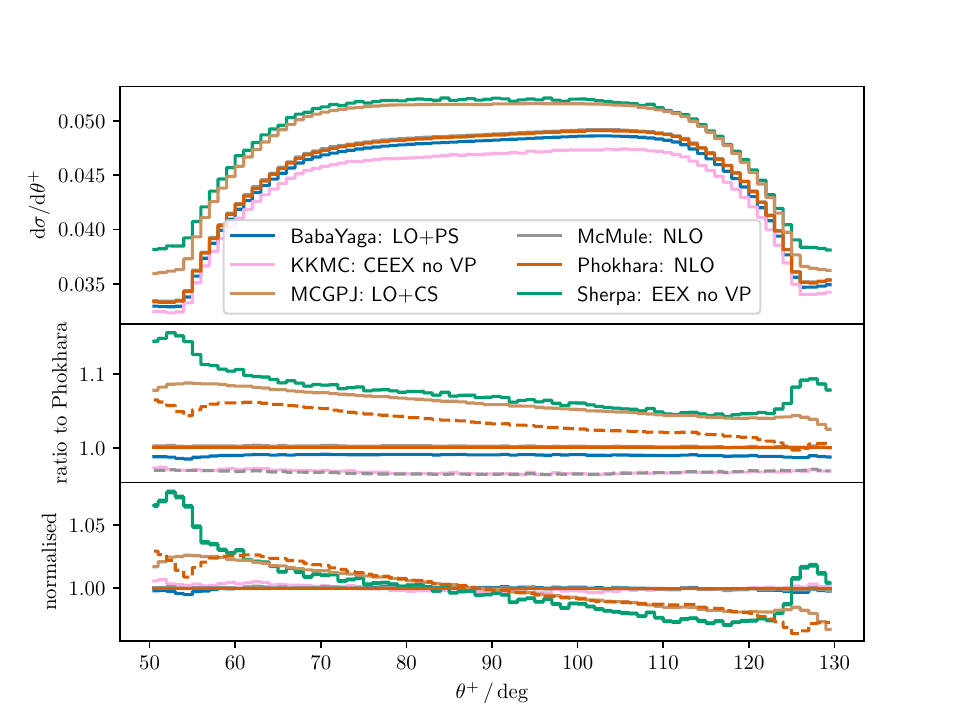}
    \caption{Distribution of the $\mu^+$ angle in the KLOE-like small-angle scenario for $e^+\,e^-\to\mu^+\,\mu^-\,\gamma$. The dashed \phokhara{} line is the LO result, the dashed \mcmule{} line is the NLO result without VPC. The bottom panel shows the ratio of the normalised distributions. }
    \label{fig:KLOESA-sth}
\end{figure}

In \figref{fig:KLOESA-sth} we show the $\theta^+$ distribution results for the process $e^+\,e^-\to\mu^+\,\mu^-\,\gamma$ of six codes. We do not include the \afkqed{} result, because the LO photon generated with the exact amplitude is almost never the main ISR photon after selection cuts and it lacks soft and virtual corrections, which would be needed in this case. In addition to the distribution and the ratios to \phokhara{} (middle panel) we also show the normalised distributions in the lower panel to focus on the shape. Indeed, at LO the $\theta^+$ distribution is symmetric w.r.t.\ $\theta^+=90^\circ$, but interference effects beyond LO induce an asymmetry. The two NLO calculations, \mcmule{} and \phokhara{} agree well, with minor differences due to the treatment of VPC. {\sc BabaYaga@NLO} and MCGPJ both include collinear effects using parton showers  on top of an LO calculation. However, in the case of MCGPJ, this does not include the angular distribution of the extra photons leading to the differences between the two codes. We note that neither \babayaga{} nor MCGPJ are designed to be used for this process. KKMC and \Sherpa{} both resum soft emissions to all orders with YFS, the former includes higher-order perturbative corrections using the CEEX formalism and the latter with EEX. There is a significant difference between the two due to the coherence effects that are included in CEEX. In the normalised ratio plot, KKMC agrees well with the NLO codes and {\sc BabaYaga@NLO}, whereas \Sherpa{} and MCGPJ have the same shape as the LO result (dashed \phokhara{} line). Also the middle panel clearly illustrates the difference in the shapes. KKMC (with currently no VPC implemented) agrees very well with the NLO result without VPC (dashed \mcmule{} line). Since VPC amount to roughly 5\% their inclusion is essential.

\subsubsection{Pion final state} \label{sec:KLOESAscP}

\begin{figure}[tb]
    \centering
    \includegraphics[width=0.9\textwidth]{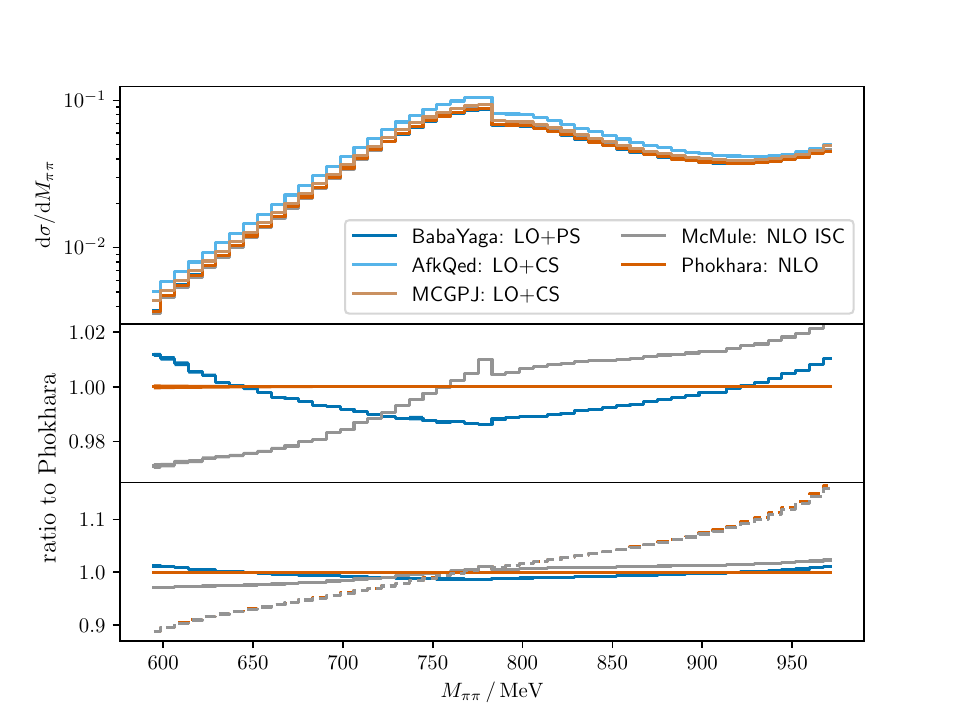}
    \caption{Distribution of $M_{\pi\pi}$ in the KLOE-like small-angle scenario for $e^+\,e^-\to\pi^+\,\pi^-\,\gamma$. The dashed lines are the LO results, with (\phokhara{}) and without (\mcmule{}) final-state photon emission. }
    \label{fig:KLOESA-mxx}
\end{figure}

For the process  $e^+\,e^-\to\pi^+\,\pi^-\,\gamma$ we consider the invariant-mass distribution $M_{\pi\pi}$ with the results shown in \figref{fig:KLOESA-mxx}. Also in this case the \afkqed{} result suffers from the problem of the photon generation being incompatible with the selection cuts. MCGPJ is also shown for illustration only as it is not designed for radiative processes. \mcmule{} includes ISC at NLO, but does not have any additional radiation off the pions. At LO, the neglect of FSC has minimal impact in this scenario, as can be seen from the comparison of the (overlapping dashed) \phokhara{} and \mcmule{} LO results in the bottom panel. However, at NLO there is a significant difference of the order of 2\%. The difference between the parton-shower improved LO results of {\sc BabaYaga@NLO} and \phokhara{} are of similar size. Hence, to obtain reliable results with an error at or below the percent level, it is imperative to include photon radiation also from the final state pions. Given that currently there is no code that includes the structure-dependent radiative corrections (see \secref{sec:comp-eepipi}) in a model-independent way, it is prudent to assume a sizeable inherent uncertainty in the NLO corrections  of this process.

\subsection{KLOE-like large-angle scenario} \label{sec:KLOElasc}

This scenario with $\sqrt{s}=1.02\,\GeV$ assumes that at least one photon with energy $E_\gamma > 20\,\MeV$ is detected. Thus, we are dealing with radiative processes $e^+\,e^-\to X^+\,X^-\,\gamma$ and the acceptance cuts are
\begin{subequations}\label{mcc:cutKLOE-la}
\begin{align}
&E_\gamma > 20\,\MeV& &\mbox{and}& 
&50^\circ\le\theta_\gamma\le 130^\circ\, , & \label{mcc:cutKLOE-la-y} 
\\
&|\vp^z_\pm| > 90\,\MeV \  \mbox{or}\  \vp^\perp_\pm > 160\,\MeV&
&\mbox{and}& &50^\circ\le\theta^\pm\le 130^\circ\, , & 
\\
&0.1\,\GeV^2 \le M_{XX}^2 \le 0.85\,\GeV^2\, , & 
\end{align}
\end{subequations}
where \eqref{mcc:cutKLOE-la-y} is to be understood that there is at least one such photon.

\subsubsection{Muon final state} \label{sec:KLAscM}

\begin{figure}[tb]
    \centering
    \includegraphics[width=0.9\textwidth]{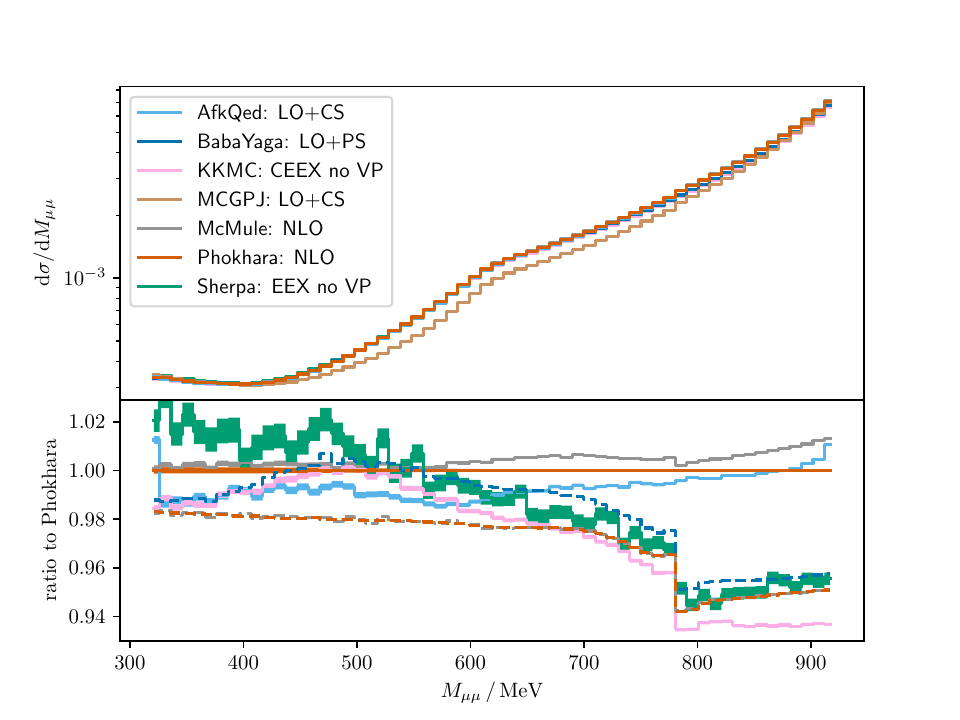}
    \caption{Distribution of $M_{\mu\mu}$ in the KLOE-like large-angle scenario for $e^+\,e^-\to\mu^+\,\mu^-\,\gamma$. The dashed {\sc BabaYaga@NLO} line shows the LO+PS result with leptonic VP, but no HVP contributions. The (overlapping) dashed \phokhara{} and \mcmule{} lines show the NLO result without any VPC.}
    \label{fig:KLA-mxx}
\end{figure}

In the upper panel of \figref{fig:KLA-mxx} we show the invariant-mass distribution for all seven codes, even though MCGPJ is not designed to be used for radiative processes. The remaining six codes -- including \babayaga{} which is also not designed for this process -- show reasonable agreement within a few percent. The main difference is due to the lack of VPC in KKMC and \Sherpa. To illustrate the impact of VPC we show the perfectly agreeing \phokhara{} and \mcmule{} NLO results without any VPC as dashed lines in the lower panel. The dashed {\sc BabaYaga@NLO} line includes leptonic but not hadronic VPC. We can infer that VPC amount to several percent and their inclusion is mandatory. The full NLO \phokhara{} and \mcmule{} results differ by up to 1\%, due to the different inclusion of VPC. \mcmule{} implements them strictly at NLO, whereas \phokhara{} performs a Dyson resummation of
them (see~\eqref{eq:mccVPresum}). The difference is formally at NNLO. Hence, if an accuracy better than 1\% is required, NNLO corrections are needed. The KKMC result does not include VPC and agrees within $1-2$\% with the corresponding \phokhara{} and \mcmule{} results. Also {\sc BabaYaga@NLO} with VPC agrees within about 2\% with the full NLO calculations. In \afkqed{}, \Photos{} for NLO final-state radiation is switched off, so that the LO photon is always detected as the main ISR photon after selection cuts. The cross section is dominated by the region $M_{\mu\mu} \gtrsim 800\,\MeV$, where the differences between the various codes are about 1\%. However, typically the full distribution is required and for smaller values of $M_{\mu\mu}$ the differences are larger.  

\subsubsection{Pion final state} \label{sec:KLAscP}

\begin{figure}[tb]
    \centering
    \includegraphics[width=0.9\textwidth]{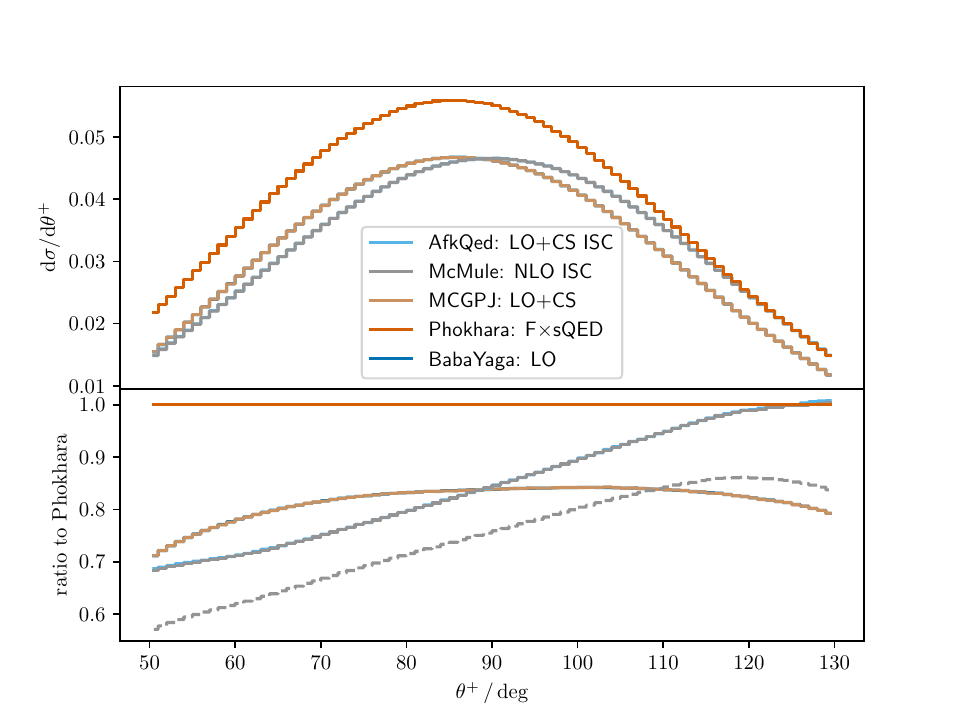}
    \caption{Distribution of $\theta^+$ in the KLOE-like large-angle scenario for $e^+\,e^-\to\pi^+\,\pi^-\,\gamma$. The dashed \mcmule{} line shows the LO result.}
    \label{fig:KLA-lth}
\end{figure}

Contrary to previous scenarios for the process $e^+\,e^-\to\pi^+\,\pi^-\,\gamma$, in the KLOE-like large-angle scenario FSC are very large. This is illustrated in \figref{fig:KLA-lth}, where we show the $\theta^+$ distribution.  At LO, \phokhara{}, {\sc BabaYaga@NLO}, and MCGPJ contain the full $2\to{3}$ matrix element using F$\times$sQED and agree perfectly. \mcmule{} and \afkqed{} only contain initial-state radiation of photons, as \Photos{} is switched off in \afkqed{} for this scenario. The neglect of FSC at LO leads to an error more than 10\%, as can be seen by comparing the dashed \mcmule{} line with the (overlapping) {\sc BabaYaga@NLO} and MCGPJ lines in the lower panel of \figref{fig:KLA-lth}. 

The effect is enhanced at NLO. \afkqed{} and \mcmule{} only contain ISC and initial-state radiation. Within this approach, the results of these two codes agree as can be seen by the (overlapping) lines in \figref{fig:KLA-lth}. 
\phokhara{} on the other hand has a full NLO result using F$\times$sQED. The FSC and mixed corrections included in \phokhara{} lead to differences of up to 30\%, in particular for small values of $\theta^+$. 
The large effect of FSC and mixed corrections in the KLOE-like large-angle scenario is also confirmed for the process $e^+\,e^-\to\mu^+\,\mu^-\,\gamma$, where they can be computed reliably in QED. Hence, it is clear that a reliable theoretical prediction of this scenario has to include FSC and mixed corrections in a solid framework. In this respect, the lack of a coherent description of structure-dependent radiative corrections in the currently available codes is concerning. 
 
\subsection{BESIII-like scenario} \label{sec:BesIIIsc}

In this scenario we deal with radiative processes at an energy of $\sqrt{s}=4\,\GeV$. In order to detect a charged particle $X^\pm$ or photon, they have to pass the selection cuts 
\begin{subequations}
\begin{align}\label{mcc:bescut}
&|\cos \theta^{\pm}|<0.93& &\mbox{and}&  &\vp_{\pm}^\perp >300\,\MeV\, ,& \\
& \big(|\cos\theta_\gamma| < 0.8 \ \mbox{and}\ E_\gamma > 25\,\MeV\big)&  &\mbox{or}& 
&\big(0.86 < |\cos \theta_\gamma| < 0.92\ \mbox{and} \ E_\gamma > 50\,\MeV\big)
\end{align}
\end{subequations}
In addition, we require that precisely one such photon has $E_\gamma \ge 400\,\MeV$.

MCGPJ is not meant to be used for this scenario and is only shown for illustration.

\subsubsection{Muon final state} \label{sec:BESscM}

\begin{figure}[bt]
    \centering
    \includegraphics[width=0.9\textwidth]{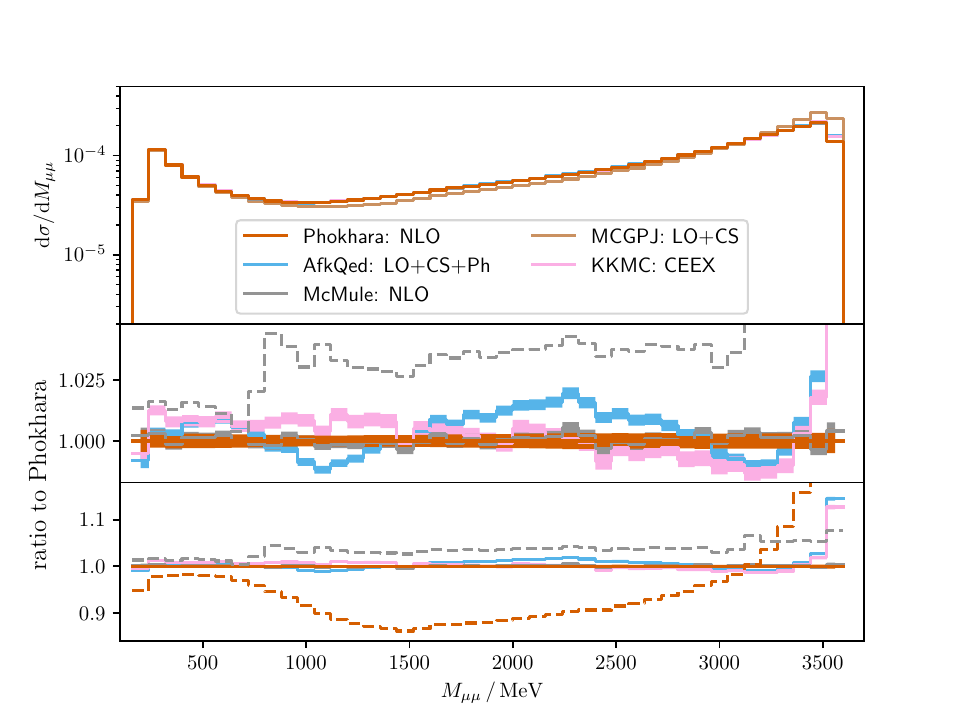}
    \caption{Distribution of $M_{\mu\mu}$ in the BESIII-like scenario for $e^+\,e^-\to\mu^+\,\mu^-\,\gamma$ without VP contributions. For comparison, the LO (dashed \phokhara{} line) and NLO with VP (dashed \mcmule{} line) results are also shown.}
    \label{fig:BES-mxx}
\end{figure}

For the process with a muon pair in the final state we present a comparison of the codes for the results of the invariant-mass distribution $M_{\mu\mu}$ without VPC. This is to be understood as a technical comparison. Indeed, as is evident from the grey dashed lines in the lower two panels of \figref{fig:BES-mxx}, the VPC are larger than 2\% and, therefore, not negligible. The NLO corrections are larger than 10\% in some regions of the distribution. Still, the NLO results (without VPC) of \phokhara{}, \afkqed{}, KKMC, and \mcmule{} agree within about 1\%. As for previous scenarios, a theoretical description of $e^+\,e^-\to\mu^+\,\mu^-\,\gamma$ with an accuracy better than 1\% requires the inclusion of NNLO corrections in a systematic way. 

\subsubsection{Pion final state} \label{sec:BESscP}

\begin{figure}[tb]
    \centering
    \includegraphics[width=0.9\textwidth]{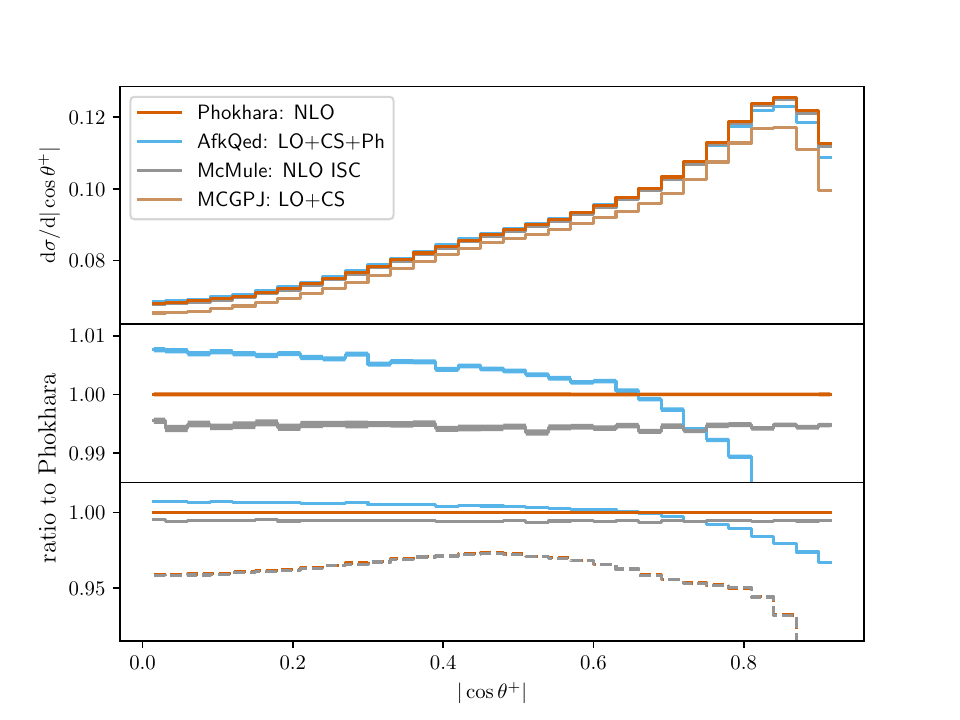}
    \caption{Distribution of $\cos\theta^+$ in the BESIII-like scenario for $e^+\,e^-\to\pi^+\,\pi^-\,\gamma$. The (overlapping) dashed \phokhara{} and \mcmule{} lines show the LO results with and without final-state radiation.}
    \label{fig:BES-cth}
\end{figure}

As for the KLOE-like scenarios, the computation of $e^+\,e^-\to\pi^+\,\pi^-\,\gamma$ leads to the issue of how to include radiation off the final-state pions. However, in the BESIII-like scenario, the ISC are dominant. In particular, the LO results for the $|\cos\theta^+|$ distribution of \phokhara{} and \mcmule{} agree within 0.1\%, as shown by the dashed lines in the lowest panel of \figref{fig:BES-cth}, even though \mcmule{} only includes ISC, whereas \phokhara{} has a complete computation using F$\times$sQED. Part of this suppression of FSR is due to the enhancement of initial-state collinear emission, which can also be noted in the process with a muon pair in the final state. However, a much larger effect is the form-factor suppression. In the case of FSR, the VFF is evaluated at $q^2=s$, whereas in the case of ISR, the typical $q^2$ is considerably smaller. The sharp fall of $F_\pi(q^2)$ with increasing $q^2$ therefore leads to a strong dominance of ISR relative to FSR.

The NLO corrections amount to roughly 5\%. Again, \phokhara{} presents a full NLO result in the F$\times$sQED framework. \mcmule{} only includes ISC, whereas \afkqed{} combines complete ISC with FSC implemented through \Photos. The impact of FSC is larger for the NLO cross section. But the various results agree to within 1\%, except for large values of $|\cos\theta^+|$. It appears that the collinear approximation used by \afkqed{} for ISC is leading to an underestimation of events in this region.

\subsection[\texorpdfstring{$B$}{B} scenario]{\boldmath $B$ scenario} \label{sec:Bsc}

This scenario is inspired by $B$ factories. We consider the radiative processes $e^+\,e^-\to X^+\,X^-\,\gamma$ with   $X\in\{e,\mu,\pi\}$   at $\sqrt{s}=10\,\GeV$ with symmetric beams. Consequently, all selection cuts are understood to be in the centre-of-mass frame. In order to be detected, we require for the charged particles and photons
\begin{subequations}
\begin{align}\label{eq:Bcut}
&0.65\,\rad\le\theta^\pm\le 2.75\,\rad&  &\mbox{and}& &\vp_\pm > 1\,\GeV\, ,& \\
&0.6\,\rad\le\theta_\gamma\le 2.7\,\rad&  &\mbox{and}& &E_\gamma>3\,\GeV\, .& \label{eq:Bcuty}
\end{align}
Furthermore, denoting the most energetic photon passing the cut \eqref{eq:Bcuty} by $\gamma^{(h)}$ and introducing $M_{XX\gamma}^2\equiv(p_+ + p_- + p_{\gamma^{(h)}})^2$ we require
\begin{align}
&\theta_{\gamma^{(h)},\widetilde\gamma} = 
\sphericalangle(\vec{p}_{\gamma^{(h)}},\vec{p}_{\widetilde\gamma}) < 0.3\,\rad&  
&\mbox{and}& &M_{XX\gamma} > 8\,\GeV\, .& \label{eq:Bcuth}
\end{align}
The second cut in \eqref{eq:Bcuth} is to suppress secondary photons. In the case of  $X=e$ we also demand 
\begin{align}
 M_{ee}>0.3\,\GeV
\end{align}
while for $X\in\{\mu,\pi\}$ no cut on the invariant mass of the charged final-state pair is made.
\end{subequations}

\subsubsection{Muon final state} \label{sec:BscM}

\begin{figure}[tb]
    \centering
    \includegraphics[width=0.9\textwidth]{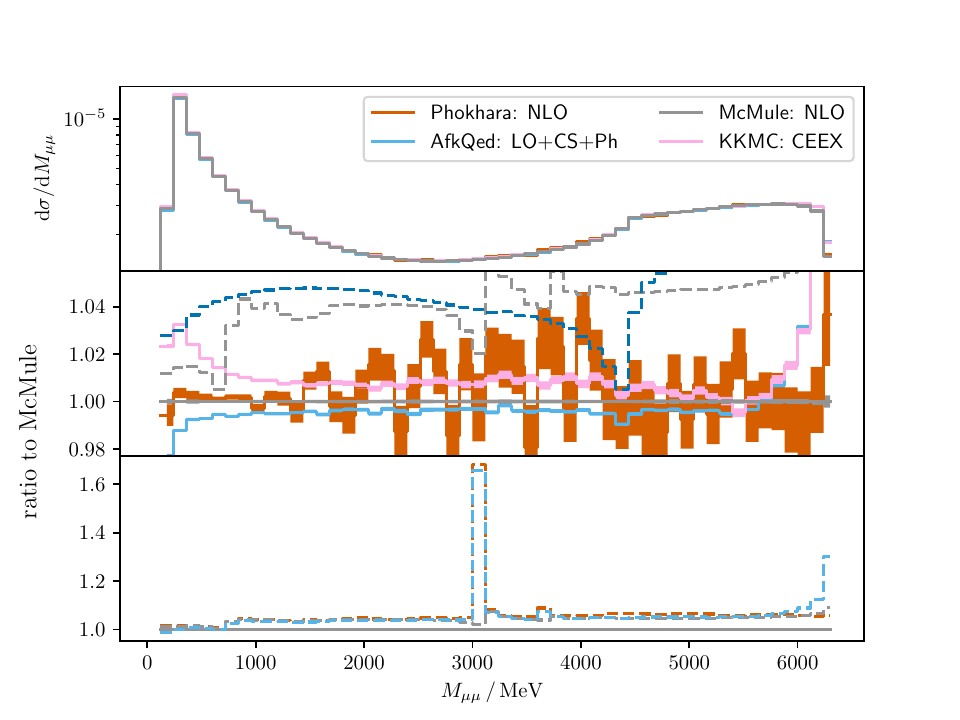}
    \caption{Distribution of $M_{\mu\mu}$ in the $B$ scenario for $e^+\,e^-\to\mu^+\,\mu^-\,\gamma$ without VPC. For comparison, the LO (dashed blue line) and NLO with VP (dashed \mcmule{} line) are shown in the middle panel. In the bottom panel we show the results with VPC as dashed lines.}
    \label{fig:Bmm-mxx}
\end{figure}

In \figref{fig:Bmm-mxx} we show the $M_{\mu\mu}$ distribution for $e^+\,e^-\to\mu^+\,\mu^-\,\gamma$. In the main plot we drop the VPC to perform a technical comparison of photonic corrections in various approaches. In fact, the VPC are about as large as the full NLO corrections and, hence, cannot be neglected at all. This is evident from the dashed lines in the middle panel that show the LO (blue) and NLO with VPC (grey) results. The full NLO results of \phokhara{} and \mcmule{} agree perfectly. Since the \phokhara{} results have rather large statistical fluctuations we show the ratio w.r.t. \mcmule{} in the lower panels. As for the BESIII-like scenario, \afkqed{} includes the first initial-state emission exactly and combines this with the collinear approximation for further ISC and \Photos{} for FSC. This agrees within 1\% with the exact NLO result, except at the very high end of the tail, where multiple emission of photons might be particularly relevant. Indeed, the \afkqed{} result there agrees well with KKMC, which also includes multiple emission through CEEX. The cross section is dominated by small values of $M_{\mu\mu}$, where there are differences larger than 2\% between KKMC and the other codes. These differences are as large as the full photonic NLO corrections.

In the bottom panel we point out the huge impact of the VPC in the region of the $J/\psi$ resonance. The \mcmule{} implementation is strictly fixed order NLO, i.e., with one insertion of $\Pi$ in the intermediate photon line. \phokhara{} and \afkqed{} resum multiple $\Pi$ insertions through a modified photon propagator, as discussed in \secref{sec:mccVP}. Formally, the difference is NNLO, but due to the resonance structure in $\Pi_h$, these differences are numerically extremely large. This region is expected to receive large NNLO corrections also due to the sensitivity to additional soft photon emission.

\subsubsection{Pion final state} \label{sec:BscP}

\begin{figure}[tb]
    \centering
    \includegraphics[width=0.9\textwidth]{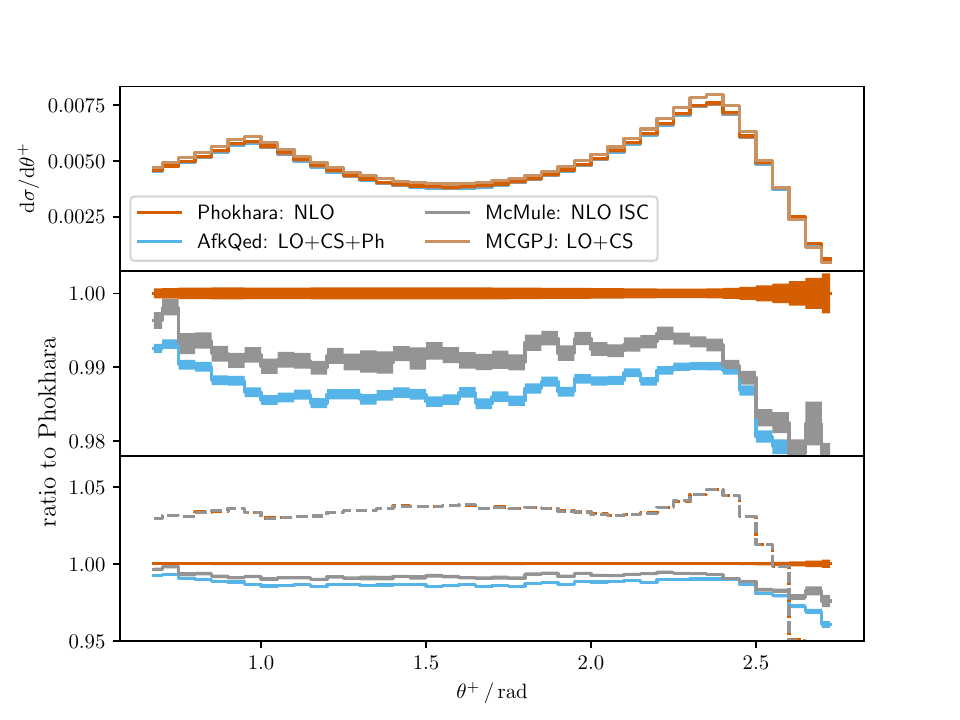}
    \caption{Distribution of $\theta^+$ in the $B$ scenario for $e^+\,e^-\to\pi^+\,\pi^-\,\gamma$. The dashed \phokhara{} and \mcmule{} lines in the lower panel show the LO results, with and without final-state radiation.}
    \label{fig:Bpp-th}
\end{figure}

As for previous scenarios, also in the $B$ scenario we consider the $\theta^+$ distribution for the process $e^+\,e^-\to\pi^+\,\pi^-\,\gamma$. The comparison of the \phokhara{} and \mcmule{} LO results, shown as dashed lines in the lowest panel \figref{fig:Bpp-th}, reveal that also in this case final-state radiation (included in \phokhara{} but not in \mcmule) is strongly suppressed at LO. In fact, the bin-by-bin differences between the LO \mcmule{} and \phokhara{} results are zero within the numerical Monte Carlo error of $\cO(10^{-4})$.

At NLO, the difference between the full F$\times$sQED computation of \phokhara{} and the results of \mcmule{} and \afkqed{} with ISC only (plus \Photos{} for \afkqed) amounts to about 1\%. As for the BESIII-like scenario, this is a sizeable fraction of the complete NLO correction that is about 4\%. As a preliminary conclusion it could be argued that the \phokhara{} result is sufficient for a description at the $1-2$\% level. But if a more precise result is required, once more, an improved treatment of structure-dependent corrections would be beneficial.

\subsubsection{Electron final state} \label{sec:BscE}

\begin{figure}[tb]
    \centering
    \includegraphics[width=0.9\textwidth]{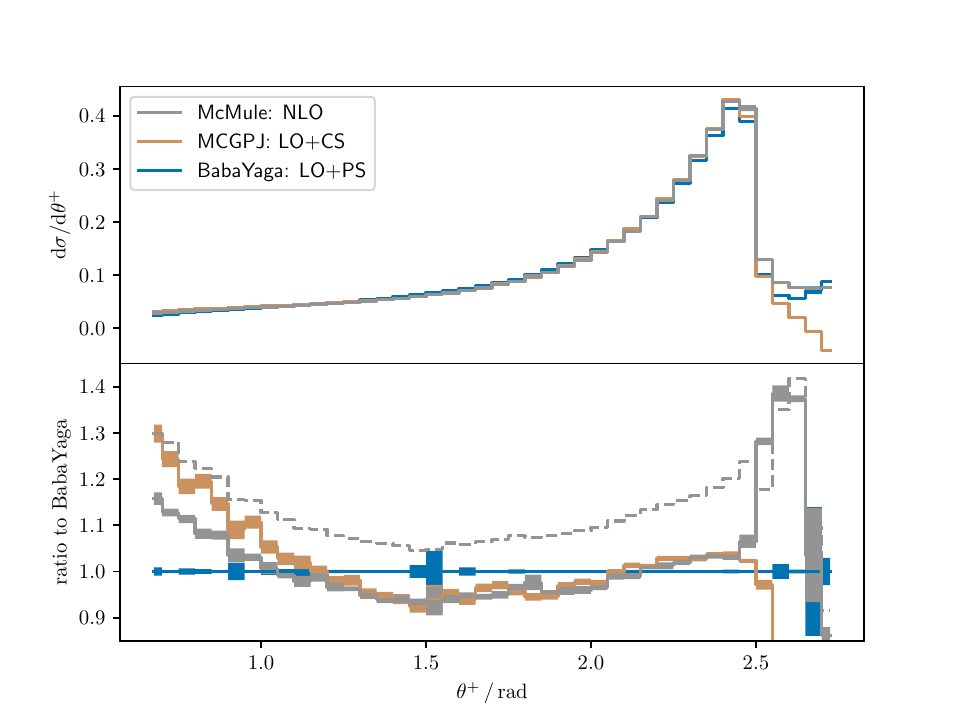}
    \caption{Distribution of $\theta^+$ in the $B$ scenario for $e^+\,e^-\to e^+\, e^-\,\gamma$. The dashed \mcmule{} line shows the LO result.}
    \label{fig:Bee-th}
\end{figure}

As a final example, in \figref{fig:Bee-th} we show results for $e^+\,e^-\to e^+\,e^-\,\gamma$, namely the $\theta^+$ distribution. For this process, currently only \mcmule{} provides a fixed-order NLO result. Comparing the LO and NLO results in the bottom panel reveals NLO corrections of about 10\%. There are also rather large differences compared to the results of MCGPJ and {\sc BabaYaga@NLO}, which provide parton-shower improved LO results. However, it should be kept in mind that these codes have not been designed for radiative processes. Still, a reliable theoretical description of radiative Bhabha scattering at the percent level in this scenario does require to include corrections beyond NLO. 

\section{Outlook}\label{sec:outlook}

The work reported in this document is meant to be Phase~I of an ongoing community effort to improve the theoretical description of scattering processes at electron--positron colliders. We have set up
a framework to coordinate the activities of several groups working on Monte Carlo codes, on hadronic physics, and on experimental aspects related to such processes.  At this stage, we have focused on the core
processes \eqref{intro:scan} and \eqref{intro:return}. We have made comparisons between various codes, either as technical validation, or as an investigation into the impact of particular contributions included in some but not
all of the codes.  All codes that have been used to obtain the results presented in this report are publicly available and further developments are foreseen.  These developments concern improvements for the core processes as well as extensions of the scope.

First, several improvements in perturbative QED computations are feasible with currently available techniques. This includes NNLO QED corrections to ISC, i.e., to $e^+\,e^-\to\gamma\,\gamma^*$ and even to
the full $2\to{3}$ processes $e^+\,e^-\to\mu^+\,\mu^-\,\gamma$ and $e^+\,e^-\to{e}^+\,e^-\,\gamma$. While it is a formidable task to obtain the two-loop integrals with full mass dependence required for the $2\to{3}$ processes, there are well established procedures to obtain sufficiently precise approximations for mass effects at
NNLO. Furthermore, such fixed-order calculations can in principle be combined with dominant effects beyond NNLO through a parton shower or YFS resummation.  Hence, for the pure QED part it is a realistic goal
to achieve a status that removes any doubts about a precision well below the percent level for arbitrary differential distributions. 

The path towards such a precision for processes with pions in the final state is much steeper.  While there is in principle a systematic way forward with a dispersive approach, currently there is no
dispersive study of the $2\pi3\gamma^{(*)}$ amplitude. Hence, at a practical level, there is a need to compromise. This involves the comparison of different approaches and approximations. A broader activity and cross talk between groups working on Monte Carlo implementations and dispersive studies will be instrumental for making further progress.

The improvements mentioned above are closely linked to obtaining not only a good theory description, but also a reliable theory error. This
is a notoriously difficult aspect and there is no unambiguous
procedure. However, with an increased number of approaches and a deeper understanding of their strengths and limitations, a more reliable error estimate is one of the future goals of this effort. 

Concerning additional extensions, an obvious next step is also to
enlarge the list of processes to be considered. In addition to
$e^+\,e^-\to\pi^+\,\pi^-\,\pi^0$ only briefly discussed in this report,
\begin{subequations}
\begin{align}\label{proc:list}
 &e^+\,e^-\to  \pi^+\,\pi^-\,\pi^0\,\pi^0\,, &
 &e^+\,e^-\to \pi^+\,\pi^-\,\pi^+\,\pi^-\,, & \\
 &e^+\,e^-\to \pi^0\,\gamma\,,&  
 &e^+\,e^-\to \pi^+\,\pi^-\,\pi^0\, \gamma\,,& \\
 &e^+\,e^-\to K^+\,K^-\,,&
 &e^+\,e^-\to K_L\,K_S\,,&  \\
 &e^+\,e^-\to \tau^+\,\tau^-\,,&
 &e^+\,e^-\to \gamma\,\gamma\,,&
\end{align}
\end{subequations}
are among the processes of particular interest. So far, only in a small subset of codes these processes have been implemented. Since (most of them) suffer from the complications related to hadronic final states, it would be
useful to extend this activity and contrast different implementations. Again, this is related to the determination of a reliable theory error.

The observables presented in this report are fairly simple in that they only include generic scenarios with some acceptance cuts. However, the Monte Carlo tools provided in connection with this work can and are being used for full experimental analyses. Through this community effort we also offer the possibility for a close collaboration between the experimental collaborations and the developers of the theoretical tools. We hope that this is of mutual benefit and also triggers further developments on the theory side. This will not only advance the physics of low-energy electron--positron collisions, but also have an impact for improved Monte Carlo tools for other experiments, such as lepton--proton scattering, the electron--ion collider, or even a future circular electron--positron collider.

\section*{Acknowledgements}

We thank V.~Druzhinin  and H.~Czy\.z for their help in setting up and running \afkqed{} and \phokhara, respectively. CCC and FP, together with EB and FU, wish to thank G.~Montagna and O.~Nicrosini for the continuous collaboration on the development of {\sc BabaYaga@NLO}. We further thank the Mainz Institute
for Theoretical Physics (MITP) of the DFG Cluster of Excellence
PRISMA$^+$ (Project ID 39083149), for supporting the topical workshop ``The Evaluation of the Leading Hadronic Contribution to the Muon $g-2$: Consolidation of the MUonE Experiment and Recent Developments in Low Energy $e^+e^-$ Data,'' and the University of Zurich for support of the 5th Workstop/Thinkstart "Radiative corrections and Monte Carlo tools for Strong 2020". These meetings were instrumental in moving this project forward.

\subsection*{Monte Carlo contacts}

The following authors are responsible for the codes: PB and LC for \afkqed, CCC and AG for \babayaga, JP and AnS for KKMC, FI for MCGPJ,  SK and MR  for \mcmule, PPR and WTB for \phokhara, and AP and LF for \Sherpa. This does not necessarily mean that they developed the code.

\subsection*{Funding information}

The work in this report is supported by the
Leverhulme Trust LIP-2021-014, the STFC (grants ST/P001246/1, ST/T000988/1, and ST/X000699/1), the DFG through the funds provided to the Sino-German Collaborative Research Center
TRR110 ``Symmetries and the Emergence of Structure in QCD'' (DFG Project-ID 196253076
-- TRR 110),  the SNSF (Project
Nos.\ 200020\_200553, 200020\_207386,  PCEFP2\_181117, PCEFP2\_194272, and TMSGI2\_211209), the Italian Ministero dell'Universit\`a e Ricerca (MUR) and European Union - Next Generation EU through the research grants 2022BCXSW9, 20225X52RA, and 2022ENJMRS under the programme PRIN 2022, the European Union STRONG2020 project under Grant Agreement Number 824093, the
{Polish National Science Centre through the Grants Nos.\ 2019/35/O/ST2/02907 and 2023/50/A/ST2/0022}, 
and by  the Priority Research Area Digiworld under the programme ``Excellence Initiative – Research University'' at the Jagiellonian University in Krakow. We gratefully acknowledge Polish high-performance computing infrastructure PLGrid (HPC Center: ACK Cyfronet AGH) for providing computer facilities and support within the computational grant no. PLG/2024/017001.

\bibliographystyle{JHEP}
\bibliography{strong2020}

\end{document}